\documentclass{aa}

\usepackage{color}
\usepackage{natbib,twoopt}
\usepackage[breaklinks=true]{hyperref} 
\bibpunct{(}{)}{;}{a}{}{,}             
\makeatletter
  \newcommandtwoopt{\citeads}[3][][]{\href{http://adsabs.harvard.edu/abs/#3}%
    {\def\hyper@linkstart##1##2{}%
     \let\hyper@linkend\@empty\citealp[#1][#2]{#3}}}
  \newcommandtwoopt{\citepads}[3][][]{\href{http://adsabs.harvard.edu/abs/#3}%
    {\def\hyper@linkstart##1##2{}%
     \let\hyper@linkend\@empty\citep[#1][#2]{#3}}}
  \newcommandtwoopt{\citetads}[3][][]{\href{http://adsabs.harvard.edu/abs/#3}%
    {\def\hyper@linkstart##1##2{}%
     \let\hyper@linkend\@empty\citet[#1][#2]{#3}}}
  \newcommandtwoopt{\citeyearads}[3][][]%
    {\href{http://adsabs.harvard.edu/abs/#3}
    {\def\hyper@linkstart##1##2{}%
     \let\hyper@linkend\@empty\citeyear[#1][#2]{#3}}}
\makeatother
\usepackage{graphicx}
\usepackage{morefloats}
\usepackage{longtable}
\usepackage{txfonts}
\begin{document} 

 \title{Spectro-interferometric observations of a sample of Be stars\thanks{Based on observations collected at the European Organisation for Astronomical Research in the Southern Hemisphere under ESO programme ID 094.D-0140}}
 \subtitle{Setting limits to the geometry and kinematics of stable Be disks}

   \author{Y.~R. Cochetti\inst{1,2}\fnmsep\thanks{Fellow of CONICET}
          \and
          C. Arcos\inst{3}
          \and
          S. Kanaan\inst{3}
          \and
          A. Meilland\inst{4}
          \and
          L.~S. Cidale\inst{1,2,3}\fnmsep\thanks{Member of the Carrera del Investigador Cient\'ifico, CONICET}
          \and
          M. Cur\'e\inst{3}
          }

   \institute{Instituto de Astrofísica de La Plata, CCT-La Plata, CONICET-UNLP,
              Paseo del Bosque s/n, CP 1900, La Plata, Buenos Aires, Argentina\\
              \email{cochetti@fcaglp.unlp.edu.ar}
              \and
              Departamento de Espectroscopía Estelar, Facultad de Ciencias Astronómicas y Geofísicas, Universidad Nacional de La Plata, Argentina
              \and
              Instituto de Física y Astronomía, Universidad de Valparaíso, Chile
              \and
              Université Côte d'Azur, Laboratoire Lagrange, Nice, 06000, France
             }

   \date{Received ..., 2018; accepted ...., 2018}

\authorrunning{Cochetti et al.}
\titlerunning{Spectro-interferometric observations of a sample of Be stars}
   
 
  \abstract
   {Be stars are rapid rotators surrounded by a gaseous disk envelope whose origin is still under debate. This envelope is responsible for observed emission lines and large infrared excess.}
   {To progress in the understanding of the physical processes involved in the disk formation, we estimate the disk parameters for
a sample of Be stars and search for correlations between these parameters and stellar properties.}
   {We performed spectro-interferometric observations of 26 Be stars in the region of the Br$\gamma$ line to study the kinematical properties of their disks through the Doppler effect. Observations were performed at the Paranal observatory with the VLTI/AMBER interferometer. This instrument provides high spectral (R$\simeq$12000) and high spatial ($\theta_{\rm{min}}=4$~mas) resolutions.}
   {We modeled 18 Be stars with emission in the  Br$\gamma$ line. The disk kinematic is described by a quasi-Keplerian rotation law, with the exception of \object{HD\,28497} that presents a one-arm density-wave structure. Using a combined sample, we derived a mean value for the velocity ratio $\overline{V/V_{\rm{c}}}=0.75$ (where $V_c$ is the critical velocity), and found that rotation axes are probably randomly distributed in the sky. Disk sizes in the line component model are in the range of  $2-13$ stellar radii and do not correlate with the effective temperature or spectral type. However, we found that the maximum size of a stable disk correlates with the rotation velocity at the inner part of the disk and the stellar mass.} 
   {We found that, on average, the Be stars of our combined sample do not rotate at their critical velocity. However, the centrifugal force and mass of the star defines an upper limit size for a stable disk configuration. For a given rotation, high-mass Be stars tend to have more compact disks than their low-mass counterparts. It would be interesting to follow up the evolution of the disk size in variable stars  to better understand the formation and dissipation processes of their circumstellar disks.} 


   \maketitle

\section{Introduction}
Classical Be stars are rapid B-type rotators of luminosity class V to III that have, or had at some time, emission lines of hydrogen (at least in H$\alpha$) and single ionized metals \citep{Jaschek1981,Collins1987}. The formation of these emission lines is attributed to the presence of a circumstellar disk-like envelope surrounding the star. 

Observations suggest the presence of two different regions in the circumstellar envelope of Be stars: a dense equatorial disk in quasi-Keplerian rotation \citep[responsible for most of the emission lines and the IR-excess;][]{Rivinius2013} and a more diluted polar wind \citep[responsible for the highly broadened UV lines;][]{Prinja1989}.

The mass density distribution of Be star disks is often modeled with a simple power law of the form $\rho(r)=\rho_0\,r^{-n}$, where $\rho_0$ is a base density and $n$ is the exponent of the radial coordinate $r$ \citep{Waters1986,Jones2008}.
Typically $\rho_0$ values are  $10^{-9} - 10^{-12}$ g\,cm$^{-3}$ and the $n$ power is about $1.5-3.5$ \citep{Cote1987,Waters1987,Silaj2010,Arcos2017,Vieira2017}.

There is a general consensus about the importance of stellar rotation on the mass ejection process necessary to form a circumstellar disk. \citet{Fremat2005} derived a mean value for the average rotational velocity of Be stars on the order of 88\% of break-up (or critical) velocity, $V_{\rm{c}}$ \footnote{Rotational velocity of a star at which the centrifugal force equals the gravitational force at the equator.}. \citet{Cranmer2005} showed that the hottest Be stars have a uniform spread of intrinsic rotation speed between 40 and 100\% of their $V_{\rm{c}}$ values, while the coolest Be stars are all quasi-critical rotators. \citet{Meilland2012} found a mean velocity ratio of $\overline{V/V_{\rm{c}}}=0.82\pm 0.08$ that corresponds to a rotation rate of $\overline{\Omega/\Omega_{\rm{c}}}=0.95\pm0.02$ \footnote{The rotational rate, $\overline{\Omega/\Omega_{\rm{c}}}$, is the ratio of the stellar angular velocity to its critical one.}. This means that only a small amount of energy would be needed to eject matter from the stellar surface, otherwise, the importance of other mechanisms (e.g., radiation pressure, pulsation, binarity) that could contribute to form the disk should be discussed.

\citet{Zorec2016} studied a set of 233 stars and transformed the distribution of observed projected rotational velocities into another distribution of true rotational velocities.
Since the $V/V_{\rm c}$ values become globally larger as stars evolve from the main sequence (MS) to later evolutionary phases, the distribution of true rotational velocities depends on the evolutionary state of the stellar sample. These authors concluded that once the mode of ratios of the true velocities of Be stars attains the value $V/V_{\rm c}=0.77$ in the MS evolutionary phase, it remains unchanged up to the end of the MS lifespan. The statistical corrections for overestimations of $V\,\sin\,i$, caused by macroturbulent motions and binarity, produce a shift of this distribution toward lower values when all MS evolutionary stages are considered together. Finally, the obtained final distribution has a mode at $V/V_{\rm c}=0.65$ with a nearly symmetric distribution, showing that the Be phenomenon is present in a wide range of true rotational velocities ($0.3<V/V_{\rm c}<0.95$). With this result, the probability that Be stars are critical rotators is very low.

\citet{Meilland2012} proposed that an efficient way to test mass-ejection processes is to constrain the geometry and kinematics of Be stars' circumstellar environments. This can only be performed using long-baseline interferometry with high spectral resolution. First detections of rotational signatures in the envelope of Be stars were carried out by \citet{Mourard1989}. Later, \citet{Meilland2007b} obtained clear evidence of Keplerian rotation in the disk of $\alpha$ Ara. Contemporaneous interferometric observations also provided information on the disk sizes \citep{Tycner2005,Meilland2009,Sigut2015} and flattenings \citep{Quirrenbach1997,Tycner2005}, whilst a clear signature of polar wind was detected in  Achernar by \citet{Kervella2006}. The presence of small-scale structures was found in $\kappa$ CMa \citep{Meilland2007a} and $\zeta$ Tau \citep{Carciofi2009} that were interpreted with the one-arm density-wave oscillation model proposed by \citet{Okazaki1997}.

In a few cases, the stellar photosphere was resolved by interferometry, allowing the determination of physical stellar parameters, as for Achernar \citep{Domiciano2003}. Long-baseline interferometry has also been a powerful tool for detecting companions and deducing physical parameters of binary systems and circumbinary disks; for example, $\delta$ Sco \citep{Meilland2011}, 48 Per and $\phi$ Per \citep{Delaa2011}, $\delta$ Cen \citep{Meilland2008}, HD 87643 \citep{Millour2009} and Achernar \citep{Kanaan2008}.

To progress in the understanding of the physical processes responsible for the mass ejection and to test the hypothesis that they depend on stellar parameters, \citet{Meilland2012} initiated a survey of the circumstellar environments of eight bright Be stars: $\alpha$ Col, $\kappa$ CMa, $\omega$ Car, p Car, $\delta$ Cen, $\mu$ Cen, $\alpha$ Ara, and $o$ Aqr. They observed the Br$\gamma$ emission line to study the kinematics within the circumstellar disk. They determined the disk extension in the line and the nearby continuum for most of the targets and constrained the disk kinematics, showing that the disk rotation law is close to a Keplerian one. However, as they did not detect any correlation between the stellar parameters and the structure of the circumstellar environments they suggested that a larger sample of Be stars was needed to obtain conclusive results. 

In this work, we present spectro-interferometric observations of a new set of 26 Be stars obtained using the Very Large Telescope Interferometer(VLTI)/Astronomical Multi-Beam Recombiner(AMBER) instrument described in Sect. 2. The results derived from our sample together with those obtained by \citet{Meilland2012} are discussed, with the aim being to understand the physical properties and origin of their circumstellar disks.

This paper is organized as follows: in Sect.  \ref{sec:obs} we present the observations. In Sect.  \ref{sec:meth} we describe the adopted model and stellar parameters used to fit interferometric data. In Sect.  \ref{sec:results} we introduce the selected sample to study and provide a qualitative analysis of AMBER data of each object together with the best-fitting model parameters. We discuss our results in Sect.  \ref{sec:discussion} and give our conclusions in Sect.  \ref{sec:conclusions}.

\section{Observations}\label{sec:obs}

Near-infrared (NIR) spectro-interferometric observations were carried out with the AMBER instrument \citep[][]{Petrov2007} installed on the VLTI at the Paranal Observatory, Cerro Paranal, Chile.

AMBER operates in the NIR part of the spectrum between 1.0 and 2.4 $\mu$m. It was designed to coherently combine three telescope beams that can either be 8.2 m unit telescopes (UTs) or 1.8 m auxiliary telescopes (ATs). These three output beams are merged in a common Airy disk containing Young’s fringes, which are dispersed by a standard long-slit spectrograph on a two-dimensional (2D) detector. Per observation, AMBER delivers the visibilities and differential phases of the science target at three different spatial frequencies and one closure phase. The fringe contrast and phase are related to the Fourier transform of the source brightness distribution on sky at the observed spatial frequency $f=\rm{B}/\lambda$, where B is the baseline distance of a telescope pair within the UT or AT configuration. The stabilisation of the fringes could be enhanced with the external fringe tracker FINITO (Fringe-tracking Instrument of NIce and TOrino), allowing longer DITs (detector integration time) up to 10 sec and  leading to a better  S/N on the spectrally dispersed fringes.
Quite often, AMBER programs include the observation of calibrator objects to obtain not only differential but also absolute visibilities, using an estimation of the sizes of the calibrators and their visibilities. However, the drawback of using AMBER in high-resolution mode (HR, with R\,$\simeq$\,12\,000) with FINITO is the difficulty in determining a stable transfer function using calibrators that usually lead to poor data calibration.

We observed 26 Be stars using the AMBER instrument in conjunction with FINITO, from October 28 to 31, 2014. Our survey does not include the observation of calibrators. The observations were performed using the ATs, and the AMBER spectrograph was setup in the HR mode and centered on the Br$\gamma$ 2.16 $\mu$m emission line. This allows us to study the circumstellar gas kinematics through the Doppler effect. The lack of a calibrator is not an issue since the differential visibility, the differential phase, and the closure phase are independent of the effects that modify the absolute visibility.

The interference pattern is related to the source brightness via a 2D Fourier transformation of the complex visibility $V(u,v)$ and for modeling purposes fairly good coverage of the $(u,v)$ plane is required.

Table~\ref{table:log} presents the log of observations: HD number, date and time, telescope baseline configuration, baselines length and position angles, and integration time of our 26 Be stars AMBER dataset. 
The projected baselines, $(u,v)$ plane, in meters for each object of the selected sample (see Subsec.~\ref{subsec:selected}) is shown in Fig.~\ref{fig:HD23630} and online Figs. \ref{fig:HD23862}-\ref{fig:HD214748}. Each observation in the ($u,v$) plane, that is, three baseline measurements, is plotted with different colors and symbols, as is explained in the caption of Fig.~\ref{fig:HD23630}.

Data were reduced using the VLTI/AMBER data-reduction software \citep[amdlib\,v3.0.9, see][for detailed information on the AMBER data reduction]{Tatulli2007, Chelli2009}.

\begin{table*}
\caption{VLTI/AMBER observing log. \label{table:log}}
\begin{tabular}{lccr@{/}lr@{/}lr@{/}lr}
\hline
\hline 
HD Number & Date \& Time & Telescope        & \multicolumn{6}{c}{Length (m) / Position angle ($^\circ$)}& DIT \\
          & UTC          &  Configuration   & \multicolumn{6}{c}{B$_i$}                                 & sec  \\
\hline
\object{HD\,23\,302} & 2014-10-29 05:57 & A1-I1-G1 &  77.5&106.4 &  38.5&54.3  & 106.5&89.6  & 5.0 \\
                     & 2014-10-31 06:24 & A1-K0-J3 & 126.8&72.0  &  38.2&-36.1 & 123.8&53.7  &  20.0\\
\hline
\object{HD\,23\,338} & 2014-10-29 05:39 & A1-G1-I1 &  78.7&108.1 &  36.8&54.5  & 105.7&91.6  &  5.0 \\
\hline
\object{HD\,23\,408} & 2014-10-29 06:36 & A1-G1-I1 &  73.7&102.8 &  41.3&53.7  & 106.3&85.5  & 5.0 \\
\hline
\object{HD\,23\,480} & 2014-10-29 06:15 & A1-G1-I1 &  76.1&104.9 &  39.9&54.0  & 106.7&87.9  & 5.0 \\
\hline
\object{HD\,23\,630} & 2014-10-29 06:55 & A1-G1-I1 &  71.3&101.2 &  42.5&53.0  & 105.3&83.6  & 5.0 \\ 
                     & 2014-10-30 07:27 & A1-G1-J3 &  65.5&97.9  &  97.6&22.0  & 133.0&51.4  & 5.0 \\ 
                     & 2014-10-31 06:45 & A1-K0-J3 & 128.1&70.3  &  36.5&-33.9 & 127.3&53.1  & 20.0\\
                     \hline
\object{HD\,23\,862} & 2014-10-30 07:47 & A1-G1-J3 &  61.9&96.0  & 100.0&23.7  & 135.2&50.3  & 5.0 \\
                     & 2014-10-31 07:06 & A1-K0-J3 & 128.8&68.4  &  34.7&-31.2 & 130.4&52.3  & 20.0 \\
\hline
\object{HD\,28\,497} & 2014-10-31 05:46 & A1-K0-J3 & 124.9&64.0  &  56.3&-32.7 & 134.9&38.1  & 10.0\\
\hline
\object{HD\,30\,076} & 2014-10-31 06:05 & A1-K0-J3 & 124.2&65.4  &  55.8&-32.7 & 132.8&39.5  & 10.0\\
\hline
\object{HD\,32\,991} & 2014-10-31 09:20 & A1-K0-J3 & 127.1&62.4  &  32.4&-18.7 & 137.8&48.6  & 5.0 \\
\hline
\object{HD\,33\,328} & 2014-10-30 07:01 & A1-G1-J3 &  79.9&111.8 & 127.2&5.3   & 135.8&42.2  &  5.0 \\
\hline
\object{HD\,35\,439} & 2014-10-31 09:02 & A1-K0-J3 & 126.6&68.6  &  47.2&-19.1 & 139.8&48.1  & 5.0 \\
\hline
\object{HD\,36\,576} & 2014-10-31 08:25 & A1-K0-J3 & 128.2&70.3  &  40.6&-31.6 & 129.5&51.4  & 5.0 \\
\hline
\object{HD\,37\,202} & 2014-10-29 07:37 & A1-G1-I1 &  78.7&108.4 &  38.2&52.6  & 106.0&90.8  & 2.0 \\
                     & 2014-10-30 08:56 & A1-G1-J3 &  69.3&101.6 &  99.5&19.0  & 131.7&51.5  & 5.0 \\
                     & 2014-10-31 08:07 & A1-K0-J3 & 126.2&72.4  &  41.0&-35.5 & 123.6&52.6  & 5.0 \\
\hline
\object{HD\,37\,490} & 2014-10-31 08:41 & A1-K0-J3 & 128.9&69.3  &  47.7&-24.8 & 137.6&48.0  & 5.0 \\
\hline
\object{HD\,41\,335} & 2014-10-31 07:50 & A1-K0-J3 & 126.6&67.4  &  54.8&-30.9 & 134.6&42.3  & 5.0 \\
\hline
\object{HD\,45\,725} & 2014-10-29 08:30 & A1-G1-I1 &  79.8&112.0 &  45.2&42.9  & 106.5&88.1  & 4.0 \\
                     & 2014-10-30 08:08 & A1-G1-J3 &  80.0&111.3 & 125.9&3.9   & 133.7&41.3  & 4.0 \\
\hline
\object{HD\,60\,606} & 2014-10-30 08:35 & A1-G1-J3 &  77.0&103.5 & 130.0&-0.6  & 139.9&33.9  & 10.0\\  
\hline
\object{HD\,66\,194} & 2014-10-31 04:37 & A1-K0-J3 & 128.3&-1.7  &  25.3&-86.5 & 134.7&-12.9 & 10.0\\
                     & 2014-10-31 05:21 & A1-K0-J3 & 128.4&9.0   &  29.8&-78.3 & 135.2&-4.3  & 10.\\
\hline
\object{HD\,68\,980} & 2014-10-30 09:16 & A1-G1-J3 &  77.1&103.8 & 130.1&-0.4  & 139.9&34.0  &  5.0 \\
\hline
\object{HD\,75\,311} & 2014-10-31 07:33 & A1-K0-J3 & 128.3&28.8  &  38.8&-60.5 & 137.0&11.7  & 5.0 \\
\hline
\object{HD\,209\,409}& 2014-10-29 00:48 & A1-G1-I1 &  77.1&112.5 &  45.8&46.3  & 106.0&88.8  & 2.0 \\ 
                     & 2014-10-29 01:27 & A1-G1-I1 &  72.6&114.1 &  46.5&47.7  & 102.3&89.0  &  2.0 \\  
                     & 2014-10-30 01:06 & A1-G1-J3 &  74.8&113.3 & 124.5&13.3  & 138.9&47.2  & 2.0 \\
                     & 2014-10-30 03:31 & A1-G1-J3 &  47.0&127.6 & 131.0&24.2  & 132.3&45.7  & 2.0 \\
                     & 2014-10-31 00:25 & A1-K0-J3 & 128.5&69.2  &  52.3&-28.2 & 136.1&45.6  & 2.0 \\
                     & 2014-10-31 03:03 & A1-K0-J3 & 110.1&66.5  &  47.1&-4.2  & 135.2&46.8  & 2.0 \\
                     & 2014-10-31 03:39 & A1-K0-J3 &  99.7&63.7  &  47.0&2.4   & 130.5&45.0  & 2.0 \\        
\hline
\object{HD\,212\,076}& 2014-10-29 02:04 & A1-G1-I1 &  66.2&106.3 &  45.9&48.9  & 100.1&83.2  & 5.0 \\  
                     & 2014-10-29 02:36 & A1-G1-I1 &  60.6&105.8 &  46.5&48.0  &  95.2&81.0  & 5.0 \\
                     & 2014-10-30 02:02 & A1-G1-J3 &  66.3&106.3 & 113.3&19.1  & 137.9&49.0  & 2.0 \\
                     & 2014-10-30 02:50 & A1-G1-J3 &  56.5&105.6 & 117.7&22.8  & 139.9&47.2  & 5.0 \\
                     & 2014-10-31 01:57 & A1-K0-J3 & 128.2&67.3  &  41.3&-22.4 & 137.7&49.0  & 5.0 \\
\hline
\object{HD\,212\,571}& 2014-10-30 01:42 & A1-G1-J3 &  72.3&112.4 & 122.4&15.1  & 139.0&47.9  & 2.0 \\
                     & 2014-10-31 01:38 & A1-K0-J3 & 128.4&69.3  &  48.5&-22.7 & 138.9&47.9  & 10.0\\ 
\hline
\object{HD\,214\,748}& 2014-10-29 03:30 & A1-G1-I1 &  68.9&137.9 &  41.2&54.6  &  86.4&108.7 & 5.0 \\  
                     & 2014-10-30 02:28 & A1-G1-J3 &  75.0&127.3 & 131.4&16.9  & 132.9&51.4  & 2.0 \\
                     & 2014-10-30 04:34 & A1-G1-J3 &  59.9&157.4 & 126.0&26.7  & 104.2&56.2  & 5.0 \\
                     & 2014-10-31 01:20 & A1-K0-J3 & 127.4&71.2  &  56.5&-23.8 & 138.7&46.1  & 5.0 \\
                     & 2014-10-31 04:15 & A1-K0-J3 &  88.5&87.5  &  55.9&1.7   & 111.1&56.2  & 5.0 \\
\hline    
\object{HD\,219\,688}& 2014-10-29 04:15 & A1-G1-I1 &  57.0&129.6 &  44.0&48.5  &  79.2&95.2  & 5.0 \\
\hline
\object{HD\,224\,686}& 2014-10-29 03:51 & A1-G1-I1 &  79.9&132.6 &  36.2&70.8  & 103.4&114.3 & 5.0 \\
\hline
\end{tabular}
\end{table*}

\section{Metodology}\label{sec:meth}

\subsection{Disk model}

Interferometric data are often analyzed using a 2D kinematic model of a rotating and/or expanding equatorial disk. This simple ``toy model'' has been used to describe circumstellar disks of classical Be stars \citep{Meilland2012,Delaa2011,Meilland2011}; see \citet{Delaa2011} for a detailed description of the model.

This model includes a central star described by a uniform disk of constant brightness $I_{\star}(x,y)$, where $x$ and $y$ are the spatial Cartesian coordinates on the sky plane, surrounded by a circumstellar envelope. The envelope emission in the line, $I_\text{l}(x,y)$, and continuum, $I_\text{c}(x,y)$, are described by two elliptical Gaussian distributions of different FWHMs, named $a_{\rm{l}}$ and $a_{\rm{c}}$, but with the same flattening $f$ due to a projection effect on the sky of a geometrically thin equatorial disk that depends on the inclination angle $i$, being $f=a_{\rm{max}}/a_{\rm{min}}=1/\text{cos}(i)$, where $a_{\rm{max}}$ is the major axis and $a_{\rm{min}}$ is the minor axis. 
        
The emission maps were combined with a 2D projected velocity map, $V_{\text{proj}}(x,y)$, of the geometrically thin equatorial disk,
\begin{equation}
V_{\text{proj}}(x,y)=(V_{\phi}\,\sin\,\phi-V_{\text{r}} \cos\,\phi)\, \sin\,i,
\end{equation}
where $V_{\phi}$ and $V_{\text{r}}$ are the rotational and radial velocity laws for the disk with the radial distance given by \citet{Hutchings1970} and \citet{Castor1975}, respectively,
\begin{align}
V_{\phi}&=V_{\text{rot}}\left(\frac{r}{R_{\star}}\right)^{j}\\
V_{\text{r}}&=V_{\text{0}}+(V_\infty-V_{\text{0}})\left(1-\frac{R_\star}{r}\right)^\beta,
\end{align}
\noindent where $R_\star$ is the stellar radius, $V_{\text{rot}}$ the rotational velocity at the disk inner radius, $V_{\rm 0}$ the expansion velocity at the photosphere, $V_{\infty}$ the terminal velocity, $\beta$ the exponent of the expansion velocity law and $j$ the exponent of the rotational velocity law.

For each wavelength $\lambda$ and within a narrow spectral band $\delta\lambda$, an iso-velocity map projected along the line of sight, $R(x,y,\lambda,\delta\lambda)$, was calculated and multiplied by the whole emission map in the line. The whole emission map for each wavelength is the weighted sum of the stellar map, the disk continuum map and the emission line map:
\begin{align}
I_\text{tot}(x,y,\lambda,\delta\lambda)= & I_{\star}(x,y)\cdot F_\star(\lambda)\ + 
          I_\text{c}(x,y)\cdot F^{\rm{env}}_{\rm c}\ + \\
          & I_\text{l}(x,y)\cdot  R(x,y,\lambda,\delta\lambda) \cdot EW,         
\end{align}
\noindent where $F_\star(\lambda)$ and $F^{\rm{env}}_{\rm c}$ are respectively the contributions of the total continuum flux from the star and the disk, while $EW$ is the line equivalent width.

The model parameters can be classified into different categories:
\begin{itemize}
\item Stellar parameters: $R_\star$, distance ($d$), inclination angle ($i$), disk major-axis position angle ($PA$)
.\item Kinematic parameters: $V_{\rm{rot}}$, $V_{\rm 0}$, $V_{\infty}$, $\beta$ and $j$.
\item Disk continuum parameters: disk FWHM in the continuum ($a_{\rm{c}}$), disk continuum flux normalized by the total continuum flux ($F^{\rm{env}}_{\rm c}$).
\item Disk emission line parameters: disk FWHM in the line ($a_{\rm{l}}$), and line equivalent width ($EW$).
\end{itemize}

We know how each parameter affects the visibility and phase variations \citep{Meilland2012}, and most parameters are unambiguously constrained using a few baselines with various orientations and lengths. Therefore, we decided to perform the model-fitting manually (instead of an automatic fit of the models).
For all targets we set the expansion velocity to zero, since usually the radial velocity in the disk is assumed negligible \citep{Miroshnichenko2003, Vinicius2006, Meilland2007a, Meilland2007b}. For the other parameters (especially $PA$, $i$, $a_{\rm{c}}$, $a_{\rm{l}}$ and $EW$, which strongly affect the interferometric observables and, therefore, are easily and unambiguously constrained), we started with rough value estimates from a preliminary fit to our interferometric data and, then, we explored the parameter space by decreasing steps to minimize $\chi^2$ (see details below).
We also explored the full range of possible model parameters using larger steps to make sure there are no other minima.
Once we had obtained the best-fitting model, we changed the parameters one by one to find the uncertainty of each one. We consider the limit of uncertainty when the model is no longer a good fit for the observations.
In what follows, we briefly describe each parameter. 

\begin{figure*}[h!]
\centering
\includegraphics[width=0.49\hsize]{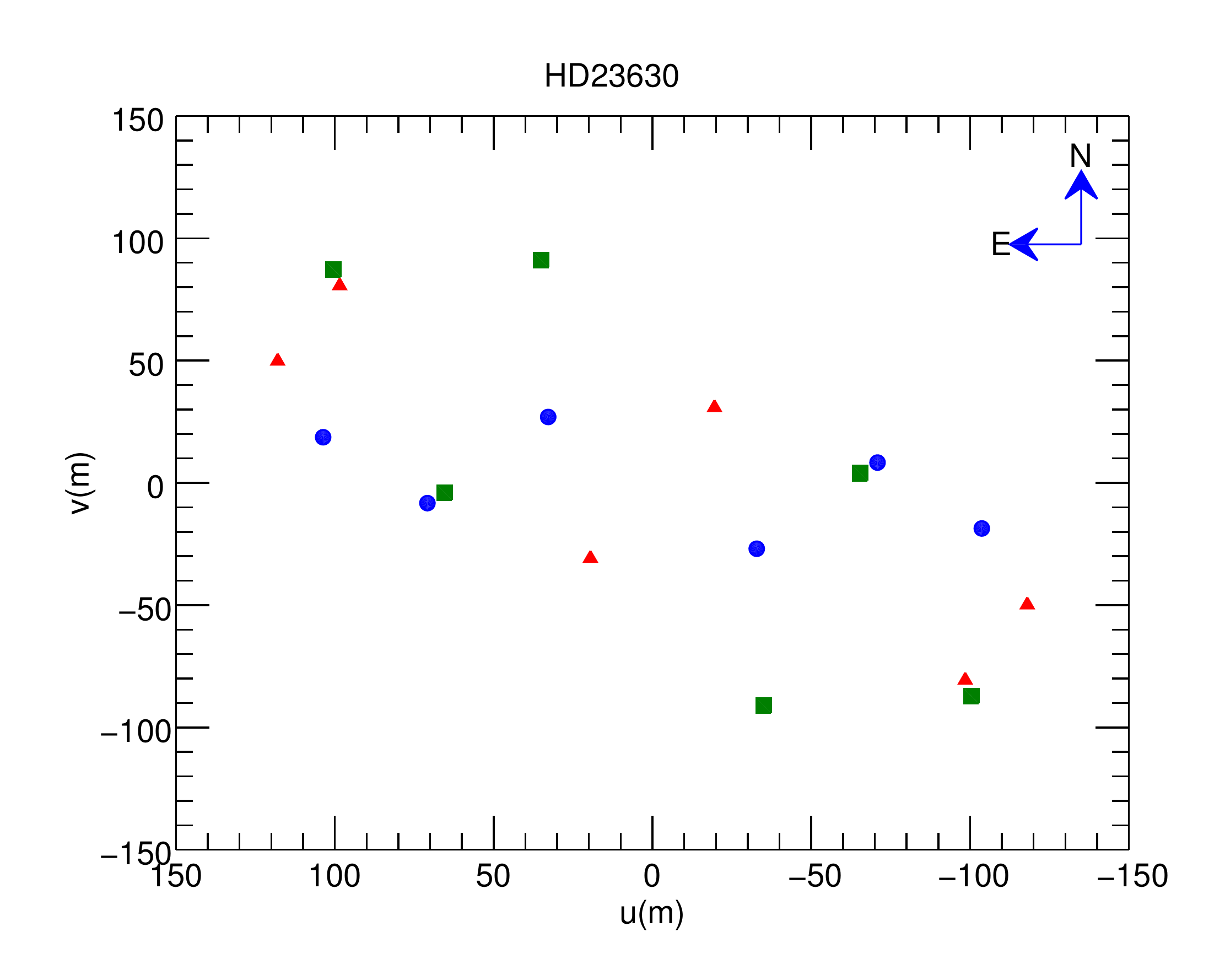}

\includegraphics[width=0.49\hsize]{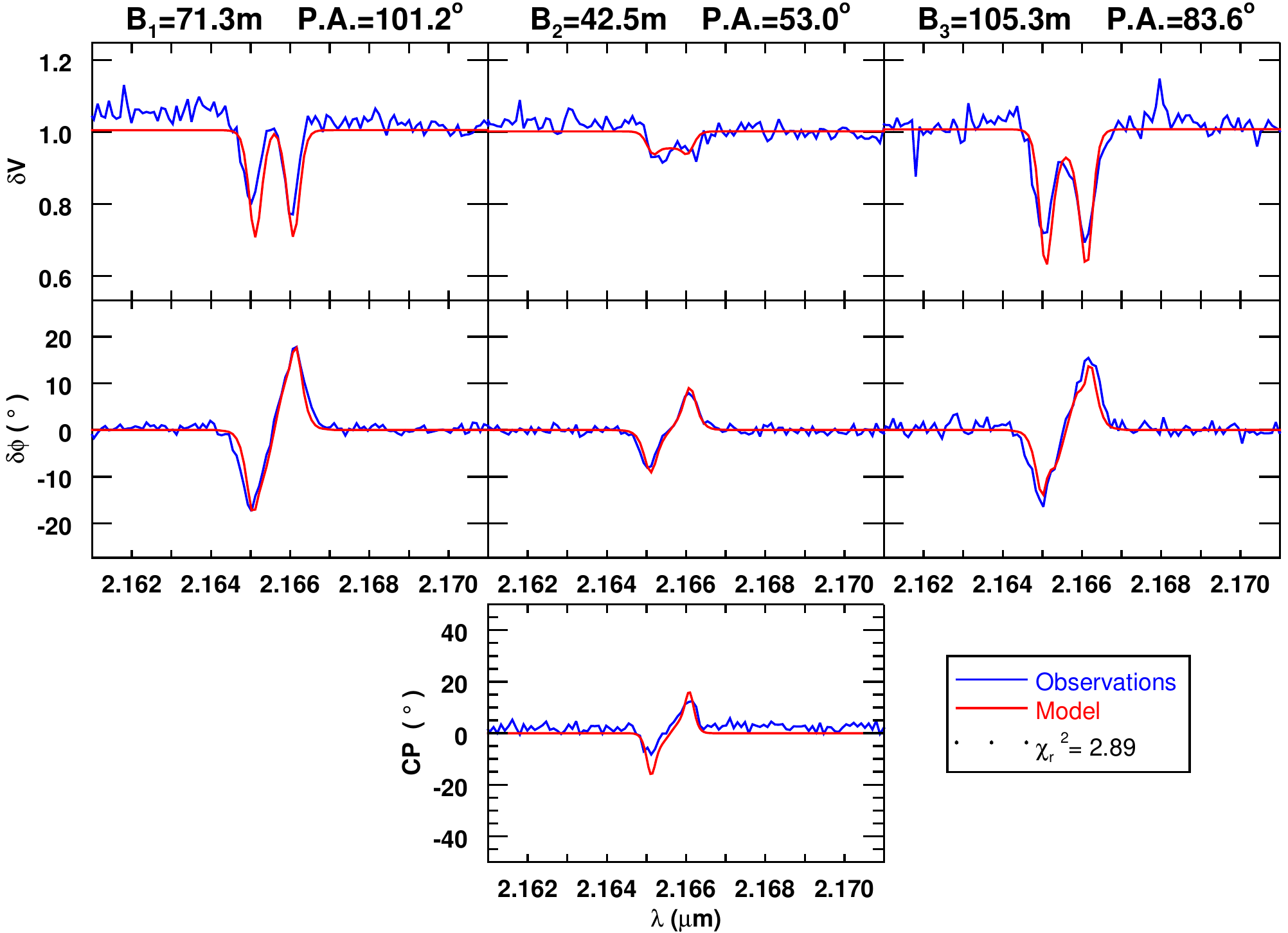}
\includegraphics[width=0.49\hsize]{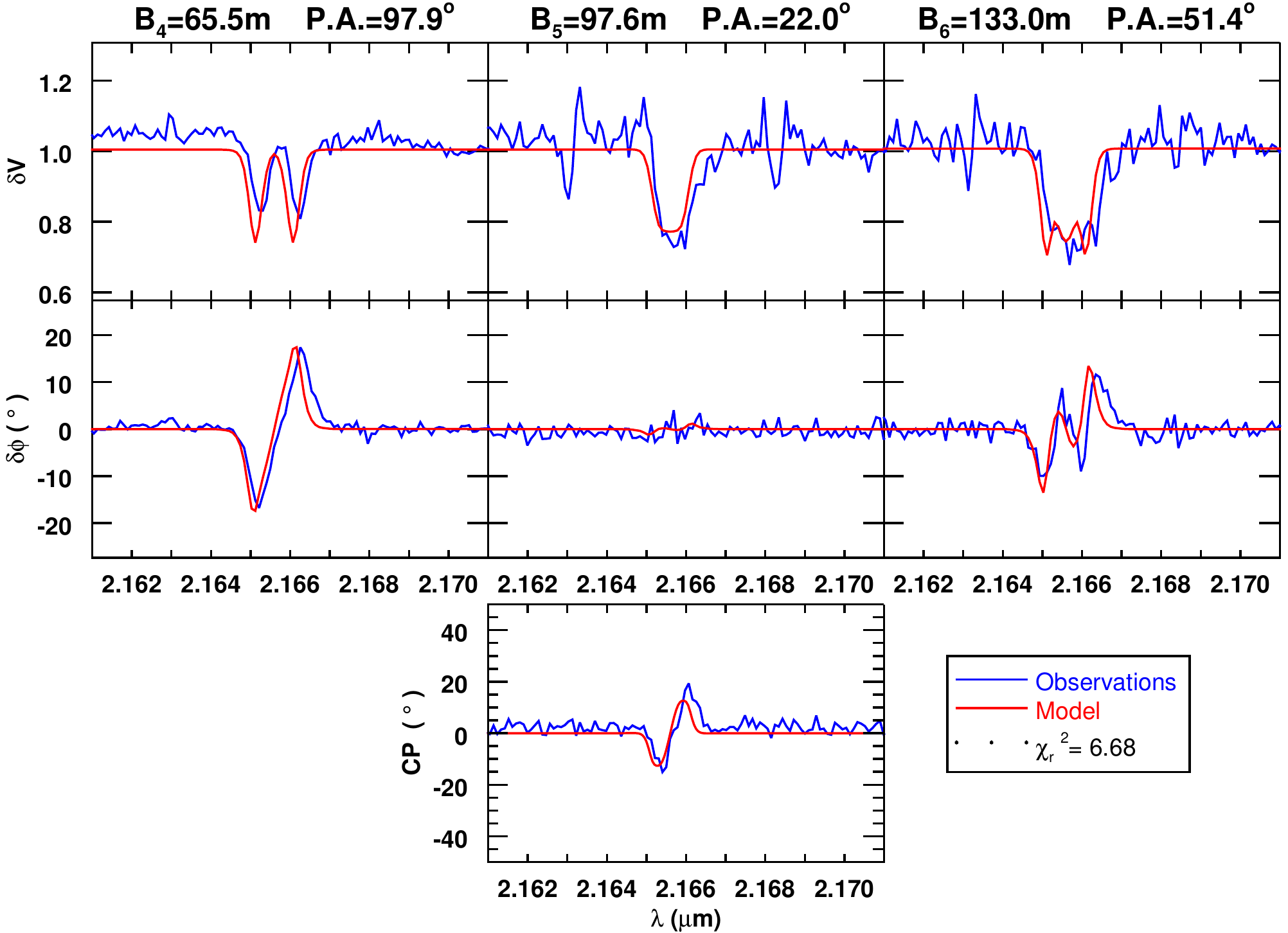}
\includegraphics[width=0.49\hsize]{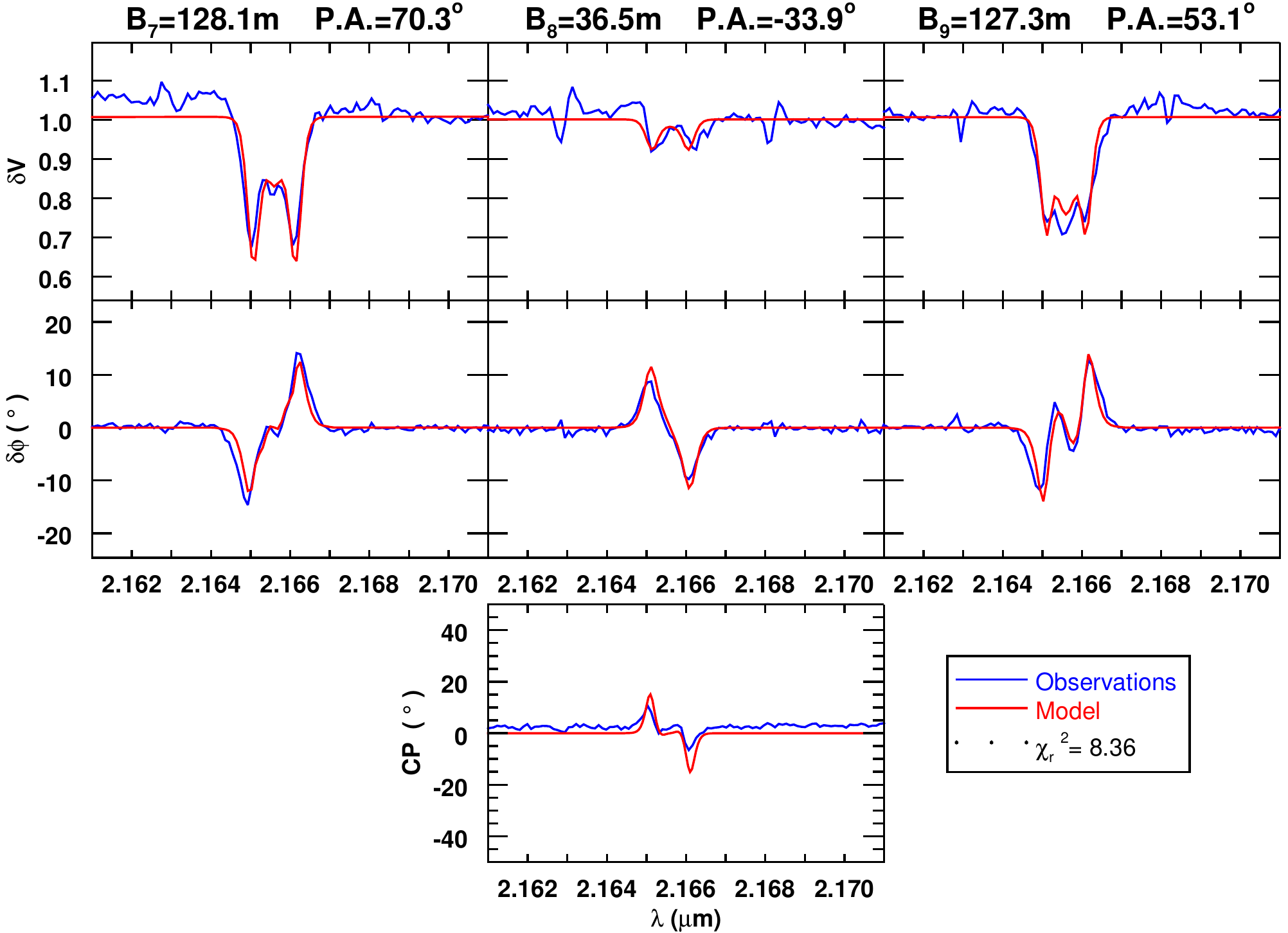}
\caption{\object{HD\,23\,630}: ($u,v$) plan coverage, differential visibilities ($\delta \text{V}$), differential phases ($\delta\phi$) and closure phases (CP) from our HR measurements. 
Each observation in the ($u,v$) plane, i.e., three baseline measurements, is plotted with a different color and symbol: 
blue circles correspond to the baselines B$_{1}$-B$_{3}$;
green squares to baselines B$_{4}$-B$_{6}$; 
red triangles to baselines B$_{7}$-B$_{9}$; 
purple diamonds to baselines B$_{10}$-B$_{12}$;
pink hexagons to baselines B$_{13}$-B$_{15}$;
cyan pentagons to baselines B$_{16}$-B$_{18}$;
orange thin diamonds to baselines B$_{19}$-B$_{21}$.
See description in Sect. \ref{sec:results}}.
\label{fig:HD23630}
\end{figure*}

\begin{itemize}
\item $PA$ (major axis position angle) modifies the amplitude of phase variations and the drop of the visibility \citep{Meilland2011}. For a nonfully resolved disk, the amplitude of the ``S'' shape variation is proportional to the baseline length and depends on its orientation. When the baseline is aligned with the major axis of the circumstellar envelope, the amplitude of the ``S'' shape is maximum and the visibility variation is ``W'' shaped, as shown in the B$_1$ baseline of Fig.~\ref{fig:HD23630}. At the same time, when the baseline is aligned with the minor axis the amplitude of the ``S'' shape is null and the shape of the drop of the visibility changes from ``W'' to ``V'' shaped, as shown in the B$_5$ baseline of Fig.~\ref{fig:HD23630}. We started the fit with 10$^\circ$ steps in the 0-360$^\circ$ range, and refined the grid using a final step of 1$^\circ$.

\item $EW$  is measured from the Br$\gamma$ line profile. To derive the model sensitivity to the $EW$, this value was varied up to $30\,\%$ using steps between 1 and 0.1 $\AA$.

\item $a_l$ (disk FWHM in the emission line in stellar diameters, $D_{\star}$) has a huge influence on the drop of the visibility along all baselines and on the amplitudes of the ``S'' shapes of the differential phases. Phase variation loses its ``S'' shape for those baselines that fully resolved the disk, because of secondary effects due to inhomogeneities of the disk or, as in most of our observations, because of a mathematical effect due to the over-resolution of the disk as shown in the B$_9$ baseline of Fig.~\ref{fig:HD23630}. 
This mathematical effect is due to a property of the Fourier transform, by which the visibility goes through zero when a simple object is fully resolved, and beyond that, its phase is inverted. This effect is more complex in the case of a multicomponent object, that is, the rotating disk around a Be star, and it generates the loss of the "S" shape of the phase variations in a fully resolved disk.
The parameter $a_l$ also affects the double-peak separation (the larger the disk, the smaller the separation). We calculated models with values in the 1-20$D_\star$ range, with steps of between 0.5 and 0.1$D_\star$.

\item $a_{\rm{c}}$ (disk FWHM in the continuum in stellar diameters, $D_{\star}$) is mainly derived from visibility measurements in the continuum. 
This parameter sets the relative flux of the disk in the continuum. As the disk emission in the continuum is larger than the stellar photosphere, this parameter affects the level of the absolute visibility in the continuum, that is, the absolute visibility in the continuum decreases when $a_{\rm{c}}$ increases. Therefore, it can modify the amplitude of the phase variations. The more resolved an object is in the continuum, the smaller the phase variations are. We explored the values in the 1-20$D_\star$ range, with steps of between 1 and 0.1$D_\star$.

\item $i$ (inclination angle) also has a significant effect on the drop of the visibility for baselines close to the polar orientation; it also has influence on the line-peak separation since it is related to the projected rotational velocity. The space parameter for $i$ is the range 0-90$^\circ$, and we explored steps of between 5$^\circ$ and 1$^\circ$.

\item $V_{\rm{rot}}$ (rotational velocity) and $j$ (exponent of the rotational velocity law) influence the double-peak separation and line width. The faster the disk rotates, the larger the peaks separate. On the other hand, as $j$ increases, the rotational velocity drops faster with distance, the line double-peak separation becomes smaller, and the line wings wider. It is quite hard to set both $V_{\rm{rot}}$ and $j$ unambiguously. The only way to do that is to study the disk kinematics using multiple lines formed in different parts of the disk. As $V_{\rm{rot}}$ and $j$ describe the global disk kinematics, they should not depend on the observed line. This has been done only once, on the Be star $\delta$~Sco \citep{Meilland2011}, using the H$\alpha$ line from the VEGA instrument, and Br$\gamma$ and the 2.06~$\mu$m \ion{He}{i} lines from AMBER.
It is worth mentioning emission line wings can also be affected by nonkinematic broadening caused by noncoherent scattering \citep{Delaa2011}.
Unfortunately, we do not have measurements in other lines to solve the ambiguity, therefore we explored the space parameter of $V_{\rm{rot}}$ and $j$ to derive rough values. For $V_{\rm{rot}}$, the space parameter was limited by $V$ (obtained from $i$ and $V\,\sin\,i$) and $V_{\rm{c}}$, and we explored this interval using initial steps of 10 km\,s$^{-1}$ and decreasing them to minimal steps of 2 km\,s$^{-1}$. For $j$, we inspected the $0.3-0.65$ range, with initial steps of 0.05 and final steps of 0.01.
\end{itemize}

\subsection{Stellar parameters from SED fittings}\label{meth:SED}

In order to derive $R_{\star}$ and $F^{\rm{env}}_{\rm c}$ for the selected sample, we performed fittings to the spectral energy distributions (SEDs).
For each target, we reconstructed its SED using data available from the VIZIER database. We fitted a \citet{Kurucz1979} model to the photometric data \citep{Vizier2000} using a minimization method implemented in the library IDL routine ``mpfit'' (details about the algorithm used can be found in Sect.~3.1 of \citet[][]{Arcos2018}. The input parameters are $T_{\rm{eff}}$, $\log\,g,$ and the reddening $R_{\rm v}$. 
The $T_{\rm{eff}}$ and $\log\,g$ values were taken from \citet{Fremat2005}. The value of $R_{\rm v}$ was set in 3.1, using the interstellar extinction law from \citet{Cardelli1989}.

For three of the sample stars (\object{HD\,32\,991}, \object{HD\,37\,202} and \object{HD\,66\,194})  $T_{\rm{eff}}$ values are not available in the literature. Therefore, we also obtained $T_{\rm{eff}}$ from the SED fitting.

To avoid contamination from the disk flux, the SED was fitted in the ultraviolet and optical wavelength ranges \citep{Meilland2009}. 
We fixed the stellar parameter from Table~\ref{table:sample}, except for $R_{\star}$ and $F^{\rm{env}}_{\rm c}$, which are left free (and $T_{\rm{eff}}$ in the cases commented before). From the best-fitting model we constrained $R_{\star}$ and $F^{\rm{env}}_{\rm c}$ for all the targets, and also the effective temperature, $T_{\rm{eff}}$, for \object{HD\,32\,991}, \object{HD\,37\,202,} and \object{HD\,66\,194}.

\begin{figure}
\resizebox{\hsize}{!}{\includegraphics{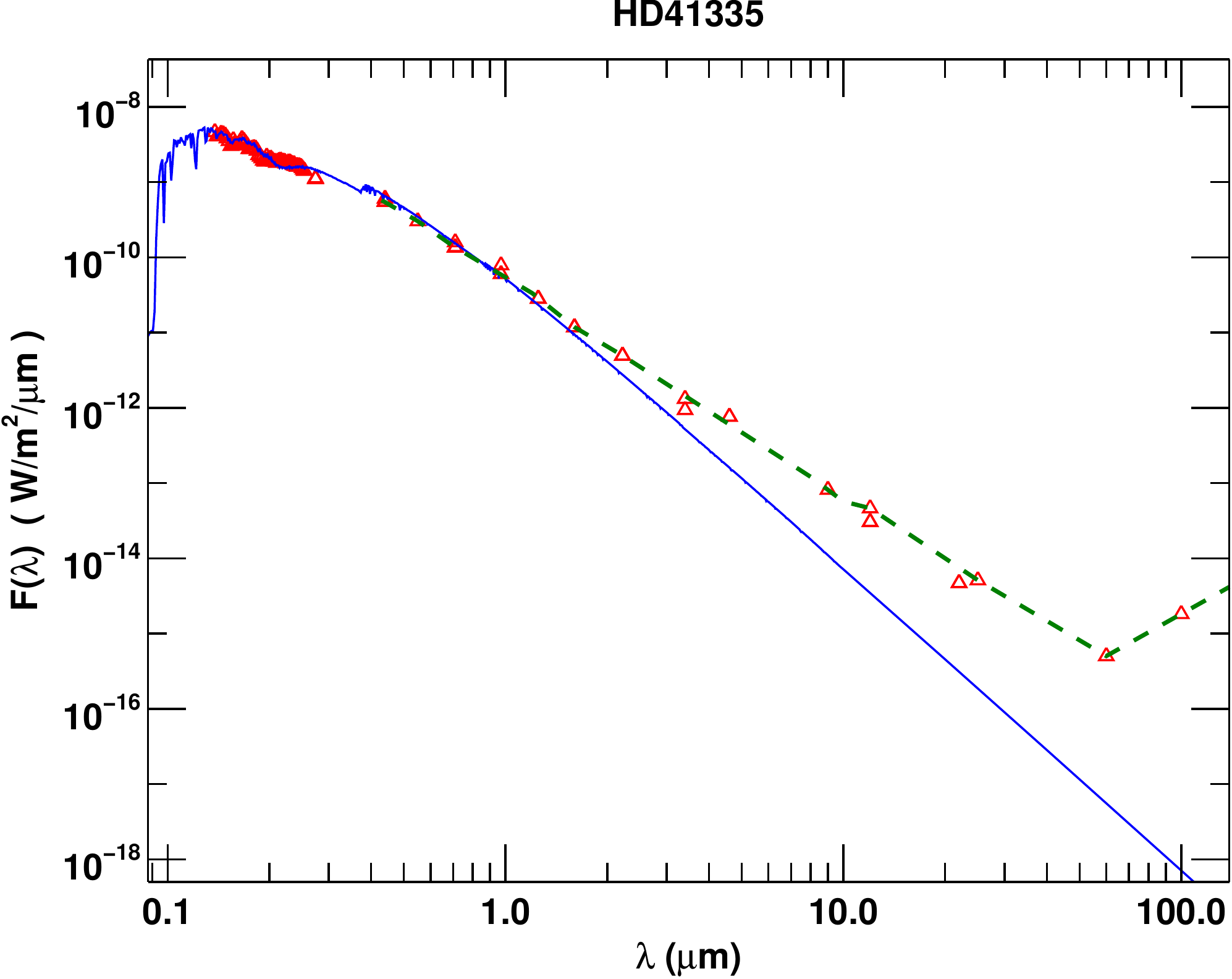}}
\caption{Example of the SED fitting model for \object{HD\,41\,335}. Red triangles represent the photometric data from VIZIER database, the solid blue line represents the best-fitting Kurucz model and the dashed green line shows the IR excess.}
\label{fig:SED}
\end{figure}

In Fig.~\ref{fig:SED} we present an example of an SED fitting model. Red triangles represent the photometric data from the VIZIER database, the solid blue line represents the best-fitting Kurucz model, and the dashed green line shows the IR excess. 

\section{Results}\label{sec:results}

\subsection{The selected sample and stellar parameters}\label{subsec:selected}

\begin{figure}[h!]
\resizebox{\hsize}{!}{\includegraphics{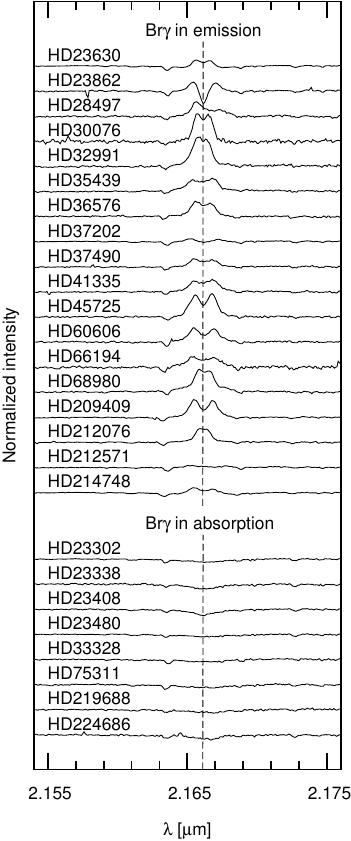}}
\caption{Br$\gamma$ line profile of Be stars observed with VLTI/AMBER instrument during our observing campaign.} 
\label{fig:brgamma}
\end{figure}

\begin{figure}
\resizebox{\hsize}{!}{\includegraphics{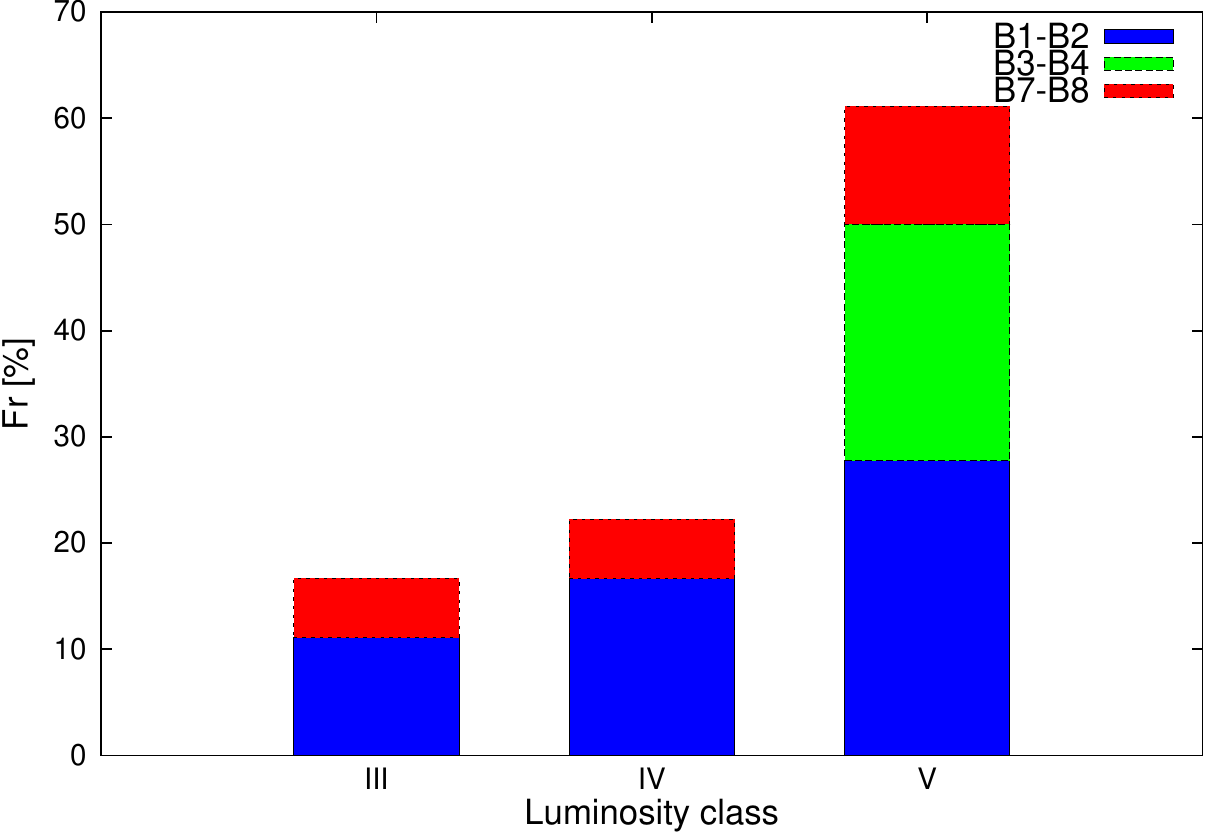}}
\caption{Histogram showing the distribution of the star sample as a function of the spectral subtype and luminosity class.}
\label{fig:TE-CL}
\end{figure}

Our sample consists of 26 Be stars observed with the VLTI/AMBER instrument. As shown in Fig.~\ref{fig:brgamma}, the majority of the stars show a double-peaked emission profile. Shell line profiles are observed in \object{HD\,23\,862} (\object{28 Tau}) and \object{HD\,37\,202} (\object{$\zeta$ Tau}). The Br$\gamma$ line is weak in \object{HD\,212\,571} (\object{$\pi$ Aqr}) while eight stars (\object{HD\,23\,302}, \object{HD\,23\,338}, \object{HD\,23\,408}, \object{HD\,23\,480}, \object{HD\,33\,328}, \object{HD\,75\,311}, \object{HD\,219\,688}, and \object{HD\,224\,686}) display an absorption line profile. These last eight stars are excluded from the present study because we want to model the circumstellar disk emission. 
Therefore, our final sample contains 18 Be stars with $m_{\rm{K}}~\le~6$, spectral types between B1 and B8, and luminosity classes between V and III, distributed according to the histogram displayed in Fig.~\ref{fig:TE-CL}. Around 55\% of our star sample have spectral types B1-B2, and the others are distributed almost uniformly between B3-B4 and B7-B8. The sample does not have stars with spectral types B5-B6. Around 60\% of the star sample are in luminosity class V. Stars with luminosity classes IV and III are a few more than 20\% and 15\% of the total sample, respectively. Luminosity classes IV and III are mostly present in spectral types B1-B2 and B7-B8, while luminosity class V is spread over all the spectral subtypes sampled.
   
Table~\ref{table:sample} lists known and calculated parameters for stars with the Br$\gamma$ line in emission: 
the HD number and star name are given in the first two columns; 
columns 3 and 4 list the $K$-band magnitude ($m_{\rm{K}}$) and spectral type (ST), respectively, taken from the CDS database.
Column 5 lists the distance to the star in pc ($d$) which was taken from \citet{vanLeeuwen2007} and \citet{Astra2016}.
Columns 6, 7, and 8 give the effective temperature ($T_{\rm{eff}}$), projected rotational velocity ($V\,\sin\,i$), and break-up velocity ($V_{\rm{c}}$) taken from \citet{Fremat2005}, with the exception of the stellar parameters for \object{HD\,32\,991}, \object{HD\,37\,490,} and \object{HD\,66\,194,} which we estimated from fittings to the SEDs using \citet{Kurucz1979} atmospheric models (see Sect. \ref{meth:SED}). For \object{HD\,32\,991}, the $V\,\sin\,i$ was taken from \citet{Abt2002}, and for \object{HD\,37\,490} and \object{HD\,66\,194} from \citet{Chauville2001}. Columns 9, 10, and 11 list the polarization angle ($P_\star$), disk continuous flux contribution relative to the total flux in the $K$-band ($F^{\rm{env}}_{\rm {c}}$), and the stellar radius ($R_{\star}$). The polarization angle, $P_\star$ was taken from \citet{Yudin2001}. $R_{\star}$ and $F^{\rm{env}}_{\rm {c}}$ were also derived from the SED by fitting a \citet{Kurucz1979} model as is explained in Sect. \ref{meth:SED}.

\begin{table*}
\caption{Adopted and calculated parameters for the sample of Be stars with emission in the Br$\gamma$ line.}
\label{table:sample}      
\centering
\resizebox{\textwidth}{!}{         
\begin{tabular}{r c c c l c l c c c c} 
\hline \hline
Name & HD Number      & $m_{\rm{K}}$$^{(a)}$ & ST$^{(a)}$  & \multicolumn{1}{c}{$d^{(b)}$}        & $T_{\rm{eff}}$$^{(c)}$    & \multicolumn{1}{c}{$V\,\sin\,i^{(c)}$}   & $V_{\rm{c}}$$^{(c)}$  & $P_\star$$^{(d)}$ & $R_{\star}$$^{(e)}$ & $F^{\rm{env}}_c$$^{(e)}$ \\
     &     & (mag)      &      & \multicolumn{1}{c}{(pc)}              & (K)             & \multicolumn{1}{c}{(km\,s$^{-1}$)}  & (km\,s$^{-1}$)  & ($^{\circ}$)     & ($R_{\sun}$)       &  \\
\hline
\object{$\eta$ Tau}     & \object{HD\,23\,630}   & 2.940      & B7III       & 127 $\pm$ 8 $^{(f)}$  & 12258 $\pm$ 505 & 149 $\pm$ 8     & 274 $\pm$ 15 & 34      & 6.6  & 0.71 \\
\object{28 Tau}         & \object{HD\,23\,862}   & 4.937      & B8Vne       & 117 $\pm$ 5           & 12106 $\pm$ 272 & 290 $\pm$ 15    & 420 $\pm$ 22 & 69      & 3.5  & 0.19 \\
\object{28 Eri}         & \object{HD\,28\,497}   & 5.798      & B2(V)ne     & 465 $\pm$ 63          & 26724 $\pm$ 427 & 342 $\pm$ 24    & 631 $\pm$ 45 & 120     & 4.1  & 0.63 \\
\object{56 Eri}         & \object{HD\,30\,076}   & 5.452      & B2(V)nne    & 822 $\pm$ 486$^{(f)}$ & 20488 $\pm$ 330 & 225 $\pm$ 12    & 456 $\pm$ 25 & 148     & 11.0 & 0.47 \\
\object{105 Tau}        & \object{HD\,32\,991}   & 4.775      & B2Ve        & 250 $\pm$  43$^{(f)}$ & 20900 $^{(e)}$  & 175 $\pm$ 18$^{(g)}$ & --      & 126     & 3.8  & 0.90 \\
\object{$\psi_{01}$ Ori}& \object{HD\,35\,439}   & 5.355      & B1Vn        & 318 $\pm$ 85          & 22134 $\pm$ 665 & 266 $\pm$ 13    & 513 $\pm$ 26 & 142     & 6.1  & 0.09 \\
\object{120 Tau}        & \object{HD\,36\,576}   & 4.777      & B2IV-Ve     & 476 $\pm$ 73          & 22618 $\pm$ 508 & 266 $\pm$ 13    & 487 $\pm$ 25 & 113     & 4.5  & 0.87 \\
\object{$\zeta$ Tau}    & \object{HD\,37\,202}   & 2.970      & B1IVe-shell & 136 $\pm$ 16          & 19310 $\pm$ 550 & 326 $\pm$ 7     & 466 $\pm$ 13 & 34      & 6.1  & 0.58 \\
\object{$\omega$ Ori}   & \object{HD\,37\,490}   & 4.805      & B3Ve        & 423 $\pm$ 52          & 14492 $^{(e)}$  & 155 $\pm$ 5 $^{(h)}$ & 319 $^{(h)}$ & 9  & 13.5 & 0.02 \\
\object{V696 Mon}       & \object{HD\,41\,335}   & 4.770      & B3/5Vnne    & 403 $\pm$ 122         & 20902 $\pm$ 610 & 376 $\pm$ 26    & 520 $\pm$ 36 & 147     & 8.2  & 0.44 \\
\object{$\beta_{01}$ Mon}& \object{HD\,45\,725}  & 4.030      & B4Ve-shell  & 207 $\pm$ 49          & 17810 $\pm$ 455 & 345 $\pm$ 21    & 486 $\pm$ 30 & 14      & 5.0  & 0.32 \\
\object{OW Pup}         & \object{HD\,60\,606}   & 5.050      & B2Vne       & 363 $\pm$ 32          & 18030 $\pm$ 332 & 285 $\pm$ 16    & 441 $\pm$ 25 & --      & 6.2  & 0.56 \\
\object{V374 Car}       & \object{HD\,66\,194}   & 5.359      & B3Vn        & 379 $\pm$ 47$^{(f)}$  & 14089 $^{(e)}$  & 200 $\pm$ 29$^{(h)}$& 367 $^{(h)}$ & 130 & 7.9  & 0.32 \\
\object{MX Pup}         & \object{HD\,68\,980}   & 4.554      & B2ne        & 284 $\pm$ 13          & 25126 $\pm$ 642 & 152 $\pm$ 8     & 534 $\pm$ 31 & --      & 6.0  & 0.40 \\
\object{$o$ Aqr}        & \object{HD\,209\,409}  & 4.661      & B7IVe       & 133 $\pm$ 4           & 12942 $\pm$ 402 & 282 $\pm$ 20    & 391 $\pm$ 27 & 6       & 4.4  & 0.17 \\
\object{31 Peg}         & \object{HD\,212\,076}  & 4.685      & B2IV-Ve     & 497 $\pm$ 70          & 19270 $\pm$ 326 & 103 $\pm$ 6     & 439 $\pm$ 26 & 58      & 8.0  & 0.70 \\
\object{$\pi$ Aqr}      & \object{HD\,212\,571}  & 5.351      & B1III-IVe   & 239 $\pm$ 16          & 26061 $\pm$ 736 & 233 $\pm$ 15    & 558 $\pm$ 37 & 177     & 5.2  & 0.30 \\
\object{$\epsilon$ PsA} & \object{HD\,214\,748}  & 4.400      & B8Ve        & 149 $\pm$ 15          & 11966 $\pm$ 356 & 205 $\pm$ 13    & 317 $\pm$ 21 & 3       & 6.6  & 0.02 \\
\hline                  
\end{tabular}
}
Values taken from:
\tablefoottext{a}{CDS database;}
\tablefoottext{b}{\citet{vanLeeuwen2007}, except noted otherwise;}
\tablefoottext{c}{\citet{Fremat2005}, except noted otherwise;}
\tablefoottext{d}{\citet{Yudin2001};}
\tablefoottext{e}{SED fittings;}
\tablefoottext{f}{\citet{Astra2016};}
\tablefoottext{g}{\citet{Abt2002};}
\tablefoottext{h}{\citet{Chauville2001}.}
\end{table*}

\subsection{Disk parameters and properties}

For the 18 stars that show the Br$\gamma$ emission line, Fig.~\ref{fig:HD23630} and online Figs. \ref{fig:HD23862}-\ref{fig:HD214748} present  the ($u,v$) plane coverages, differential visibilities, differential phases and closure phases as a function of wavelength, together with the best-fitting model. Differential visibilities, differential phases and closure phases from our observations are plotted in blue, and the best-fitting kinematic model is overplotted in red.

In almost all cases, the data show a drop of visibility in the emission line caused by variations in either the disk extension or the relative flux between the line and the continuum \citep{Meilland2007b}. In these cases, the visibility drop is ``W'' shaped and the differential phase is ``S'' (or more complex) shaped. Both characteristics bring out a rotating equatorial disk \citep{Meilland2011}. 

\begin{figure}
\resizebox{\hsize}{!}{\includegraphics{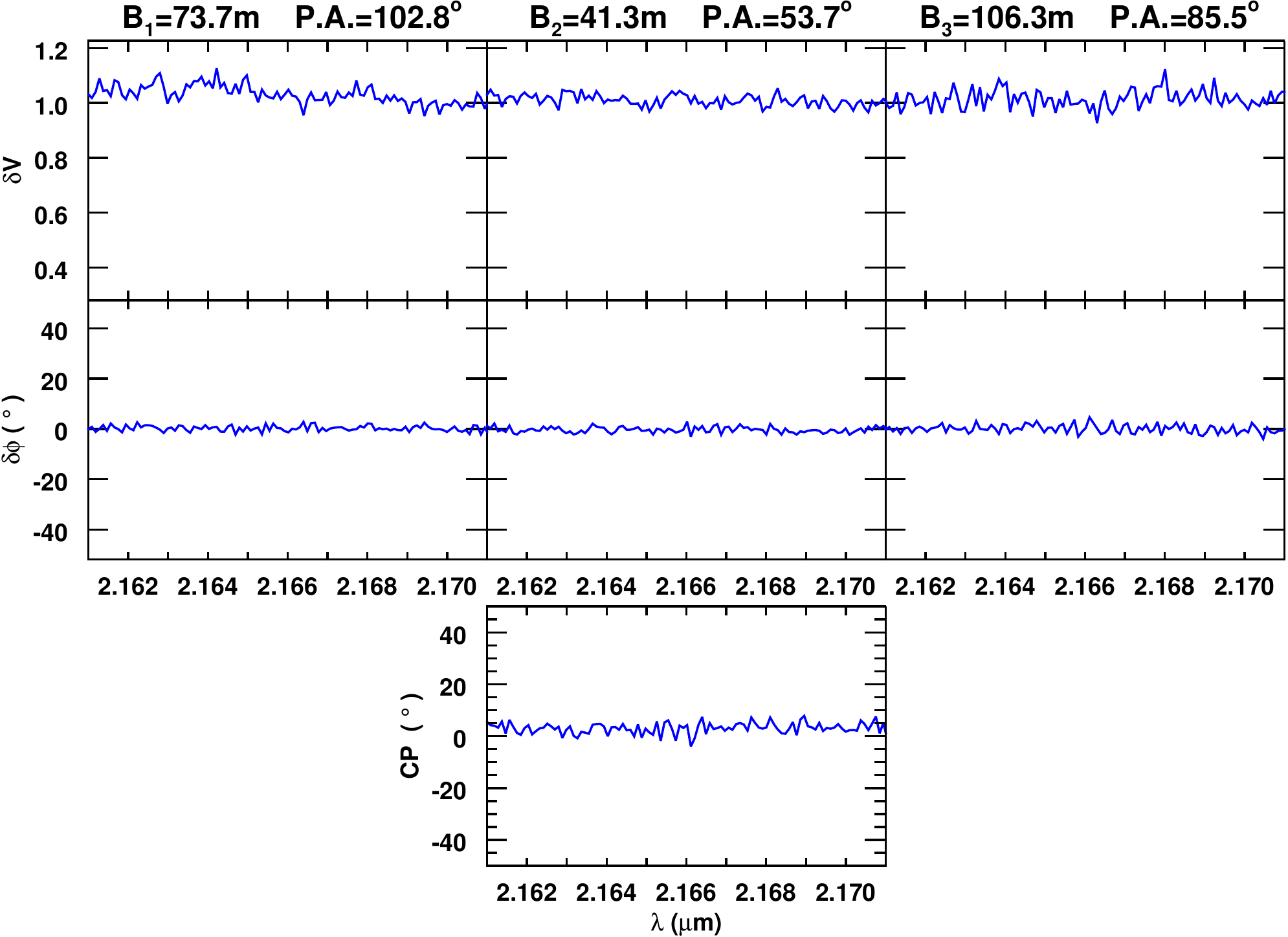}}
\caption{Differential visibility, differential phase and closure phase corresponding to the interferometric observations of \object{HD\,23\,408}, with the Br$\gamma$ line in absorption.}
\label{fig:HD23408}
\end{figure}

For the stars with the Br$\gamma$ line in absorption, Fig.~\ref{fig:HD23408} shows, as an example, the differential visibility, differential phase, and the closure phase as a function of wavelength for the unresolved disk of \object{HD 23408}.

The upper rows of Table~\ref{table:fit} summarize the parameters of the best-fitting model obtained from our axisymmetric kinematic model for each star with the Br$\gamma$ 
line in emission. Column 1 gives the HD number of each star; columns 2 and 3 list global parameters of the disk (inclination angle $i$ and the position angle $PA$); columns 4 and 5 give the kinematical parameters of the disk ($V_{\rm{rot}}$ and $j$); columns 6 and 7 provide values for the  major axis of the disk in the continuum ($a_{\rm{c}}$) and in the line ($a_{\rm{l}}$), respectively; columns 8 and 9 give Br$\gamma$ EW and V/R ratio; and column 10 lists the reduced $\chi^2$ of the fit.

To derive $a_{\rm{c}}$ we modeled the differential visibility normalized to the continuum, and the uncertainties (see Table~\ref{table:fit}) were obtained following the procedure described in Sect.~\ref{sec:meth}. In some cases, the uncertainties are not listed because the $a_{\rm{c}}$ values were not properly constrained. Therefore only a rough value (marked with $^*$ in Table~\ref{table:fit}) is given, which corresponds to the parameter obtained from a fitting model that minimizes the $\chi^2$ value. 
Despite the large uncertainty on the parameter $a_{\rm{c}}$, the parameters of the disk at the line frequency are still quite accurate, since they are determined independently of the $a_{\rm{c}}$ values (i.e., using the drop of visibility in the line).

\begin{table*}
\caption{Best-fitting parameters, with their respective errors, obtained from our axisymmetric kinematic model. In the case of \object{HD\,28497} we modified the used model to include a spiral above the Gaussian distribution (see Sect. \ref{sec:object}). Upper rows correspond to the disk parameters determined in this work, and the lower rows show the values from \citet{Meilland2012}}             
\label{table:fit}      
\centering          
\begin{tabular}{c|r@{$\pm$}lr@{$\pm$}l|r@{$\pm$}lr@{$\pm$}l|c|r@{$\pm$}lr@{$\pm$}lc|c} 
\hline \hline
& \multicolumn{4}{c|}{Global geometric} & \multicolumn{4}{c|}{Global kinematic} & K-band continuum & \multicolumn{5}{c|}{Br$\gamma$}    & $\chi^2$\\
& \multicolumn{4}{c|}{parameters}       & \multicolumn{4}{c|}{parameters}       & disk geometry    & \multicolumn{5}{c|}{disk geometry} &  \\
\hline
Parameter & \multicolumn{2}{c}{$i$}          & \multicolumn{2}{c|}{$PA$}         & \multicolumn{2}{c}{$V_{\rm{rot}}$}  & \multicolumn{2}{c|}{$j$} & $a_{\rm{c}}$         & \multicolumn{2}{c}{$a_l$}         & \multicolumn{2}{c}{$EW$} & V/R & - \\
          & \multicolumn{2}{c}{($^{\circ}$)} & \multicolumn{2}{c|}{($^{\circ}$)} &  \multicolumn{2}{c}{(km\,s$^{-1}$)} & \multicolumn{2}{c|}{-}   & ($D_{\star}$) & \multicolumn{2}{c}{($D_{\star}$)} & \multicolumn{2}{c}{($\AA$)} & - & - \\
\hline
\object{HD\,23630}  &  53& 5 & 110&5  & 270&25 &                  -0.45&0.05 &  1.3$\pm$1.0 & 7.0 &1.0 &  5.80&0.8 & 1.02 & 5.86 \\
\object{HD\,23862}  &  85& 5 & 333&10 & 410&50 &                  -0.47&0.05 &  4.0*        & 11.0&1.0 &  8.86&1.0 & 1.00 & 2.31 \\
\object{HD\,28497}  &  44&10 & 30 &10 & 600&50 &                  -0.35&0.05 &  6.0$\pm$2.0 & 11.0&2.0 & 14.00&3.0 & 1.29 & 0.96 \\
\object{HD\,30076}  &  38& 5 & 73 &4  & 400&20 &                  -0.45&0.05 &  6.0$\pm$1.0 & 10.0&0.5 & 20.67&3.0 & 1.06 & 0.70 \\
\object{HD\,32991}  &  40&10 & 157&5  & 500&50 &                  -0.50&0.05 &  2.0*        & 13.0&2.0 & 20.00&2.0 & 1.07 & 0.62 \\
\object{HD\,35439}  &  55& 5 & 45 &20 & 397&50 &                  -0.50&0.05 &  1.7*        & 4.5 &0.3 & 20.00&5.0 & 0.96 & 1.01 \\
\object{HD\,36576}  &  60& 5 & 198&10 & 380&50 &                  -0.40&0.02 &  2.0$\pm$1.0 & 13.0&1.0 & 15.00&3.0 & 1.01 & 1.60 \\
\object{HD\,37202}  &  67& 5 & 127&5  & 400&30 &                  -0.45&0.03 &  4.7*        & 4.8 &0.4 &  7.43&2.0 & 1.02 & 3.19 \\
\object{HD\,37490}  &  49& 5 & 305&5  & 308&30 &                  -0.50&0.05 &  1.0*        & 2.5 &0.5 & 11.62&3.0 & 1.02 & 0.97 \\
\object{HD\,41335}  &  68& 5 & 55 &5  & 400&30 &                  -0.60&0.03 &  2.0*        & 5.0 &0.4 & 17.00&4.0 & 0.95 & 1.98 \\
\object{HD\,45725}  &  72& 5 & 101&5  & 440&40 &                  -0.50&0.03 &  1.2*        & 12.0&1.0 & 22.70&2.0 & 0.97 & 2.71 \\
\object{HD\,60606}  &  70&10 & 240&10 & 390&20 &                  -0.50&0.05 &  2.0*        & 7.5 &0.5 & 17.00&2.0 & 1.03 & 1.23 \\
\object{HD\,66194}  &  55& 5 & 220&20 & 350&50 & \multicolumn{2}{c|}{-0.40*} &  3.0*        & 8.0 &1.0 & 10.73&2.0 & 1.01 & 0.78 \\
\object{HD\,68980}  &  22& 3 & 310&5  & 410&40 &                  -0.50&0.05 &  1.5*        & 3.0 &0.5 & 25.00&4.0 & 1.03 & 1.16 \\
\object{HD\,209409} &  70& 5 & 290&5  & 355&50 &                  -0.45&0.03 &  3.0*        & 8.0 &0.5 & 18.12&2.0 & 1.02 & 2.50 \\
\object{HD\,212076} &  22& 5 & 202&20 & 400&60 &                  -0.50&0.05 &  3.0*        & 4.0 &0.5 & 12.00&2.0 & 1.01 & 1.16 \\
\object{HD\,212571} &  34&10 & 87 &15 & 440&40 &                  -0.50&0.1  &  1.5*        & 2.0 &0.3 &  6.00&2.0 & 1.00 & 1.01 \\
\object{HD\,214748} &  73&10 & 67 &10 & 244&20 &                  -0.46&0.03 &  2.0*        & 3.9 &0.4 &  5.00&0.5 & 1.04 & 1.85 \\
\hline
\object{HD\,37795}  &  35&5  & \multicolumn{2}{c|}{10} & 350&10 & -0.5&0.1   & 2.0$\pm$0.5  &  5.5&0.3 & 7.0&0.5  & 1.04 & 4.0\\
\object{HD\,50013}  &  35&10 &  25&10 & 480&40 &                  -0.5&0.2   & 3.5$\pm$0.5  &  6.2&2.0 & 13.0&2.0 & 1.13-1.10 & 6.8\\
\object{HD\,89080}  &  65&10 &   5&5  & 300&20 &                  -0.45&0.1  & 3.0$\pm$1.0  &  6.5&1.0 & 5.8&0.5  & 1.04 & 1.1\\
\object{HD\,91465}  &  70&10 & -25&10 & 400&30 &                  -0.45&0.1  & 2.0$\pm$0.5  & 11.0&2.0 & 10.0&1.0 & $<$1 & 2.5\\
\object{HD\,105435} &  35&15 &  40&10 & 500&50 &                  -0.5&0.3   & 2.0$\pm$1.0  &  9.0&2.0 & 19.2&2.0 & $\sim$1& 2.3\\
\object{HD\,120324} &  25&5  &  80&15 & 510&20 &                  -0.5&0.1   & $<$3.0       &  4.0&1.0 & 5.6&0.3  & 1.03 & 1.3\\
\object{HD\,158427} &  45&5  &  88&2  & 480&20 &                  -0.5&0.1   & $<$2.0       &  5.8&0.5 & 14.5&1.0 & 1.01 & 1.7\\
\hline                  
\end{tabular}
\end{table*}

\subsection{Comments on individual objects}\label{sec:object}

\begin{itemize}

\item {\bf \object{HD\,23630} (\object{$\eta$ Tau})} is a Be star of spectral type B7\,III that belongs to the the Pleiades cluster.  It exhibits a H$\alpha$ line profile without significant changes on long timescales \citep{Tycner2005}. Using interferometric observations and a Gaussian fit for the H$\alpha$ emission, \citet{Tycner2005} found an angular diameter of 2.08$\pm$0.18 mas for the major axis of an elliptical Gaussian model. \citet{Grundstrom2006} obtained, using \citet{Tycner2005} data, an inclination angle $i~=~44^\circ$, and an interferometric ratio between the disk radius in the continuum $R_{\rm d}$ and the stellar radius $R_{\star}$ of $R_{\rm d}/R_{\star}=2.9\pm0.3$, while the disk extension is $5.4\pm0.7$ derived from the H$\alpha$ line EW.

\citet{Silaj2010} fitted the observed H$\alpha$ double-peaked emission line using a central star of B8 spectral type and a power-law mass density distribution for the disk, $n=3.5$. These authors also derived the following parameters: inclination angle $i~=~20^\circ$ and base density at the stellar equator $\rho_{0}= 10^{-11}\,$g\,cm$^{-3}$.

\citet{Touhami2013} made a K-band continuum survey using the CHARA Array long-baseline interferometer and found that the circumstellar disk of $\eta$ Tau was unresolved. 
 
For this object we performed three interferometric measurements with very high S/Ns using the VLTI/AMBER interferometer (see Fig.~\ref{fig:HD23630}). The best-fitting model provides the following geometrical and kinematical parameters for the disk: $PA~=~110^\circ$, $i~=~53^{\circ}$, $V_{\rm{rot}}~=~270$\,km\,s$^{-1}$, $j~=~-0.45$, $a_{\rm c}~=~1.3\,D_{\star}$, and $a_{\rm{l}}~=~7\,D_{\star}$. We noted that although the baselines were not oriented along the measured $PA$, the differential phase lose the ``S'' shape for those baselines that fully resolved the disk.
The derived inclination from the interferometric data is in agreement with the expected shape for the Br$\gamma$ double-peaked line profile (shown in Fig.~\ref{fig:brgamma}) that suggests a disk seen at intermediate inclination angle.\\


\item {\bf \object{HD\,23862} (\object{28 Tau})} is a Be star of spectral type B8\,V,  member of the Pleiades cluster. It is a single-lined spectroscopic binary with a low-mass companion \citep{Nemravova2010}. It has an orbital period of 218 days, a semi-amplitude of 5.9 km\,s$^{-1}$, and a large orbital eccentricity $e=0.60$ \citep{Katahira1996a,Katahira1996b}. The orbital solution proposed by \citet{Nemravova2010} has a semimajor axis between 223.5 $R_{\sun}$ and 225.5 $R_{\sun}$, and a periastron separation between 53.0 $R_{\sun}$ and 53.4 $R_{\sun}$, depending on the orbital inclination of the system.

The star underwent several B, Be, and Be shell transition phases with a period of 34 years \citep{Sadakane2005}. From the analysis of the long-term polarimetric observations, \citet{Hirata2007} interpreted the variation of the polarization angle in terms of the precession of the disk caused by the companion. 

The star disk was marginally resolved with the CHARA array. \citet{Touhami2013} derived a disk-to-star radius ratio of $1.879$, adopting a $PA~=~159^{\circ}$  and axial ratio $r=0.438$. 
Fitting the H$\alpha$ line profile, \citet{Silaj2014} obtained a model with an inclination angle $i~=~76^\circ$, a disk base density of $6.2\times10^{-12}\,$g\,cm$^{-3}$, and a power-law index of 2.5.

For this object we obtained two interferometric measurements with a good S/N (see Fig.~\ref{fig:HD23862}). The best-fitting model provides the following geometrical and kinematical parameters for the disk: $PA~=~333^{\circ}$, $i~=~85^{\circ}$, $V_{\rm{rot}}~=~410$\,km\,s$^{-1}$, $j~=~-0.47$, $a_{\rm c}~=~4\,D_{\star}$, and $a_{\rm{l}}~=~11\,D_{\star}$. The Br$\gamma$ line profile  exhibits a double-peaked emission with a shell absorption line at its center (Fig.~\ref{fig:brgamma}), typically seen in Be stars at a high inclination angle. \\


\item {\bf \object{HD\,28497} (\object{228\,Eri})} is a Be star of spectral type B2. This star has shown variable emission and shell episodes. \citet{Andrillat1990} derived a size for the H$\alpha$ region between 1.79 and 4.79 stellar radii. \citet{Silaj2010} fitted the H$\alpha$ double-peaked emission using as a best-fitting model a central star with spectral type B0 and the following disk parameters $i~=~45^{\circ}$, $\rho_0~=~10^{-10}\,$g\,cm$^{-3}$, and $n=4$, while \citet{Vieira2017} obtained $\log\,\rho_0=-10.7\pm0.3$, and $n=3.4\pm0.1$ from the IR continuum radiation.

We obtained one interferometric measurement with a high S/N (see Fig.~\ref{fig:HD28497}). The ``W'' and `S'' shapes from the differential visibility and differential phases were very asymmetrical, being the amplitude smaller in the red than the blue part of the spectrum. Both features were impossible to fit with a simple model. The Br$\gamma$ line was also clearly asymmetrical, with a $V/R \sim 1.29$ (see Fig.~\ref{fig:brgamma}). To improve the fit of our data we considered a disk model with one-arm over-density, like the one developed by \citet{Okazaki1997}. We considered an antisymmetric spiral-like intensity distribution similar to the one given by \citet[][in their Eq. 5]{Stee2013}. The best-fitting model provides the following geometrical and kinematical parameters for the disk: $PA~=~30^{\circ}$, $i~=~44^{\circ}$, $V_{\rm{rot}}~=~600$\,km\,s$^{-1}$, $j~=~-0.35$, $a_{\rm c}~=~6.0\,D_{\star}$, and $a_{\rm{l}}~=~11\,D_{\star}$. The arm was oriented with an angle $\phi=50^{\circ}$ from the major axis of the envelope.\\


\item {\bf \object{HD\,30076} (\object{56\,Eri})} is a B2\,V Be star. \citet{Andrillat1990} derived a size for the H$\alpha$ region of between 1.37 and 6.0 stellar radii. \citet{Silaj2010} fitted the H$\alpha$ double-peaked emission line using a central star of spectral type B2 and disk parameters with $i~=~20^{\circ}$, $\rho_0~=~5\times10^{-10}\,$g\,cm$^{-3}$, and $n=3.5$. From a statistical study, \citet{Jones2011} found that this star is variable with a confidence level of 95\%.

We performed one interferometric measurement (see Fig.~\ref{fig:HD30076}). The disk parameters for the best-fitting model are: $PA~=~73^{\circ}$, $i~=~38^{\circ}$, $V_{\rm{rot}}~=~400$\,km\,s$^{-1}$, $j~=~-0.45$, $a_{\rm c}~=~6.0\,D_{\star}$, and $a_{\rm{l}}~=~10\,D_{\star}$. The position angle of the B$_1$ baseline ($PA_1=65.4^\circ$) was oriented close to the measured $PA$ of the disk, thus the ``W'' and ``S'' shaped features in the visibility drop and differential phase were clearly present. The observed double-peaked Br$\gamma$ line profile (Fig.~\ref{fig:brgamma}) is consistent with the line-of-sight inclination angle obtained from the interferometric model.\\


\item {\bf \object{HD\,32991} (105\,Tau)} is an early-type Be shell star (B2\,V). The star exhibits photometric variability as $\gamma$ Cas type variables \citep{Lefevre2009}. The best-fitting model parameters derived by \citet{Silaj2010} from the H$\alpha$ emission line are: a central B2 type star plus a disk with $i~=~20^\circ$, $\rho_{0}=5\times10^{-10}\,$g\,cm$^{-3}$, and $n=3.5$. This object has not previous disk size determinations.

We obtained one interferometric measurement (see Fig.~\ref{fig:HD32991}). The best-fitting model provides the following set of disk parameters: $PA~=~157^{\circ}$, $i~=~40^{\circ}$, $V_{\rm{rot}}~=~500$\,km\,s$^{-1}$, $j~=~-0.5$, $a_{\rm c}~=~2.0\,D_{\star}$, and $a_{\rm{l}}~=~13\,D_{\star}$. As the observed baseline B$_2$ (with $PA~=~-18.7^{\circ}$), oriented along the major axis of the disk, was the shortest one, the ``W'' and ``S'' shaped features in differential visibility and differential phase are very weak. 
The Br$\gamma$ line single-peaked profile (Fig.~\ref{fig:brgamma}) is consistent with the small inclination angle obtained using our interferometric model.\\


\item {\bf \object{HD\,35439} (\object{$\psi_{01}$\,Ori})} is a variable star of spectral type B1\,V, passing from Be $\rightarrow$ B $\rightarrow$ Be phases.  \citet{Andrillat1990} derived a size for the H$\alpha$ region of between 1.31 and 3.66 stellar radii. \citet{Silaj2010} fitted a double-peaked H$\alpha$ line profile using a B0-type central star with the following disk parameters $i~=~45^\circ$, $\rho_{0}=5\times10^{-11}\,$g\,cm$^{-3}$, and $n=3.5$. The value of $\rho_{0}$ is greater than the one derived by \citet{Vieira2017} from the WISE SED ($\log\,\rho_0=-11$ and $n=3.4$). On the other hand, \citet{Arcos2017} found two different set of parameters: $i~=~50^\circ$, $\rho_{0}=2.5\,10^{-11}\,$g\,cm$^{-3}$, $n=2.5$ by modeling observations of the H$\alpha$ line taken in 2012 and $i~=~50^\circ$, $\rho_{0}=5\times10^{-12}\,$g\,cm$^{-3}$, and $n~=~2$ from data taken in 2015.

We performed one interferometric measurement (see Fig.~\ref{fig:HD35439}). The best-fitting disk model parameters are: $PA~=~45^{\circ}$, $i~=~55^{\circ}$, $V_{\rm{rot}}~=~397$\,km\,s$^{-1}$, $j~=~-0.5$, $a_{\rm c}~=~1.75\,D_{\star}$, and $a_{\rm{l}}~=~4.5\,D_{\star}$. The B$_3$ baseline had an orientation close to the $PA$ of the major axis of the disk, therefore pronounced ``W'' and ``S'' shaped features were observed in the differential visibility and differential phase. The double-peaked Br$\gamma$ line profile shown in Fig.~\ref{fig:brgamma} is consistent with the shape expected from a disk with inclination angle derived from the interferometric model.\\


\item {\bf \object{HD\,36576} (\object{120\,Tau})} is a B2\,IV-V Be star. This star presents multiperiodicity consistent with multimodal nonradial pulsations \citep{Bossi1989}. The size of the H$\alpha$ line forming region was estimated to be between 1.28 and 3.59 stellar radii by \citet{Andrillat1990}. \citet{Silaj2010} modeled the single-peaked H$\alpha$ line emission using a power-law density distribution for a central B2 type star plus a disk with parameters: $i~=~45^{\circ}$, $\rho_0~=~5\times10^{-10}\,$g\,cm$^{-3}$, and $n=3.5$. However, from the SED model, \citet{Vieira2017} obtained $\log\,\rho_0=-11.4$, and $n=2.36$.

We performed one measurement using the VLTI/AMBER interferometer (see Fig.~\ref{fig:HD36576}). The best-fitting model for the disk provides the following geometrical and kinematical parameters: $PA~=~198^{\circ}$, $i~=~60^{\circ}$, $V_{\rm{rot}}~=~380$\,km\,s$^{-1}$, $j~=~-0.4$, $a_{\rm c}~=~2.0\,D_{\star}$, and $a_{\rm{l}}~=~13.0\,D_{\star}$. Although the baselines were not oriented close to the measured $PA$ of the disk, the ``W'' shape in differential visibility and ``S'' shape in the differential phase are clearly seen in the largest baseline. The double peaked Br$\gamma$ line profile is consistent with the shape expected from a disk with an intermediate angle orientation.\\


\item {\bf \object{HD\,37202} (\object{$\zeta$ Tau})} is one of the most observed Be stars. It has a spectral type B1\,IV. It has been known as a spectroscopic binary for decades. \citet{Harmanec1984} derived an orbital period of 132.9735 days and an eccentricity $e\approx 0.15$. 

Complex variations in the H$\alpha$ shell spectrum and long-term V/R variations were observed \citep{Ballereau1987,Tycner2005}. As the star is close enough \citep[$d=136\pm16$\,pc,][]{vanLeeuwen2007}, the circumstellar disk can be resolved with the actual instruments, and spectro-interferometry extends this resolution to the disk dynamics. The limb-darkened photospheric diameter was estimated at 0.4 mas \citep{Tycner2004, Gies2007}, and this corresponds to a radius of 5.5 to 6 $R_{\sun}$ at the Hipparcos distance.

Using interferometric observations and fitting Gaussians from the H$\alpha$ emission disks, \citet{Tycner2005} found a value for the major axis of 3.14$\pm$0.21 mas. Based on \citet{Tycner2005} data, \citet{Grundstrom2006} derived an inclination angle $i~=~75^\circ$, and interferometric radius of $R_{\rm d}/R_{\star}=7.3\pm0.5$, while from the H$\alpha$ line EW they calculated a disk extension of $R_{\rm d}/R_{\star}= 8.7\pm1.5$.
\citet{Touhami2011} presented an analysis of $K$-band observations of the continuum emission from the circumstellar gas disk obtained with CHARA Array. Using a radiative transfer code for a parameterized version of the viscous decretion disk model they derived the following parameters: $\rho_0=1.4\times10^{-10}\,$g\,cm$^{-3}$, $n=2.9$, $i=87^{\circ}$, and disk temperature $T_d~=~2/3\,T_{\rm {eff}}$.

\citet{Gies2007}, using CHARA in the $K$-band, measured the semi-major axis of the disk to 1.79 mas or 4.5 $R_{\star}$. This value is lower than the estimated Roche-lobe radius for the primary star of the system, which \citet{Tycner2004} computed as 144 $\pm$ 12 $R_{\sun}$ (5.3 mas), at a binary separation of 254 $\pm$ 20 $R_{\sun}$ (9.2 mas). Considering this separation between both components the secondary star would be out of the circumstellar disk, in agreement with other models \citep{Castle1977}.
\citet{Stefl2009} studied three V/R cycles between 1997 and 2008. They found that after each minimum in V/R, the shell absorption weakens and splits into two components, leading to three emission peaks. Also, the phasing of the Br$\gamma$ emission showed that the photocenter of the line-emitting region lied within the plane of the disk but was offset from the continuum source. The plane of the disk remained stable throughout the observed V/R cycles.
\citet{Schaefer2010} resolved the circumstellar disk between 2007 and 2009 with the MIRC beam combiner at the CHARA array. They fit a nearly edge-on disk, with a FWHM major axis $\sim$\,1.8\,mas at the $H$-band. In addition, they found a correlation between the position angle of the disk and the spectroscopic V/R ratio, suggesting that the disk is precessing.

For this object, we obtained three interferometric measurements, two with a quite high S/N ratio and one with a lower S/N (see Fig.~\ref{fig:HD37202}). The best-fitting model provides the following geometrical and kinematical parameters for the disk: $PA~=~127^{\circ}$, $i~=~67^{\circ}$, $V_{\rm{rot}}~=~400$ km\,s$^{-1}$, $j$=-0.45, $a_{\rm c}~=~4.7\,D_{\star}$, and $a_{\rm{l}}~=~4.8\,D_{\star}$. The line profile is typical for a  Be star seen at high inclination angle (double-peaked with a shell absorption line at its center), in agreement with its Be shell classification. For the observations with good S/N, the ``W'' shape in the drop of the differential visibility and ``S'' shape of the differential phase are clear. \\


\item {\bf \object{HD\,37490} (\object{$\omega$ Ori})} is a classical B3\,V emission line star, which has been studied over many years. 
It presents short (~1 day) and mid-term (~11 months) photometric variability \citep{Balona1987, Bergin1989, Balona1992}. These variations were associated to nonradial pulsations, and the increase or decrease of the disk size. \citet{Andrillat1990} estimated a size for the H$\alpha$ region of between 1.07 and 2.27 stellar radii.
\cite{Neiner2002} determined the stellar parameters employing two methods: astrophysical formulae and the Barbier-Chalonge-Divan (BCD) spectrophotometric system.
With the stellar parameters derived from the BCD system, $\log T_{\rm eff}$=4.306 $\pm$ 0.016 and $\log$ g=3.48 $\pm$ 0.03, they determined $\Omega$=0.83$\Omega_{c}$, $i~=~32\pm15^\circ$, $R_{\star}=6.84\pm0.25R_{\odot}$, and $M_{\star}=8.02\pm0.25M_{\odot}$ using evolutionary tracks from \citet{Schaller1992}.

\cite{Balona2002} found a $V\,\sin\,i =173 \pm 2$ km\,s$^{-1}$ from nonemission helium lines, and $V\,\sin\,i = 226 \pm 7$ km\,s$^{-1}$ from the metal lines. They suggested that even the helium lines with weak emission are likely contaminated by the emission from the gas disk. \cite{Neiner2003} found evidence for the presence of a weak magnetic field, sinusoidally varying with a period of 1.29 days. However, new observations were unable to confirm the presence of this magnetic field \citep{Neiner2012}.
\citet{Silaj2010} fitted the double-peaked H$\alpha$ emission line using a central star of spectral type B2, $i~=~20^{\circ}$, $\rho_0~=~10^{-11}\,$g\,cm$^{-3}$, and $n=3.5$. \citet{vanBelle2012} suggested that this object is a possible target for interferometric observations, with an estimated angular size of 0.5 mas and flattening $Rb/Ra -1$=0.13.

For this star we obtained one interferometric measurement (see Fig.~\ref{fig:HD37490}) and derived the following parameters for the disk: $PA~=~305^{\circ}$, $i~=~49^{\circ}$, $V_{\rm{rot}}~=~308$\,km\,s$^{-1}$, $j~=~-0.5$, $a_{\rm c}~=~1.0\,D_{\star}$, and $a_{\rm{l}}~=~2.5\,D_{\star}$. The shape of the double-peaked line profile indicates that the object is seen from an intermediate inclination angle, consistent with the value we derived and a more precise determination of $i~=~42\pm7^{\circ}$ obtained by \citet{Neiner2003}.\\


\item {\bf \object{HD\,41335} (\object{V696 Mon})} is an interacting binary system \citep{Peters1972}. \cite{Peters1983} measured the radial velocity and obtained an orbital period of 80.86 days. The system consists of a primary star B1.5IV-V of $8-11\,M_{\sun}$ and a secondary of $1\,M_{\sun}$. They measured a separation between the components of 0.84 UA ($\sim\,180\,R_{\sun}$).
Studying the system SED, \cite{Waters1991} showed that the companion could be a hot subdwarf (Sd) star but not a cool giant star. Based on UV and H$\alpha$ observations, \cite{Peters2016} was able to detect the weak signal of the spectral lines corresponding to a hot Sd star, and proposed a primary star with 9 $M_{\sun}$ and a secondary with 0.7 $M_{\sun}$. They proposed that this Sb  star creates a one-armed spiral feature, a tidal wake in the disk of the Be star \citep[][named Be+SdO binary system or $\phi$ Per type]{Thaller1995}.
\citet{Andrillat1990} derived a size for the H$\alpha$ circumstellar envelope of between 1.66 and 3.43 stellar radii.
From the SED, \citet{Vieira2017} obtained $\log\,\rho_0=-10.3\pm0.3$, and $n=3.0\pm0.1$. Furthermore, \citet{Arcos2017} derived $\rho_0=5\times10^{-12}\,$g\,cm$^{-3}$, $n=2.0$, and $i~=~80^\circ$.

We performed one interferometric measurement for this object (Fig.~\ref{fig:HD41335}). The best-fitting model provides the following geometrical and kinematical parameters of the disk: $PA~=~55^{\circ}$, $i~=~68.5^{\circ}$, $V_{\rm{rot}}~=~400$ km\,s$^{-1}$, $j$=-0.6, $a_{\rm c}~=~2.0\,D_{\star}$, and $a_{\rm{l}}~=~5.0\,D_{\star}$. The double-peaked line profile, the drop of the differential visibility and phase variations present small asymmetries. The largest baselines (B$_1$ and B$_3$) were oriented with an angle close to the disk $PA$. Therefore, the ``W'' shape in the drop of differential visibility and the ``S'' shape of the differential phase are clear. The double-peaked line profile is consistent with the line shape seen at an intermediate inclination angle. 
Although \citet{Peters2016} proposed the presence of a one-armed spiral structure, we have not needed to consider an overdensity to fit our data.\\


\item {\bf \object{HD\,45725} (\object{$\beta_{01}$ Mon})} is a Be-shell star of spectral type B4\,V. This object has been catalogued as a triple star by \citet{Struve1925}. 
\citet{Abt1984} measured separations of 7.1'', 2.8'', and 25.9'' between the system components. Assuming that these separations were the semimajor axis of the orbits, they estimated periods of 10\,000, 2\,600, and 100\,000 years, respectively. \citet{Outmaijer2010} catalogued the system as a binary and fit bidimensional Gaussians to the spatial profiles, finding a separation between the components of 7.14 arcsec and a position angle of 133$^\circ$. 
\citet{Andrillat1990} derived a size for the H$\alpha$ region of between 1.86 and 3.64 stellar radii.

\citet{vanBelle2012} suggested that \object{HD\,45\,725} is a good candidate for interferometric observations, with estimations for the projected rotational velocity of $V\,\sin\,i=325$~km\,s$^{-1}$, an angular size of 0.70~mas, and a flattening of $Rb/Ra -1$=0.17. 
\citet{Arcos2017} obtained the following disk parameters $\rho_0~=~5\times10^{-12}\,$g\,cm$^{-3}$, n= 2, and $i~=~70^{\circ}$.
   
We obtained two interferometric measurements, one with a relatively high S/N and the other with a low value (see Fig.~\ref{fig:HD45725}). The best-fitting disk model provides the following geometrical and kinematical parameters: $PA~=~101^{\circ}$, $i~=~72^{\circ}$, $V_{\rm{rot}}~=~440$ km\,s$^{-1}$, $j$=-0.5, $a_{\rm c}~=~1.2\,D_{\star}$, and $a_{\rm{l}}~=~12.0\,D_{\star}$. 
We note that the Br$\gamma$ line does not present the shell absorption profile expected for our high value of the inclination angle. The baselines B$_1$ and B$_4$ were very similar and were oriented close to the $PA$. The baselines B$_2$ and B$_6$  had almost the same orientation, but different lengths, and therefore the shape of the differential visibility and differential phase are more evident in the B$_6$ base. \\


\item {\bf \object{HD\,60606} (\object{OW Pup})} is a B2V Be star. \citet{Lefevre2009} catalogued this object as a $\gamma$ Cas-type variable. The SED was modeled by \citet{Vieira2017} who derived $\log\,\rho_0=-11.7\pm0.1$, and $n=2.5\pm0.1$, while from the H$\alpha$ line \citet{Arcos2017} obtained the following disk parameters: $\rho_0~=~1\times10^{-10}\,$g\,cm$^{-3}$, $n= 3$, and $i~=~70^{\circ}$. There are no disk size determinations  available for this object.
   
Based on one interferometric measurement (Fig.~\ref{fig:HD60606}), the best-fitting model provides the following geometrical and kinematical parameters for the disk: $PA~=~240^{\circ}$, $i~=~70^{\circ}$, $V_{\rm{rot}}~=~390$ km\,s$^{-1}$, $j$=-0.5, $a_{\rm c}~=~2.0\,D_{\star}$, and $a_{\rm{l}}~=~7.5\,D_{\star}$. The longest baseline was also the closest to the modeled $PA$, and therefore the ``W'' shape of the differential visibility and the ``S'' shape of the differential phase are well defined.\\ 


\item {\bf \object{HD\,66194} (\object{V374 Car})} is a Be star of spectral type B3V and is a member of the open cluster \object{NGC 2516}. It has been identified as a blue straggler star by \citet{Ahumada2007}. \citet{Gonzales2000} reported some variability in the measured radial velocities and suggested that this object could be a spectroscopic binary with mass transfer in a close system. The star exhibits photometric variability as $\gamma$ Cas-type variables \citep{Lefevre2009}. This object has not previous disk size determinations.

We performed two interferometric observations with a low S/N ratio (Fig.~\ref{fig:HD66194}). The best-fitting model provides the following geometrical and kinematical parameters for the disk: $PA~=~220^{\circ}$, $i~=~55^{\circ}$, $V_{\rm{rot}}~=~350$\,km\,s$^{-1}$, $j~=~-0.4$, $a_{\rm c}~=~3.0\,D_{\star}$, and $a_{\rm{l}}~=~9.0\,D_{\star}$. Both observations were carried out with a very similar instrumental configuration. Therefore, none of them show clear signs of either a drop in the differential visibilities or the typical ``S'' shape in the differential phase. The Br$\gamma$ line profile is consistent with the intermediate inclination angle obtained from our interferometric model (see Fig.~\ref{fig:brgamma}). \\


\item {\bf \object{HD\,68980} (\object{MX Pup})} is a spectroscopic binary of spectral type B1.5\,IVe \citep{Slettebak1982} with a period of 5.1526 days and a very excentric orbit \citep{Carrier2002}. Its light curve also shows long-term variability  with  a  time  scale  of  about  9  years  accompanied by $V/R$ variations \citep{Mennickent1997,Hanuschik1995}.
\citet{Silaj2010} used a simple power-law model to fit the one-peaked emission H$\alpha$ line profile, using a central star of B2 spectral type with the following disk parameters: $i~=~20^{\circ}$, $\rho_0~=~2\times10^{-10}\,$g\,cm$^{-3}$, and $n=3.5$. \citet{Arcos2017} obtained two set of independent parameters for the disk, $\rho_0~=~5\times10^{-12}\,$g\,cm$^{-3}$, $n= 2$, and $i~=~50^{\circ}$ using observations taken in 2013, and $\rho_0~=~2.5\times10^{-11}$, $n= 2.5$, and $i~=~50^{\circ}$ with data acquired in 2015. Whilst \citet{Vieira2017} derived from the SED a $\log\,\rho_0~=~-11.5$, and $n=2.8$. There are no disk size determinations  available for this object.

We obtained one interferometric measurement (Fig.~\ref{fig:HD68980}). The best-fitting model provides the following geometrical and kinematical parameters for the disk: $PA~=~310^{\circ}$, $i~=~22^{\circ}$, $V_{\rm{rot}}~=~410$\,km\,s$^{-1}$, $j~=~-0.5$, $a_{\rm c}~=~1.5\,D_{\star}$, and $a_{\rm{l}}~=~3.0\,D_{\star}$. The longest B$_3$ baseline was oriented almost perpendicular to the $PA$ of the disk. The differential visibility has a ``V'' shape and the differential phase has a small amplitude ``S'' shape. The Br$\gamma$ line profile is in emission with a small asymmetry (see Fig.~\ref{fig:brgamma}). Its shape is consistent with a line profile seen at the small inclination angle found. \\
 

\item {\bf \object{HD\,209409} (\object{$o$ Aqr})} is a rapidly rotating Be shell star classified as a B7\,IV. The star has a stable H$\alpha$ emission and does not show $V/R$ variations \citep{Rivinius2006}. The star has been fully resolved by interferometry in many studies. \citet{Touhami2013} derived a disk-to-star radius ratio $R_d/R_s=5.776$, and $PA~=~107.5^{\circ}$, while \citet{Cyr2015} found an axis ratio of 0.25 in the $K$-band. \citet{Sigut2015} combined H$\alpha$ and interferometric data covering a period of 7 years and concluded that all the observations were consistent with a circumstellar disk seen at an inclination of $75^{\circ}$ and $PA~=~110^{\circ}$ with $\rho_{0}=6.6\times10^{-11}\,$g\,cm$^{-3}$, and $n=2.7$. \citet{Meilland2012} obtained the following parameters for the disk: $PA~=~120^{\circ}$, $i~=~70^{\circ}$, $V_{\rm{rot}}~=~400$\,km\,s$^{-1}$, $j~=~-0.5$, $a_{\rm c}<10\,D_{\star}$, and $a_{\rm{l}}~=~14\,D_{\star}$. Alternatively, \citet{Yudin2001} found $PA~=~177^{\circ}$ for the observed polarization.
From line profile fittings, \citet{Silaj2010} modeled the double-peaked H$\alpha$ line profile with a central B8 type star and disk parameters with $i~=~45^\circ$, $\rho_{0}=5\times10^{-11}\,$g\,cm$^{-3}$, and $n=3.5$, while \citet{Arcos2017} obtained a lower base density and index power ($\rho_{0}=5\times10^{-10}\,$g\,cm$^{-3}$, and $n=2.0$), and two different inclinations: $i~=~25^{\circ}$ (in 2013), and $i~=~50^{\circ}$ (in 2015). 

For this target we obtained seven interferometric measurements with high S/Ns (Fig.~\ref{fig:HD209409_1} and \ref{fig:HD209409_2}). From the  best-fitting disk model we derived: $PA~=~290^{\circ}$, $i~=~70^{\circ}$, $V_{\rm{rot}}~=~355$\,km\,s$^{-1}$, $j~=~-0.45$, $a_{\rm c}~=~3.0\,D_{\star}$, and $a_{\rm{l}}~=~8.0\,D_{\star}$. The baselines B$_1$, B$_4,$ and B$_7$ were oriented close to the $PA$ of the disk, and therefore the ``W'' shape of the drop of the differential visibility and the ``S'' shape of the differential phase are well defined. The B$_{11}$ baseline was one of the largest and was oriented almost perpendicularly to the measured $PA$, so the drop of the differential visibility has a ``V'' shaped structure and the ``S'' shaped feature of the differential phase has a small amplitude. The double-peaked Br$\gamma$ line profile, the ``W'' shaped feature in the drop of the differential visibility and the ``S'' shaped feature of the differential phase are quasi-symmetric. This indicates that the object is seen from a high inclination angle and it does not present major inhomogeneity in the disk.\\


\item {\bf \object{HD\,212076} (\object{31\,Peg})} is a Be star with a spectral type B2\,IV-V. The star disk was marginally resolved with the CHARA array. To derive a disk-to-star radius ratio of $1.852$, \citet{Touhami2013} should adopt a $PA~=~148^{\circ}$ and axial ratio $r=0.955$. \citet{Silaj2010} fitted the H$\alpha$ emission line assuming a circumstellar disk surrounding a B2 type-star with $i~=~20^{\circ}$, $\rho_0~=~10^{-11}\,$g\,cm$^{-3}$, and $n=2.5$. These disk parameters differ from those found by \citet[][$n=2$, $i~=~30^{\circ}$, $\rho_0~=~2.5\,10^{-11}\,$g\,cm$^{-3}$ in 2012, and $\rho_0~=~2.5~\times~10^{-12}\,$g\,cm$^{-3}$ in 2015]{Arcos2017} and by \citet[][$\log\,\rho_0=-11.7$, and $n=2.13$.]{Vieira2017}. 

We obtained five interferometric observations, three with high S/N and two with low S/N ratios. The best-fitting model corresponds to an equatorial disk with $PA~=~202^{\circ}$, $i~=~22^{\circ}$, $V_{\rm{rot}}~=~400$\,km\,s$^{-1}$, $j~=~-0.5$, $a_{\rm c}~=~3.0\,D_{\star}$, and $a_{\rm{l}}~=~4.0\,D_{\star}$. The shapes of ``W'' in the visibility drop and ``S'' in the differential phase are seen in a few baselines (generally the largest), as shown in Fig.~\ref{fig:HD212076}.\\


\item {\bf \object{HD\,212571} (\object{$\pi$ Aqr})} is a Be star of spectral type B1\,III-IV. From the analysis of the H$\alpha$ line profile, \citet{Bjorkman2002} suggested that \object{HD\,212571} may be a binary system with an orbital period of 84.1 days, consisting of stars with masses of $M_{1}\sin^3\,i=12.4\,$ M$_{\sun}$, and $M_{2}\sin^3\,i=2.0$ M$_{\sun}$. They also suggested that photometric, spectroscopic, and polarimetric variations observed during the second half of the twentieth century might be due to variable mass transfer between the binary components.
\citet{Wisniewski2010} found that the timescale of disk-loss events corresponds to almost 29 complete orbits of the binary companion. They also found that the position angle of the intrinsic polarization is $P_{\star}=166.7^\circ$, indicating that the disk is oriented on the sky at a position angle of $PA~=~76.7^\circ$.
  
\citet{Silaj2010} modeled the H$\alpha$ double-peaked emission line using a B0-type central star and an equatorial disk with $\rho_0=10^{-11}\,$g\,cm$^{-3}$, and $n=3.5$, seen with an inclination angle $i~=~45^{\circ}$. \citet{Arcos2017} derived three different sets of parameters, using spectra taken at different epochs: $\rho_0~=~2.5\,10^{-11}\,$g\,cm$^{-3}$, $i~=~60^{\circ}$, and $n=2.5$ for the epochs 2012 and 2015, and $\rho_0~=~7.5\,10^{-12}\,$g\,cm$^{-3}$, $i~=~60^{\circ}$, and $n=2.5$ for 2013, while \citet{Vieira2017} obtained  $\log\,\rho_0=-12.1$ and $n=3.5  $ from the SED. The disk size has not been determined for this object.

We performed two interferometric measurements with good S/N (Fig.~\ref{fig:HD212571}) and derived the following geometrical and kinematical parameters for the disk: $PA~=~87^{\circ}$, $i~=~34^{\circ}$, $V_{\rm{rot}}~=~440$\,km\,s$^{-1}$, $j~=~-0.5$, $a_{\rm c}~=~1.5\,D_{\star}$, and $a_{\rm{l}}~=~2.0\,D_{\star}$. The ``W'' shape in the drop of the differential visibility is barely perceptible in the B$_3$ and B$_4$ baselines. The ``S'' shape of the differential phase is present only in a few baselines. The double-peaked Br$\gamma$ line profile is very weak, as shown in Fig.~\ref{fig:brgamma}.\\


\item {\bf \object{HD\,214748} (\object{$\epsilon$ PsA})} is a fast-rotating B-type star with a stellar classification B8\,Ve. \citet{Arcos2017} modeled the H$\alpha$ line emission using a power-law density distribution with $\rho_0~=~2.5\,10^{-10}$gr\,cm$^{-3}$, $n=3.5$, and $i~=~50^{\circ}$. This disk base density is higher than the one calculated by \citet{Vieira2017} ($\log\,\rho_0=-11.9$, and $n=2.7$) based on a pseudo-photosphere model for the disk continuum emission.

We performed five interferometric measurements with a high S/N (see Fig.~\ref{fig:HD214748}). The  best-fitting model provides the following geometrical and kinematical parameters for the equatorial disk: $PA~=~67^{\circ}$, $i~=~73^{\circ}$, $V_{\rm{rot}}~=~244$\,km\,s$^{-1}$, $j=-0.46$, $a_{\rm c}~=~2.0\,D_{\star}$, and $a_{\rm{l}}~=~3.9\,D_{\star}$. The B$_{10}$ baseline was oriented close to the observed $PA$ of the disk. The Br$\gamma$ line is in emission and exhibits a strong asymmetry (see Fig.~\ref{fig:brgamma}).

\end{itemize}


\subsection{Polarization angle}

\begin{figure}
\resizebox{\hsize}{!}{\includegraphics{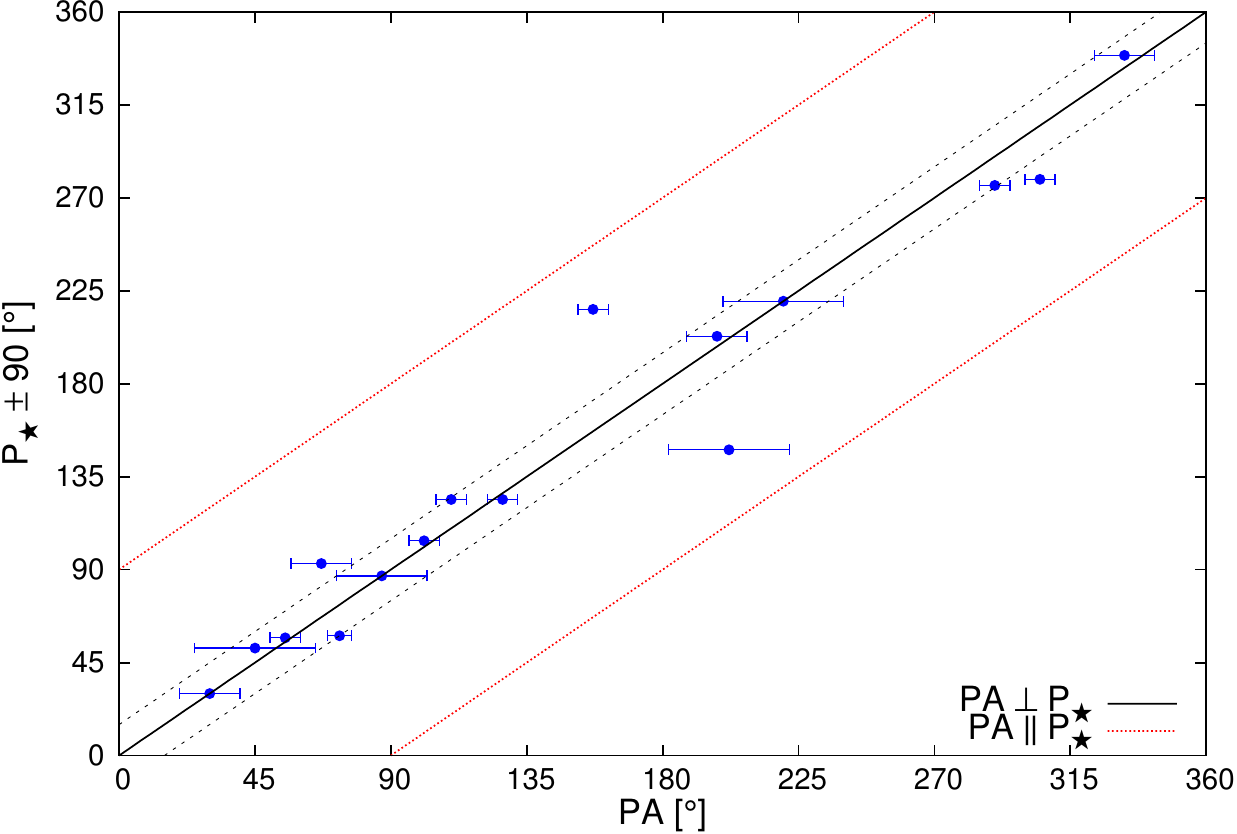}}
\caption{Position angle of the disk $PA$ vs. polarization angle $P_\star$. The solid black line corresponds to the case where $PA$ is perpendicular to the polarization measurement ($PA$=$P_\star \pm 90$). The solid red line corresponds to the situation when $PA$ is parallel to the polarization measurement ($PA$=$P_\star \pm 180$). The dotted lines correspond to an uncertainty of $15^{\circ}$.}
\label{fig:PA}
\end{figure}

The polarization position angle in an optically thin axisymmetric disk is usually perpendicular to the plane of the disk since it mainly comes from electron scattering \citep{Brown1977}. However, \citet{Delaa2011} reported that for \object{48 Per} the major axis is oriented parallel to the polarization angle, and proposed that this might be due to the disk optical thickness. In almost all cases the disk major axes of our star sample seem to be roughly perpendicular to the polarization measurement, as shown in Fig.~\ref{fig:PA}. Exceptions are \object{HD\,32991}, and \object{HD\,212076}. Both objects have small inclination angles that make the determination of the interferometric position angles more uncertain, especially when the observations have a bad S/N, as in \object{HD\,32991,} or reveal the presence of a small disk, like in \object{HD\,212076}. 

For \object{HD\,23\,862}, the $PA$ is almost perpendicular to the polarization angle measured by \citet{Yudin2001}, but different from previous  reported values. This difference could be explained by disk precession caused by a companion \citep{Hirata2007}. 

\subsection{Global properties}

In most cases, our simple kinematic model reproduced our measurements. However, in one case (\object{HD\,28\,497}) we needed to add a ``one-armed'' over-density structure to improve the fit process of our data. The disk kinematics is dominated by rotation, with a law close to a Keplerian rotation in almost all targets. The largest discrepancies are found in \object{HD\,36\,576} ($j=-0.4$), \object{HD\,41\,335} ($j=-0.4$), \object{HD\,66\,194} ($j=-0.6$), and \object{HD\,28\,497} ($j=-0.35$). The last one corresponds to the object with the ``one-armed" over-density.

Using the $V\,\sin\,i$ and $V_{\rm{c}}$ values taken from the literature and our own $i$ values,  we obtained the $V/V_{\rm{c}}$ ratios presented in Table~\ref{table:VVc}. We found a mean ratio of $\overline{V/V_{\rm{c}}}=0.71\pm0.06$.

The rotational rate is defined in \citet{Fremat2005} by the ratio of the stellar angular velocity to its critical one, $\frac{\Omega}{\Omega_{\rm {c}}}~=~\frac{V}{V_{\rm {c}}}\frac{R_{\rm eqc}}{R_{\rm eq}}$, where $R_{\rm eqc}$ and $R_{\rm eq}$ are the equatorial radii (in polar radii) for stars rotating at $V_{\rm{c}}$ and $V$, respectively. Values of $\Omega/\Omega_{\rm{c}}$ are listed in Table~\ref{table:VVc}. The mean rotational rate from our sample of Be stars is $\Omega/\Omega_{\rm{c}}=0.86\pm0.04$. 

\begin{table}
\caption{$V/V_{\rm{c}}$ ratio and rotational rate $\Omega/\Omega_{\rm{c}}$ of our Be stars.}    
\label{table:VVc}      
\centering          
\begin{tabular}{c|r@{$\pm$}l|r@{$\pm$}l} 
\hline \hline
 & \multicolumn{2}{c}{$V/V_{\rm{c}}$} & \multicolumn{2}{c}{$\Omega/\Omega_{\rm{c}}$} \\
\hline
\object{HD\,23630}  & 0.68&0.12 &  0.87&0.09  \\ 
\object{HD\,23862}  & 0.69&0.08 &  0.87&0.06  \\
\object{HD\,28497}  & 0.78&0.25 &  0.93&0.13  \\
\object{HD\,30076}  & 0.79&0.18 &  0.94&0.09  \\
\object{HD\,32991}  & \multicolumn{2}{c|}{-} & \multicolumn{2}{c}{-} \\
\object{HD\,35439}  & 0.63&0.10 &  0.83&0.09  \\
\object{HD\,36576}  & 0.63&0.09 &  0.83&0.08  \\
\object{HD\,37202}  & 0.76&0.07 &  0.92&0.04  \\
\object{HD\,37490}  & 0.64&0.07 &  0.84&0.06  \\ 
\object{HD\,41335}  & 0.78&0.14 &  0.93&0.07  \\
\object{HD\,45725}  & 0.75&0.11 &  0.91&0.07  \\
\object{HD\,60606}  & 0.69&0.12 &  0.87&0.09  \\
\object{HD\,66194}  & 0.67&0.14 &  0.86&0.11  \\
\object{HD\,68980}  & 0.76&0.18 &  0.92&0.10  \\
\object{HD\,209409} & 0.77&0.13 &  0.93&0.07  \\
\object{HD\,212076} & 0.62&0.21 &  0.82&0.18  \\
\object{HD\,212571} & 0.75&0.29 &  0.91&0.17  \\
\object{HD\,214748} & 0.68&0.12 &  0.87&0.09  \\
\hline                  
\end{tabular}
\end{table}

\subsection{Statistical analysis of the combined sample}

\begin{figure}
\resizebox{\hsize}{!}{\includegraphics{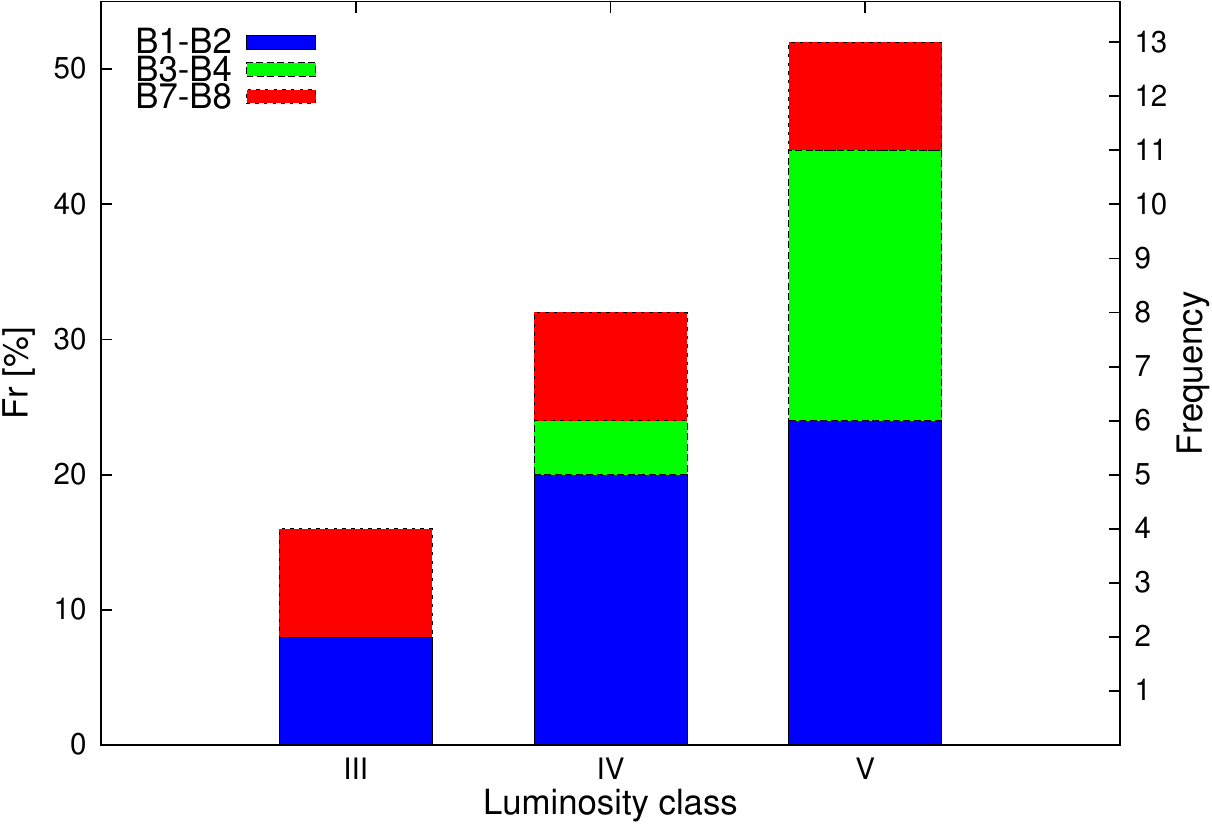}}
\caption{Distribution of Be stars per spectral type and luminosity class for the combined sample.}
\label{fig:TE-CL-all}
\end{figure}

With the aim of studying a larger and more representative sample of Be stars and gaining a better understanding of the physical properties and origin of the circumstellar envelopes, we combine our results with those obtained by \citet{Meilland2012}. The star \object{HD\,209409} was excluded from the \citet{Meilland2012} sample, and disk parameters were derived from our new observation, which has a higher S/N, improving the model fitting. The disk parameters obtained by \citet{Meilland2012} are listed in the lower rows of Table~\ref{table:fit}. Therefore, the combined star sample contains 25 Be stars, with spectral types between B1 and B8, and luminosity classes between V and III, distributed according to the histogram displayed in Fig.~\ref{fig:TE-CL-all}.

To search for the optimal bin number and bin size in each histogram, we adopted a bin-width optimization method that minimizes the integrated squared error \citep{Shimazaki2007}. 

\begin{figure}
\resizebox{\hsize}{!}{\includegraphics{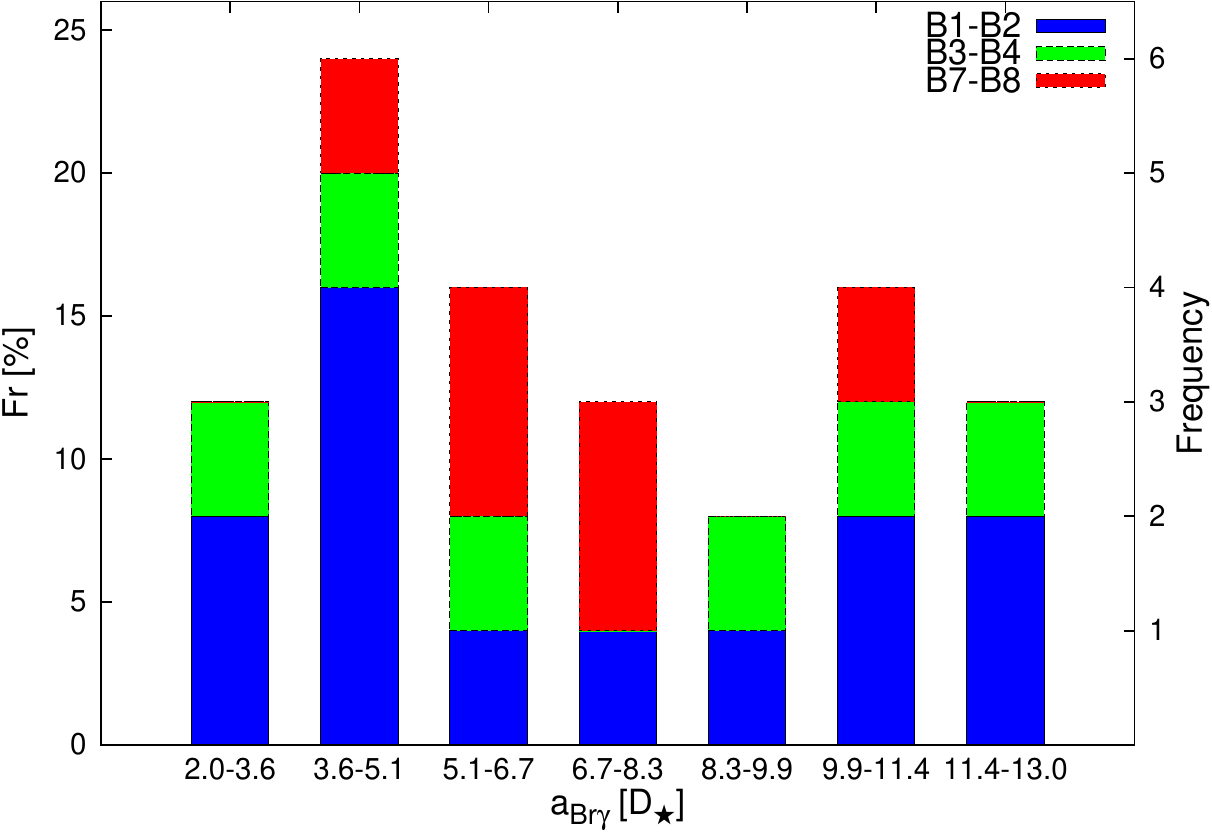}}
\resizebox{\hsize}{!}{\includegraphics{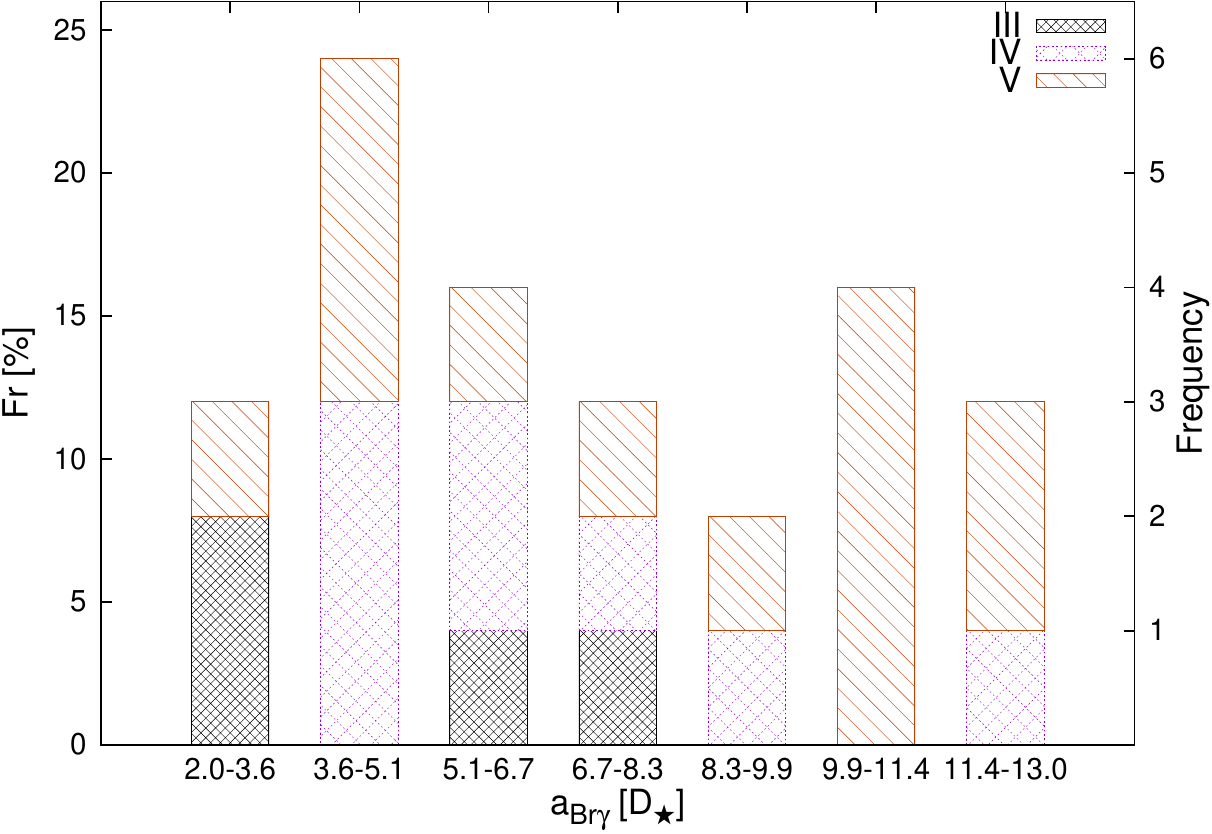}}
\caption{Distribution of the disk extension in the line per spectral type and luminosity class for the combined sample.}
\label{fig:al_all}
\end{figure}

The distribution of the disk extension in the line per spectral type and luminosity class is presented in Fig.~\ref{fig:al_all}. This distribution is bimodal, with maxima at $3.6-5.1\,D_{\star}$ and $9.9-11.4\,D_{\star}$.
About a third of the star sample shows small disks, with $a_{\rm{l}}$ between $2$ and $5.1\,D_{\star}$. 
The largest measured for the disk size is 13 $D_{\star}$. 
Stars with luminosity class V present disks with different sizes, whilst stars with luminosity class III only present disks smaller than $8.3\,D_{\star}$. 

\begin{figure}
\resizebox{\hsize}{!}{\includegraphics{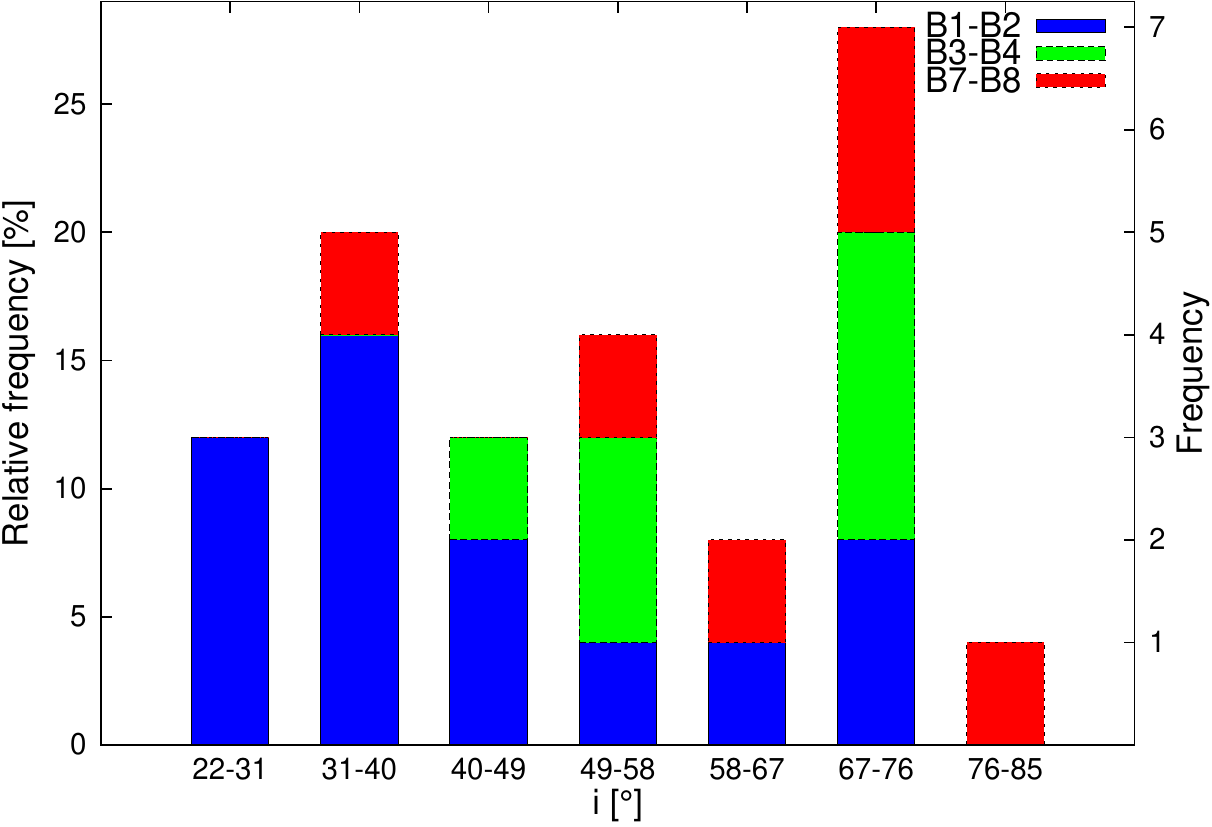}}
\resizebox{\hsize}{!}{\includegraphics{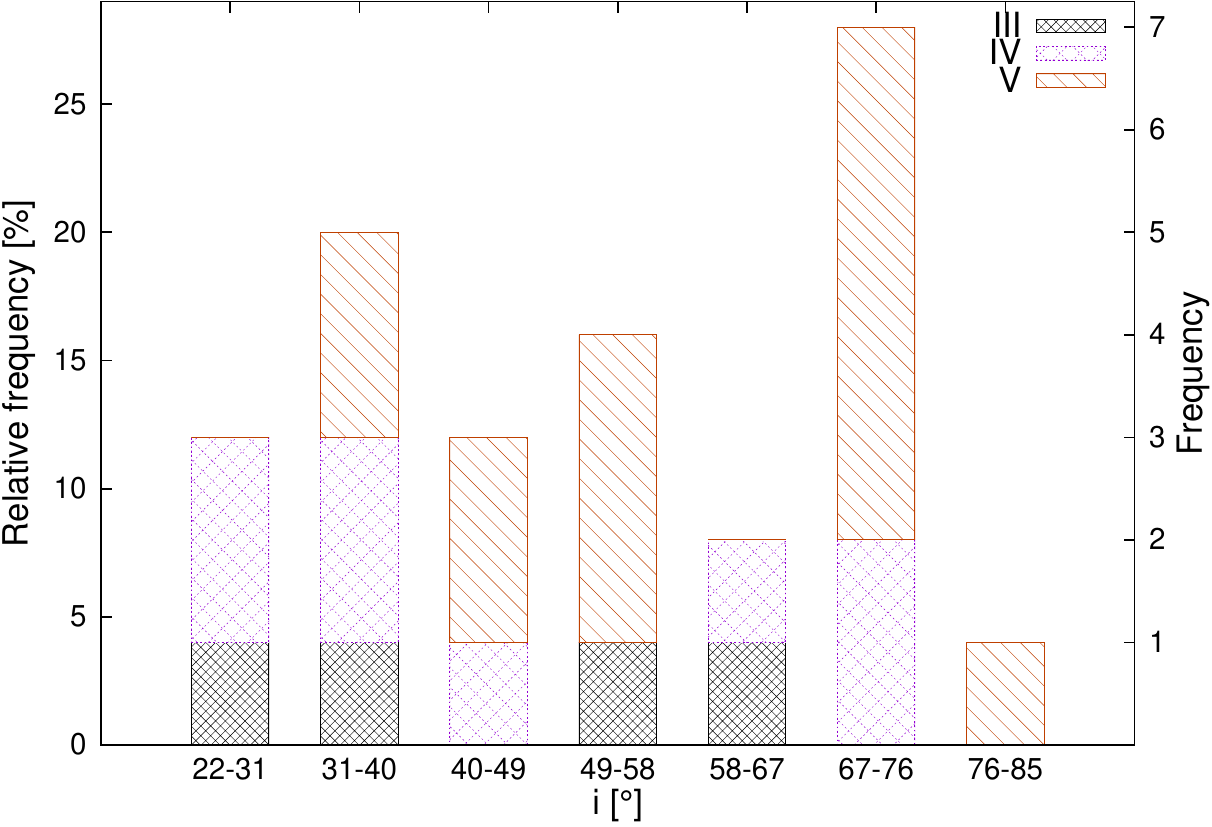}}
\caption{Distribution of inclinations along the line-of-sight per spectral type and luminosity class for the combined sample.}
\label{fig:i_all}
\end{figure}

In Fig.~\ref{fig:i_all} we present the distribution of the observed inclination per spectral type and luminosity class. The distribution presents a peak at the interval 67-76$^{\circ}$. A few late B-type stars were observed near equator-on ($5\%$). Stars seen at small angles (lower than 40$^{\circ}$) are mostly of spectral type B1-B2, and all luminosity classes are seen. For intermediate values, the distribution is almost uniform for the different spectral types. Our sample does not have stars of luminosity class III seen at high inclination angles.

\begin{figure}
\resizebox{\hsize}{!}{\includegraphics{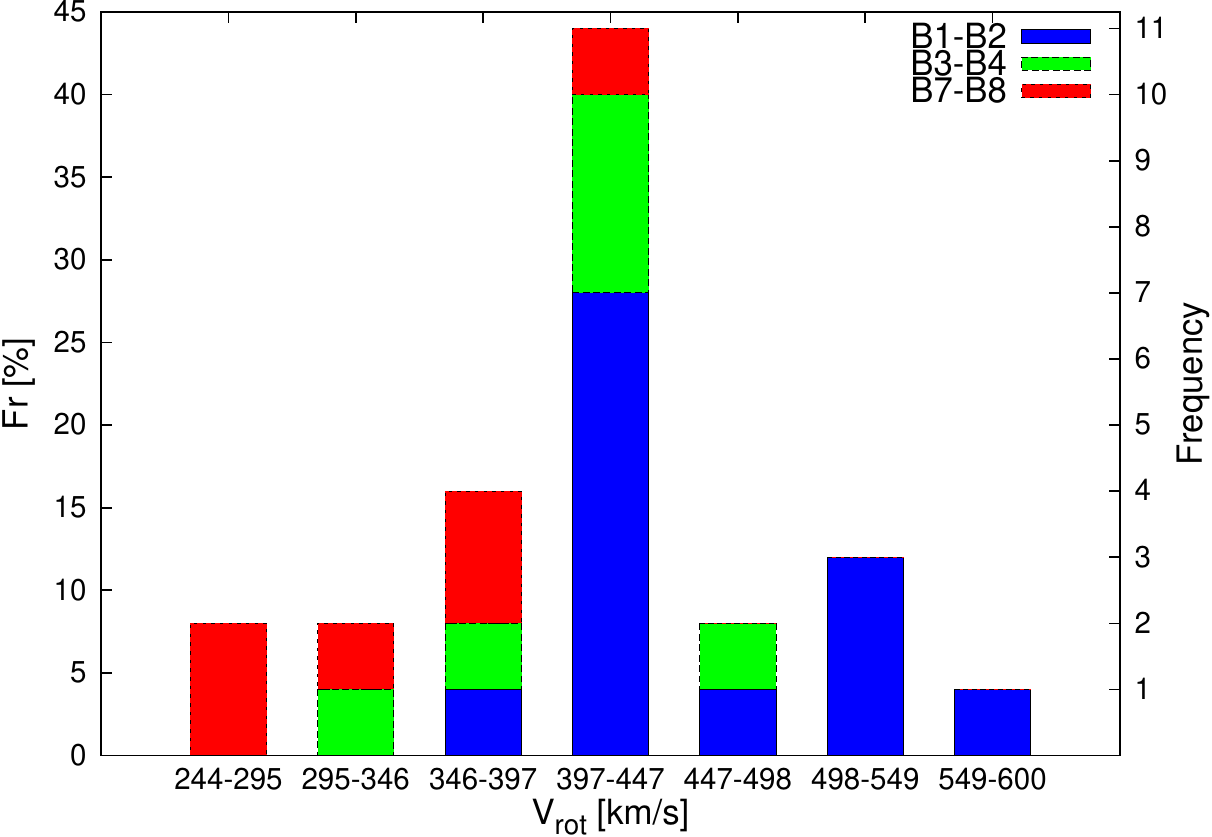}}
\resizebox{\hsize}{!}{\includegraphics{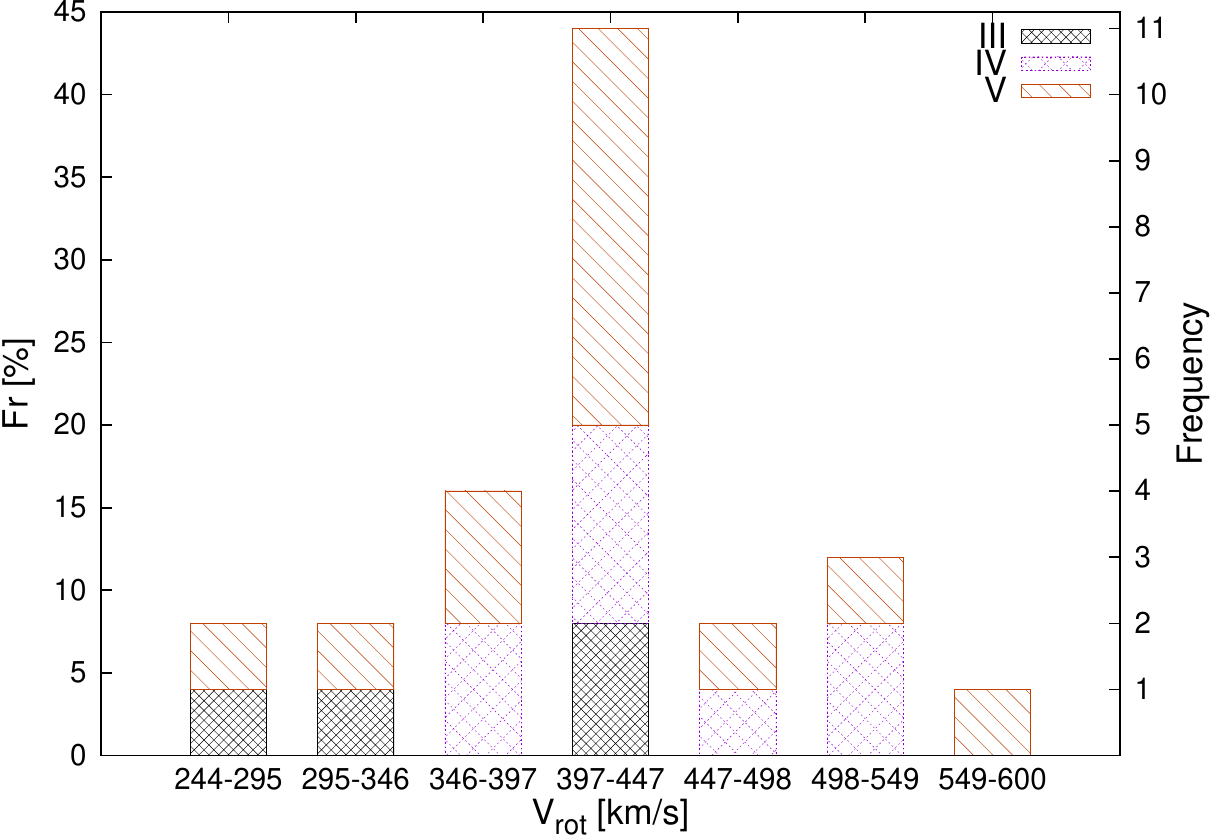}}
\caption{Distribution of $V_{\rm{rot}}$ per spectral type and luminosity class for the combined sample.}
\label{fig:Vrot_all}
\end{figure}

In Fig.~\ref{fig:Vrot_all} we present the distribution of the rotational velocity at the disk inner radius ($V_{\rm{rot}}$) per spectral type and luminosity class. This distribution has a peak in the bin $397-447$\,km\,s$^{-1}$. The peak contains stars of all spectral types and luminosity classes. Early spectral-type stars have disks with high rotational velocities, whilst late B-type stars show disks with low rotational velocities. As regards the luminosity class, all stars exhibit more or less the same distribution, with a maximum at $V_{\rm rot}\sim400$\,km\,s$^{-1}$. Only one star (\object{HD\,28497}) has an extremely high value of $V_{\rm{rot}}$, that reaches $\sim600$\,km\,s$^{-1}$.

\begin{figure}
\resizebox{\hsize}{!}{\includegraphics{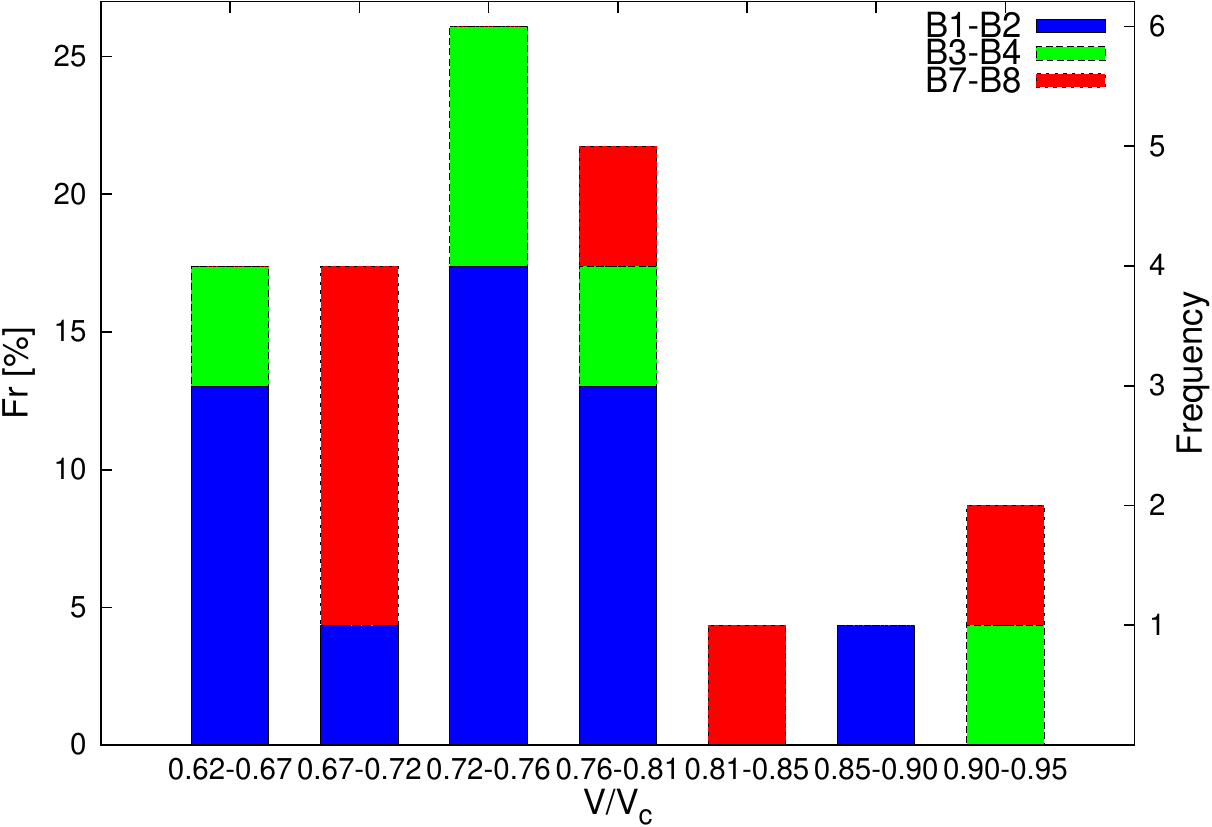}}
\resizebox{\hsize}{!}{\includegraphics{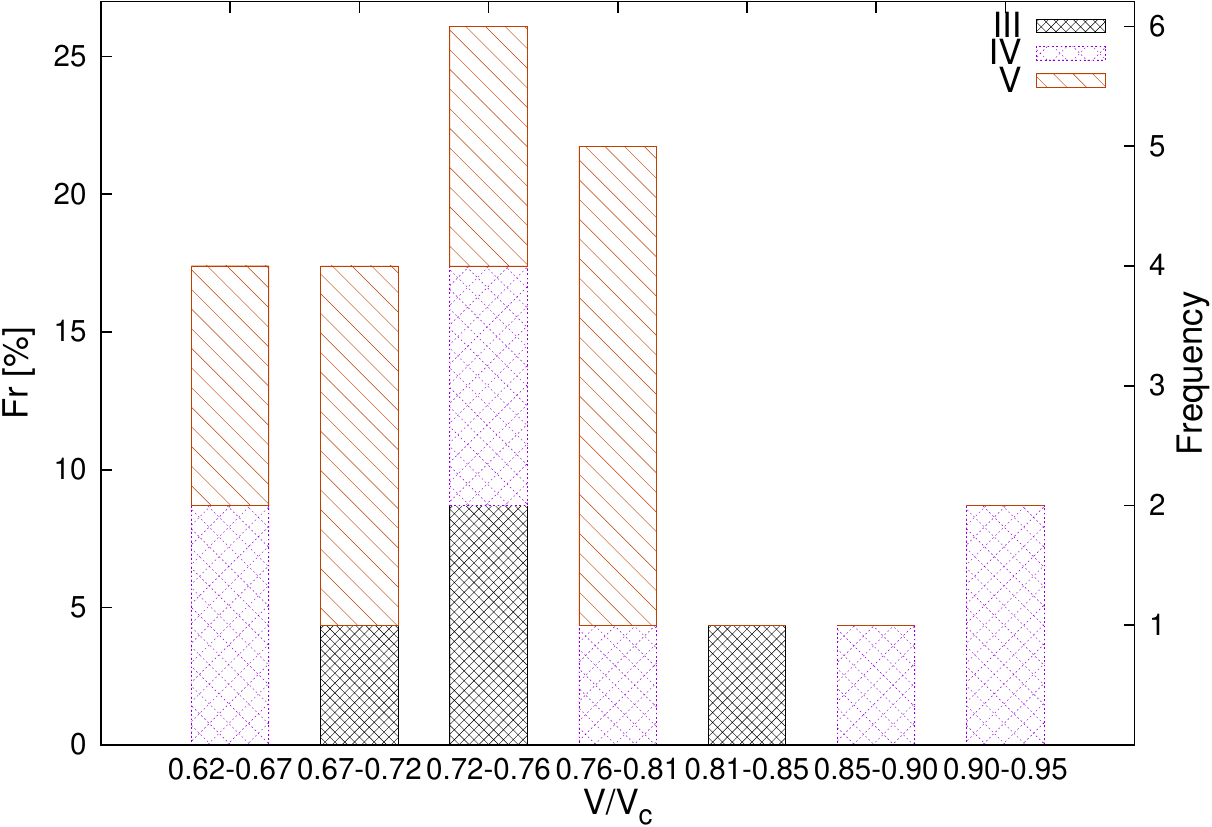}}
\caption{Distribution of $V/V_{\rm{c}}$ per spectral type and luminosity class for the combined sample.}
\label{fig:VVcrit_all}
\end{figure}

\begin{figure}
\resizebox{\hsize}{!}{\includegraphics{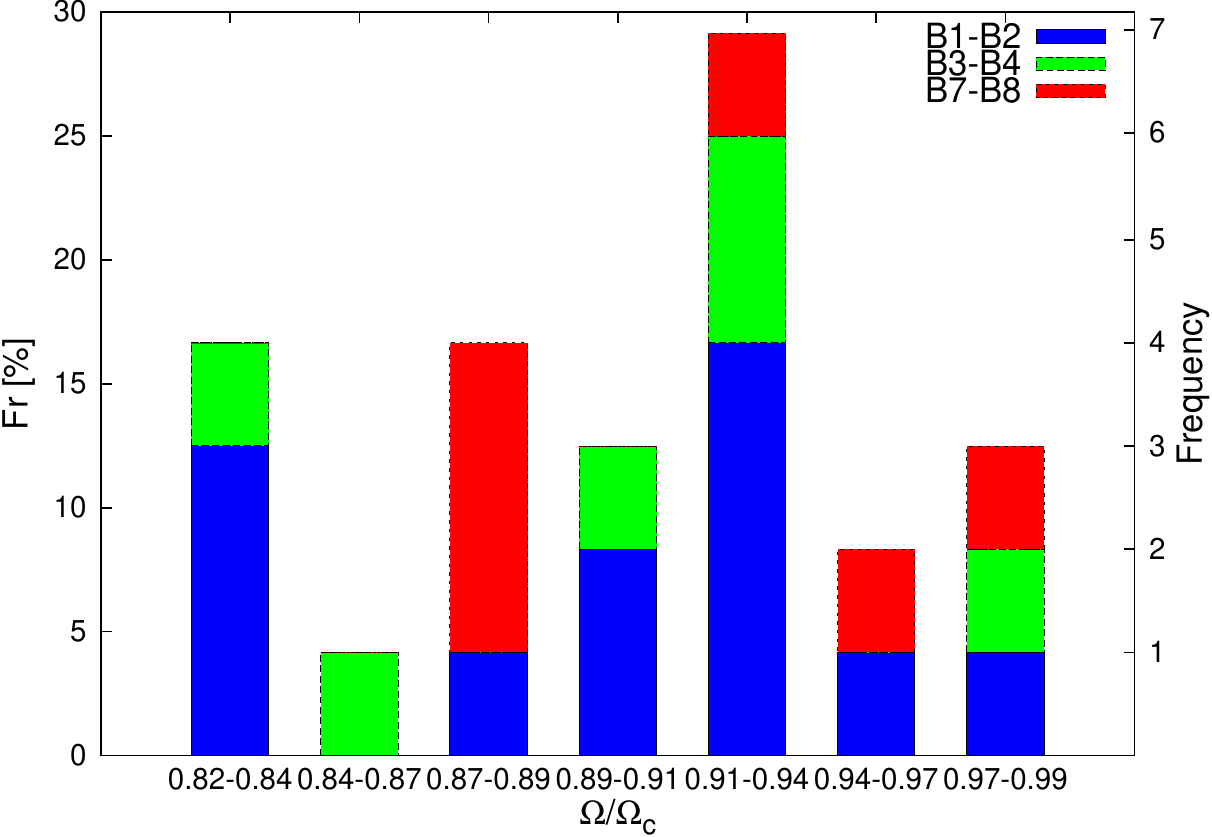}}
\resizebox{\hsize}{!}{\includegraphics{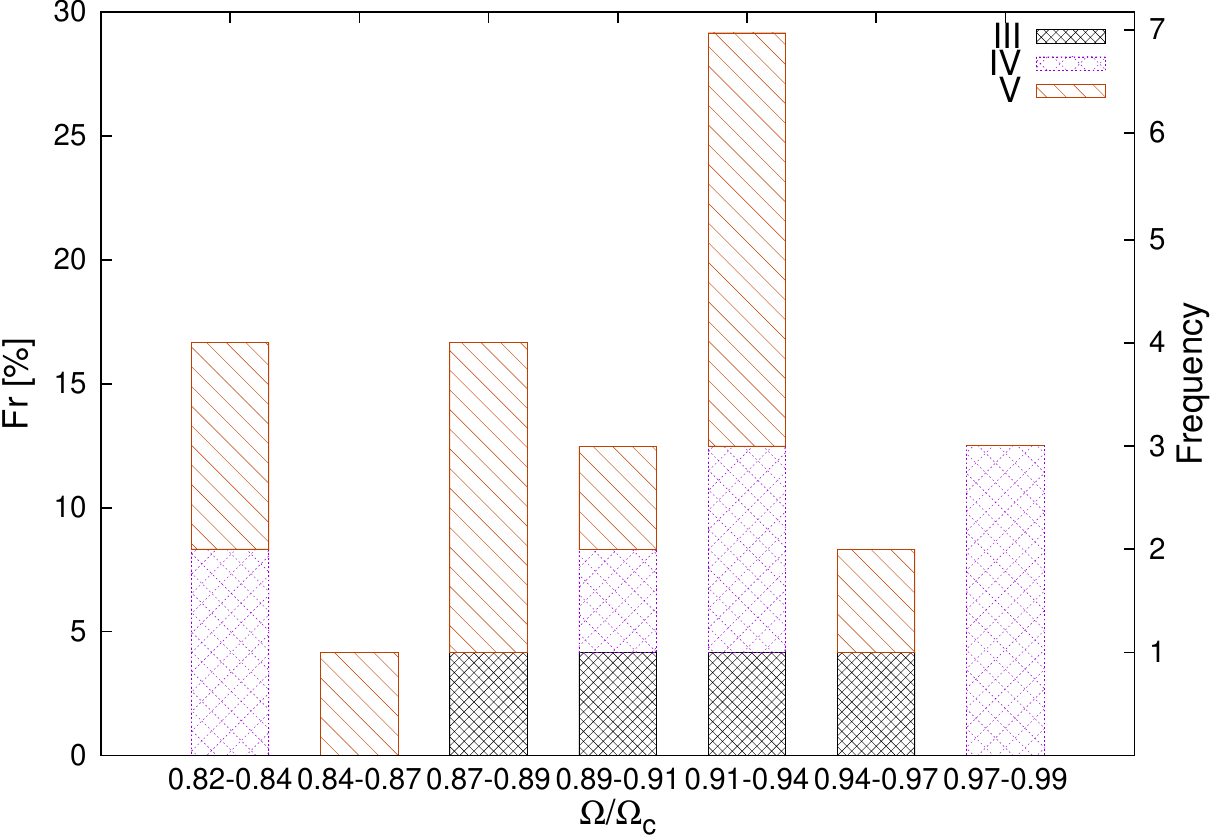}}
\caption{Distribution of $\Omega/\Omega_{\rm{c}}$ per spectral type and luminosity class for the combined sample.}
\label{fig:OOc_all}
\end{figure}

The distribution of $V/V_{\rm{c}}$ ratio per spectral type and luminosity class is presented in the Fig.~\ref{fig:VVcrit_all}. Between 0.62 and 0.81, the distribution is almost constant and includes $\sim80\%$ of the sample. The $\sim20\%$ left is distributed between 0.81 and 0.99. 
We do not see any correlation with the spectral type or the luminosity class. In Fig.~\ref{fig:OOc_all} we present the distribution of $\Omega/\Omega_{\rm{c}}$, which shows values between 0.82 and 0.99. The distribution is similar for stars with different spectral types and luminosity classes. 

Although, there is no clear correlation between the disk extension in the line and $V_{\rm{rot}}$, we observe that disk extensions larger than 10\,$D_{\star}$ are only present when $V_{\rm rot} > 400$\,km\,s$^{-1}$. This outcome is discussed in a broader context in Sect. \ref{sec:discussion}.

Using the obtained values for $R_{\star}$ and $V_{\rm{c}}$, we estimated the stellar masses from the expression $V_{\rm{c}}=\sqrt{G\,M_{\star}/R_{\star}}$, where $G$ is the gravitational constant. At first glance, no correlation was found between the stellar mass and the disk size.

\section{Discussion}\label{sec:discussion}

\subsection{Rotational rate}

Considering the combined sample, we found a mean ratio $\overline{V/V_{\rm{c}}}~=~0.75\pm0.08$ and a mean rotational rate $\overline{\Omega/\Omega_{\rm{c}}}~=~0.90\pm0.05$. These values agree with the ones determined by \citet{Fremat2005} from a sample of 130 stars, that is, $\overline{V/V_{\rm{c}}}\simeq0.75$ and $\overline{\Omega/\Omega_{\rm{c}}}\simeq0.88$, and by \citet[][$\overline{V/V_{\rm{c}}}=0.77$]{Zorec2016}. 
However, when this ratio is corrected by the effects of gravitational darkening, macroturbulence, and binarity, its value drops to $\overline{V/V_{\rm{c}}}\simeq0.65$, according to \citet{Zorec2016}. As our sample contains seven binary stars, our $\overline{V/V_{\rm{c}}}$ value, which is not corrected for the mentioned effects, should be an upper limit.
On the bases of this result, we conclude that the Be stars of our sample, on average, do not rotate at their critical velocity.

\begin{figure}
\resizebox{\hsize}{!}{\includegraphics{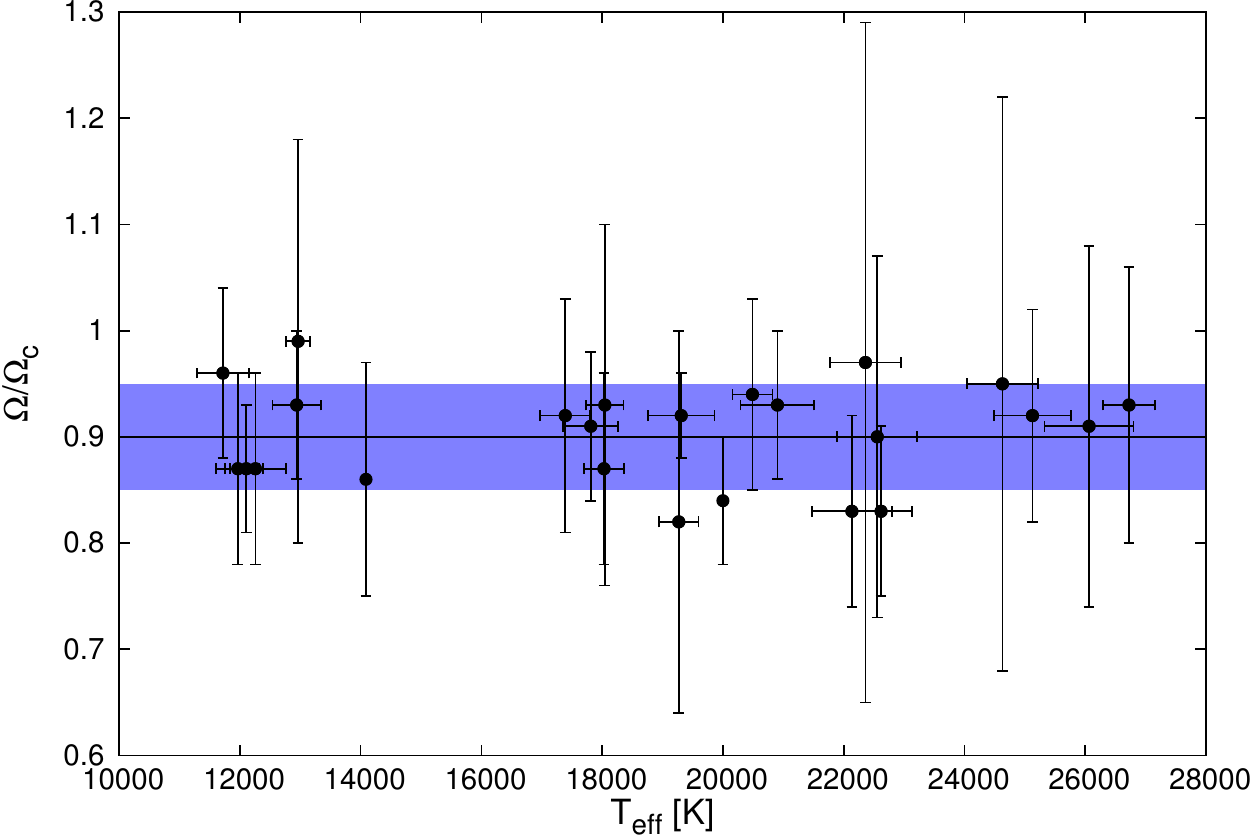}}
\resizebox{\hsize}{!}{\includegraphics{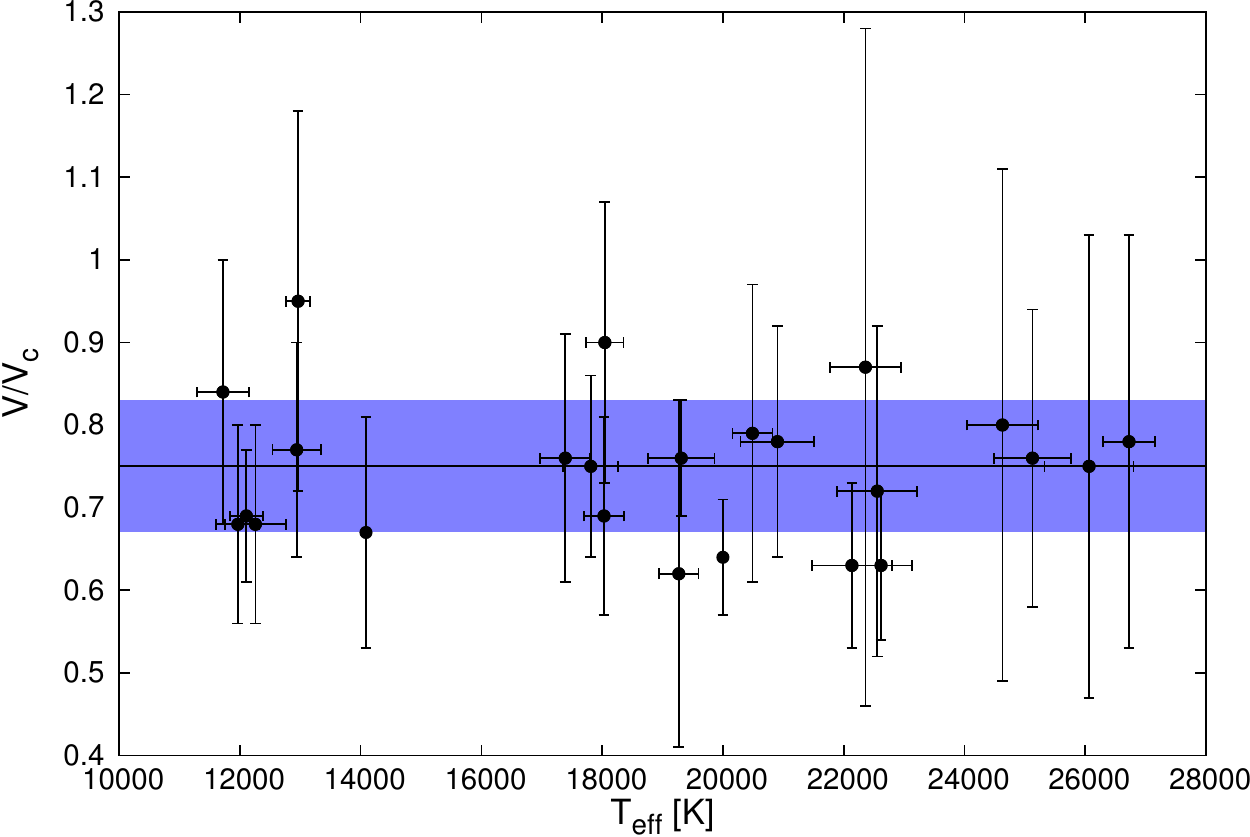}}
\caption{Distributions of $V/V_{\rm{c}}$ and $\Omega/\Omega_{\rm{c}}$ with the effective temperature. There is a lack of correlation.}
\label{fig:cranmer}
\end{figure}

\citet{Cranmer2005} proposed that the rotational rates have different limits depending on the effective temperature. To test his hypothesis, we plotted the values of the rotational rates against $T_{\rm eff}$ (see Fig.~\ref{fig:cranmer}) but found no relationship.

\subsection{Stellar rotation axes}

To see if the rotational axes are randomly distributed, we compared the distribution of projected angles from our data with a theoretical distribution \citep[see, e.g.,][]{Cure2014}.

Let us define the 3D vector $\mathbf v$, and $s=\sin\,i$, the projection of $\mathbf v$ to the plane normal to the line of sight, with $i$ being the angle between $\mathbf v$ and the line of sight. If we assume that $\mathbf v$ is uniformly distributed over the sphere, then the probability distribution function (PDF) verifies that $f(s)=s/\sqrt{1-s^2}$ \citep{Cure2014}.

\begin{figure}
\resizebox{\hsize}{!}{\includegraphics{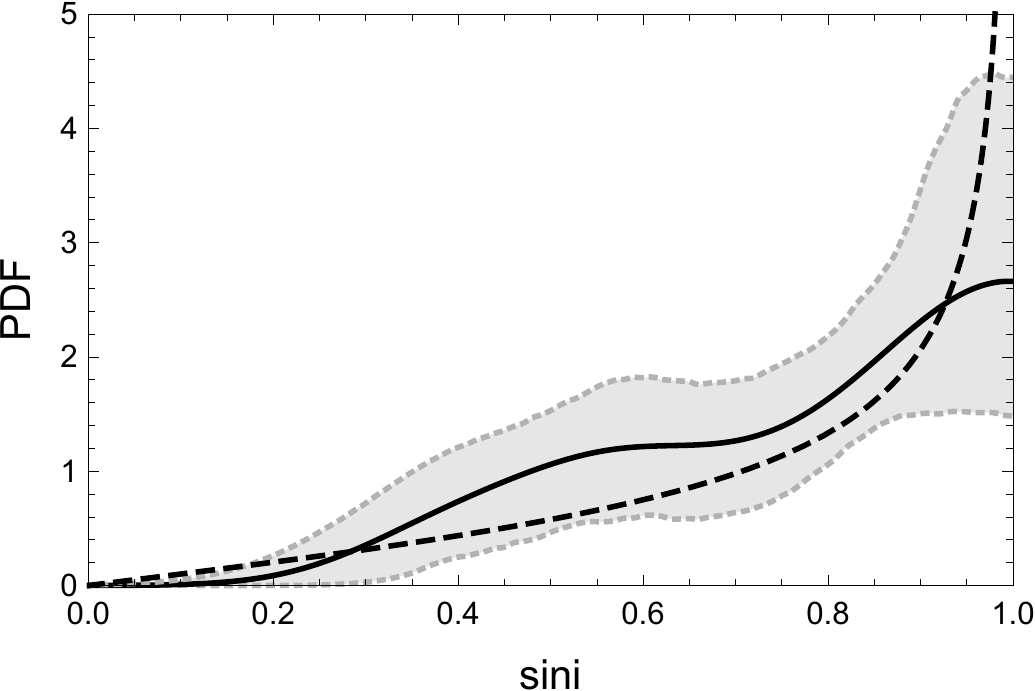}}
\caption{PDF (solid black line) generated from our $N=25$ data sample. It was calculated using a standard KDE algorithm with a Gaussian kernel and a Silverman's bandwidth of 0.1029. The $95\%$ confidence interval is shown in the shaded gray region and was calculated using the quantiles from $2000$ bootstrap samples of length $N=25$. The theoretical PDF $f(s)$ function is shown as a dashed line.}
\label{fig:bootstrap}
\end{figure}

In Fig.~\ref{fig:bootstrap} we present, in black solid line, the kernel density estimation (KDE)  of the observed $\sin\,i$ values and the theoretical PDF $f(s)$. Furthermore, we performed a Bootstrap study, generating $2000$ random samples of our $\sin\,i$ values, each with a sample length of 25 data. The resulting confidence interval using the 0.025 and 0.975 quantiles is also plotted in the figure (shaded gray region), showing that the theoretical PDF lies almost entirely inside the confidence interval. In view of this finding, we performed Kolmorogov-Smirnov and Anderson-Darling tests to verify if the observed data are distributed according to the expected PDF$-f(s)$. We obtained a $p$-value of 0.0686 and 0.0997, respectively. Both tests do not reject the null hypotesis, meaning that the sample might come from a random orientation of stellar rotational axes, but in view of the small $p$-values, a detailed study with a larger data sample is necessary to elucidate the underlying distribution.

\subsection{Correlation between $a_{l}$ and $V_{\rm{rot}}$}

The disk size in the Br$\gamma$ line ranges between 2\,$D_{\star}$ and 13\,$D_{\star}$, and is independent of the effective temperature of the stars.

\begin{figure}
\resizebox{\hsize}{!}{\includegraphics{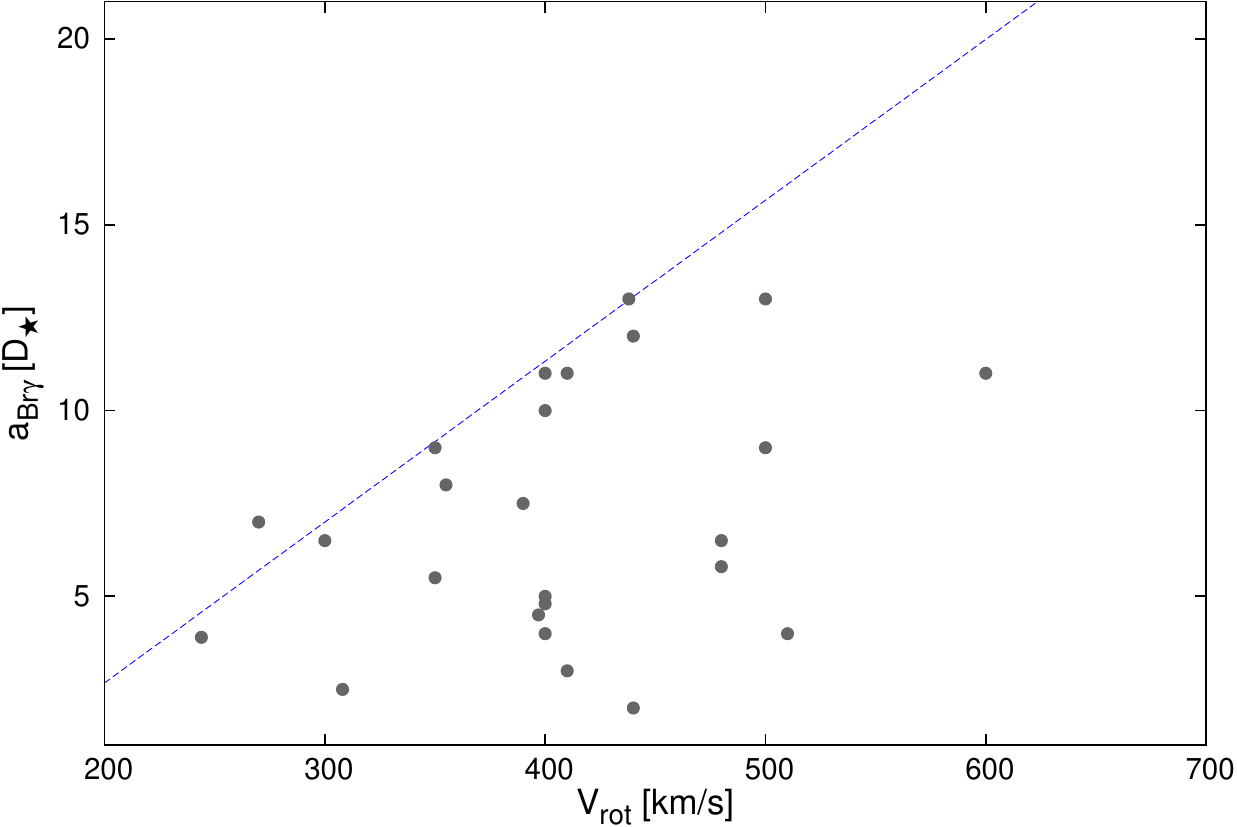}}
\caption{Relation between $V_{\rm{rot}}$ and the disk extension in the Br$\gamma$ line for the star sample of this work and those collected from \citet{Meilland2012}. For a given $V_{\rm{rot}}$, the blue dashed line could be an upper limit for the disk extension.}
\label{fig:al_limit}
\end{figure}

To discuss the origin of the circumstellar envelope and the role of the stellar rotation, we search for a relation between the disk extension in the Br$\gamma$ line and $V_{\rm{rot}}$. Figure~\ref{fig:al_limit} shows the whole sample of stars and highlights that there are no Be stars with low values for $V_{\rm{rot}}$ and large disk size. Furthermore, we observe that all the stars are located below the blue dashed line. This could mean that there is an upper limit for the disk size according to a given rotational velocity.

\begin{figure}
\resizebox{\hsize}{!}{\includegraphics{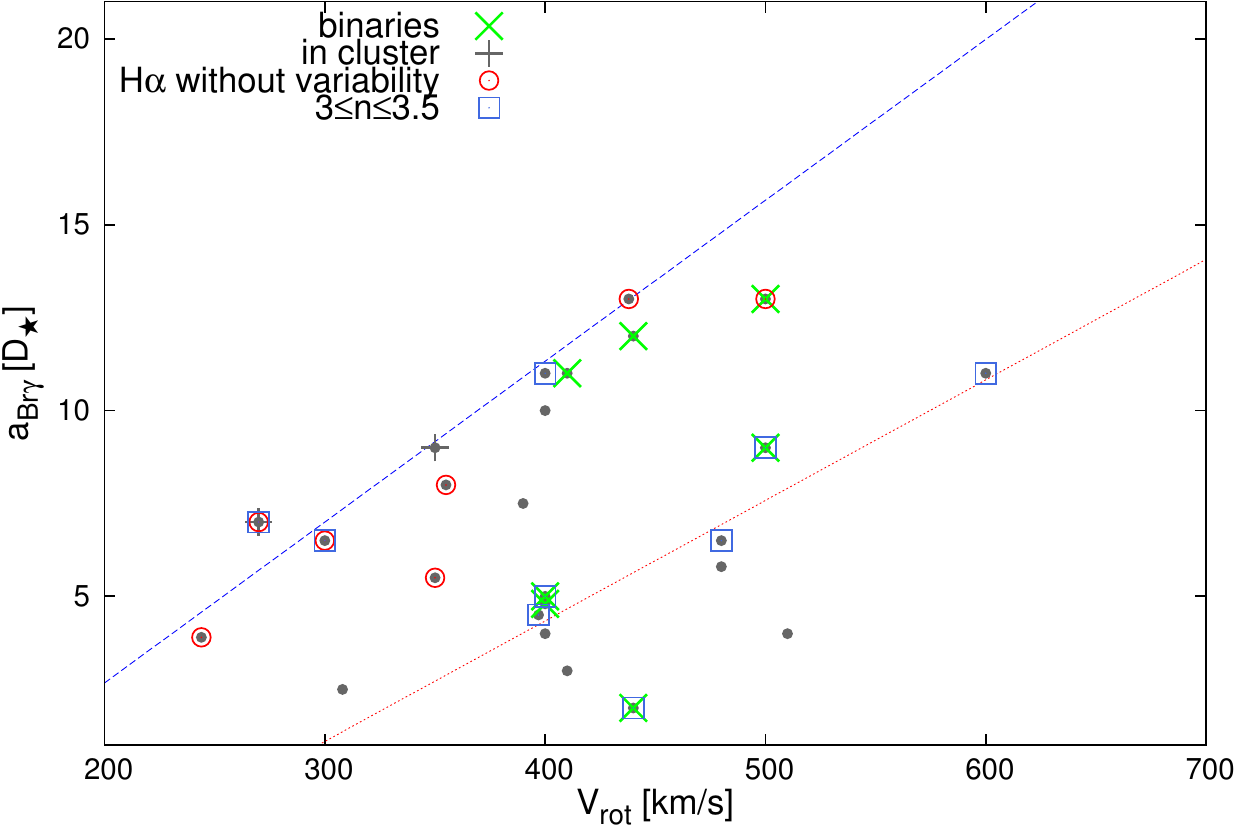}}
\caption{Relation between $V_{\rm{rot}}$ and the disk extension in the line for the star sample (black dots) included in this work and \citet{Meilland2012}. The colored lines symbols are given in the top left of the figure. The blue dashed line from Fig.~\ref{fig:al_limit} could state an upper limit for the disk size of the stars indicated as ``without variability''. The orange dotted line seems to oultine a second linear regretion from stars with stable disks.}
\label{fig:al_limit_est}
\end{figure}

Considering the spectral variability of Be stars, we look for a criterion to determine which stars have stable disks. To this purpose, we compared the strength and spectral appearance of the H$\alpha$ line for each star using observations published by \citet{Doazan1991}, \citet{Silaj2010,Silaj2014}, \citet{Arcos2017} and \citet{Arcos2018inprep}.
Thus, we differentiate the objects that present a stable intensity in the H$\alpha$ line from objects that present variations. Based on the H$\alpha$ criterion, the objects that we  consider as stable  are: \object{HD\,23\,630}, \object{HD\,32\,991}, \object{HD\,36\,576}, \object{HD\,37\,795}, \object{HD\,89\,080}, \object{HD\,209\,409}, and \object{HD\,214\,748}. These objects are identified as ``without variability'' in Fig.~\ref{fig:al_limit_est}, and the blue dashed line could state an upper limit for the disk size of these stars.  

Another criterion was based on the results found by \citet{Vieira2017} who analyzed the evolution of the disks using the viscous decretion disk (VDD) model, and found that the timescale for disk growth is shorter than the timescale for disk dissipation. Also, they found a correlation between $\rho_0$ and $n$ values and determined distinct regions in the $n-log\rho_0$ plane. In this diagram, the $n\lesssim3$ region is associated with dissipating disks; the strip between $n\simeq3$ and $n\simeq3.5$ corresponds to the steady-state zone, and the region where $n\gtrsim3.5$ is related to disks in formation. Based on this new criterion, \citet{Vieira2017} found that \object{HD\,23630}, \object{HD\,28497}, \object{HD\,35439}, \object{HD\,41335}, \object{HD\,50013}, \object{HD\,89080}, \object{HD\,91465}, \object{HD\,105435} and \object{HD\,212571} have steady disks. In Fig.~\ref{fig:al_limit_est} these objects are identified as ``$3\le n\le3.5$''. Some of these stars fall in the linear relationship found before, but others seem to outline a second linear regression, below the previous one. This second linear relationship is plotted with an orange dotted line and could be related with another stability condition for the disk extension for stars of a given spectral type or stellar mass.

\begin{figure}
\resizebox{\hsize}{!}{\includegraphics{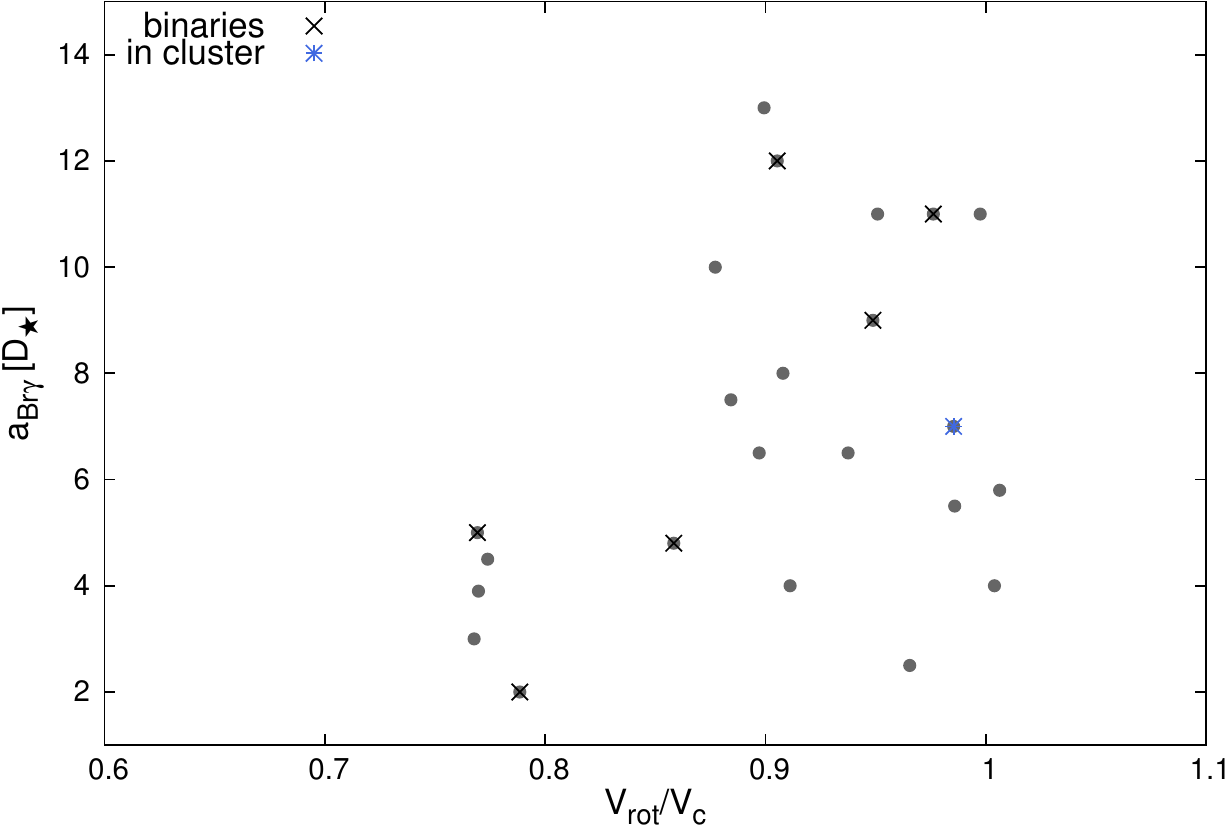}}
\caption{Relation between $V_{\rm{rot}}/V_{\rm c}$ and the disk extension in the Br$\gamma$ line for the combined sample. Stars with the largest sizes have $V_{\rm{rot}}$ close to its critical rotation velocity.}
\label{fig:vrotvcrit_sizel}
\end{figure}

\begin{figure}
\resizebox{\hsize}{!}{\includegraphics{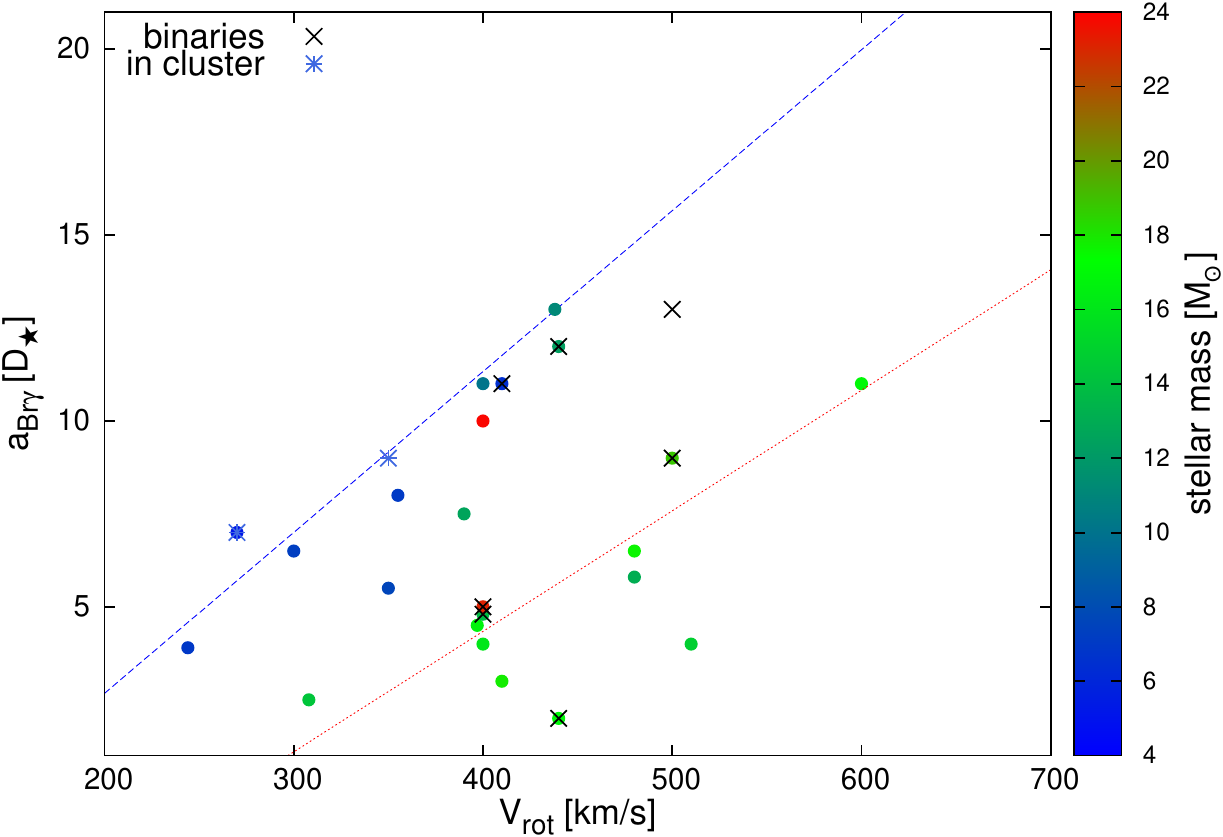}}
\caption{Relation between $V_{\rm{rot}}$ and disk extension in the line for the star sample studied in this work and in \citet{Meilland2012}. For a given $V_{\rm{rot}}$, the blue dashed line traces the location of low-mass Be stars with stable disks and the orange dotted line identifies the location of high-mass Be stars with stable disks.}
\label{fig:al_limit_mass}
\end{figure}

Figure~\ref{fig:vrotvcrit_sizel} presents our findings for disk size versus $V_{\rm{rot}}/V_{\rm{c}}$. We observe that the stars with the largest disks present values of $V_{\rm{rot}}/V_{\rm{c}}\gtrsim0.9$, however no correlation with the stellar type was observed, as shown in Fig.~\ref{fig:cranmer}.

Since it has often been observed that late-type Be stars are less variable than early-type ones \citep{Hubert1998,Jones2011,Barnsley2013,Labandie2017}, it would be possible that the relations for stable disks are connected to the mass of the star.
Nevertheless, it is interesting to highlight that, for a given $V_{\rm{rot}}$, the least massive  B-type stars are close to the upper size limit while the most massive stars are near the lower limit, as is illustrated in Fig.~\ref{fig:al_limit_mass}. This result points out that Be stars with high stellar masses have denser and more compact disks than their low-mass counterparts. As the former are more luminous, compact and dense disks would be brighter in the Br$\gamma$ line. This could explain why the peak of the Be phenomenon is around the B1-B3 spectral types. Evidence supporting a scenario where disks of early Be stars are more massive and smaller than those of late-Be types was previously reported by \citet{Arcos2017}, who used a large sample of H$\alpha$ observations.

Here, it is worth mentioning that \object{HD\,212\,571} is a binary star that is located below the orange linear regression, and, according to \citet{Vieira2017}, has a steady disk. For this object, the timescale of the disk-loss events could be important, since the observations used by \citet{Vieira2017} were performed from January to November 2010 and our interferometric observations were carried out in October 2014. \citet{Arcos2017} observed the same object one year later, and found an $n$ value that locates it in the dissipation region. Therefore, it is possible that the disk size we found does not correspond to the disk size on a steady phase.

\subsection{Comparison of disk sizes in the H$\alpha$ line and the Br$\gamma$ line}

Disk sizes observed in the Br$\gamma$ line are systematically larger than those obtained from the $K$-band with the CHARA interferometer \citep{Gies2007, Touhami2013}. However, we do not find any clear correlation between the sizes of the disk in the Br$\gamma$ and H$\alpha$ lines. Using 48 northern Be stars, \citet{Catanzaro2013} found that the radii of the disks measured in the H$\alpha$ line have values between 2 $R_{\star}$ and 14 $R_{\star}$, with a maximum concentration in the interval of 6-8 $R_{\star}$. The values are similar to the disk size we measured in the Br$\gamma$ line.

 Measurements of disk size in H$\alpha$ are available for thirteen of the stars in our sample \citep{Hanuschik1988, Andrillat1990, Grundstrom2006}.
The Br$\gamma$ size measured from \object{HD\,23\,630} is larger than the H$\alpha$ size obtained by \citet{Grundstrom2006} but smaller than the one obtained by \citet{Hanuschik1988}, even though this star has a stable H$\alpha$ profile. 
For \object{HD\,37\,202}, \object{HD\,158\,427,} and \object{HD\,214\,748}, the Br$\gamma$ sizes are smaller than the ones measured in H$\alpha$, as was found in $\delta$~Sco by \citet{Meilland2011}.  For \object{HD\,37\,490}, the Br$\gamma$ and H$\alpha$ \citep{Andrillat1990, Hanuschik1988} lines lead to similar measurements for the disk size, contrary to \object{HD\,28\,497}, \object{HD\,30\,076}, \object{HD\,35\,439}, \object{HD\,36\,576}, \object{HD\,41\,335} and \object{HD\,45\,725,} for which the Br$\gamma$
line presents larger disk sizes than the ones obtained from H$\alpha$. It should be noted that all these objects have variable H$\alpha$ profiles, possibly indicating that their disk sizes have increased.

Considering the variability of the Be stars, and the different methods used to derive the sizes of their circumstellar disks, it would be useful to perform simultaneous multiwavelength observations.

\section{Conclusions}\label{sec:conclusions}

We analyzed a sample of 26 Be stars observed with the VLTI/AMBER instrument. We were able to obtain geometrical and kinematic disk parameters for 18 of these that present the Br$\gamma$ line in emission, by fitting interferometric data with a 2D kinematic model for a rotating equatorial disk. For 5 stars of our sample the kinematical properties of the disks have been obtained for the first time.

Disk sizes measured in the Br$\gamma$ line are similar to the ones obtained from the H$\alpha$ line, and they range between 2\,$D\star$ and 13\,$D\star$. We do not observe any correlation between disk size and $T_{\rm eff}$  (or spectral type) of the central star. One third of the stars in our sample have sizes lower than 6\,$D\star$ and stars with luminosity class III tend to present the smallest disks. 
Nevertheless, using a combined sample of Be stars \citep[our 18 Be star sample plus 7 stars studied by][]{Meilland2012} we found an upper limit for the disk extension in the Br$\gamma$ line that correlates with the rotational velocity in the inner part of the disk and the stellar mass. For a given  $V_{\rm{rot}}$, high-mass Be stars have more compact and dense disks than their low-mass counterparts. In almost all cases, the star disks rotate in Keplerian fashion, with the exception of \object{HD\,28497, which} required a one-arm over-density model.

The distribution of stellar rotational velocities was found to be relatively symmetrical and has a maximum at $V_{\rm rot}\sim 365$ km\,s$^{-1}$. Stars with early spectral types tend to have high rotational velocities, whilst late B-type stars have low rotational velocities. We derived a mean ratio $\overline{V/V_{\rm{c}}}=0.75\pm0.08$ and a mean rotational rate
$\overline{\Omega/\Omega_{\rm{c}}}=0.90\pm0.05$, in agreement with those authors that propose that, on average, Be stars do not rotate at their critical velocity \citep{Cranmer2005,Fremat2005,Zorec2016}. On the other hand, we did not observe any correlation between $V/V_{\rm{c}}$ values and spectral types, as was previously suggested by \citet{Cranmer2005}.

It is interesting to remark that the rotational axes seem to be uniformly distributed in the sky. However, this last result is not conclusive due to the low quantile value obtained. Therefore, more observations should be carried out to confirm this tendency.

It would be interesting to follow up the evolution of disk size in variable Be stars to better understand disk formation and dissipation processes of their circumstellar disks. To this aim simultaneous observations at different wavelengths are necessary. Furthermore, it would be interesting to improve radiative transfer models with various density laws in order to decipher whether or not the maximum disk size found (about 13\,$D_{\star}$) in the Br$\gamma$ emission line has a physical meaning in terms of density and ionization structure of a ``fully'' built Be disk.

\begin{acknowledgements}
We would like to thank our anonymous referee for his/her valuable comments and suggestions, which helped to improve our manuscript. 
We thank M. L. Arias for her careful reading and comments. 
C.A thanks becas de doctorado nacional de CONICYT 2016-2017. 
M.C. and C.A. acknowledge support from Centro de Astrof\'isica de Valpara\'iso.  
L.C. acknowledges financial support from CONICET (PIP 0177), the Agencia Nacional de Promoci\'on Cient\'ifica y Tecnol\'ogica (PICT
2016-1971) and the Programa de Incentivos (G11/137) of the Universidad Nacional de La Plata (UNLP), Argentina.
L.C. and M.C. thank support from the project CONICYT + PAI/Atracci\'on de capital humano avanzado del extranjero (folio PAI80160057).
\end{acknowledgements}

\bibliographystyle{aa} 
\bibliography{biblio} 


\begin{appendix} 
\section{}\label{app:uvmodel}
For each target, we present the  ($u,v$) plan coverage, with the AMBER HR measurements. The best-fitting kinematics models are overplotted. 

   \begin{figure*}
   \centering
   \includegraphics[width=0.49\hsize]{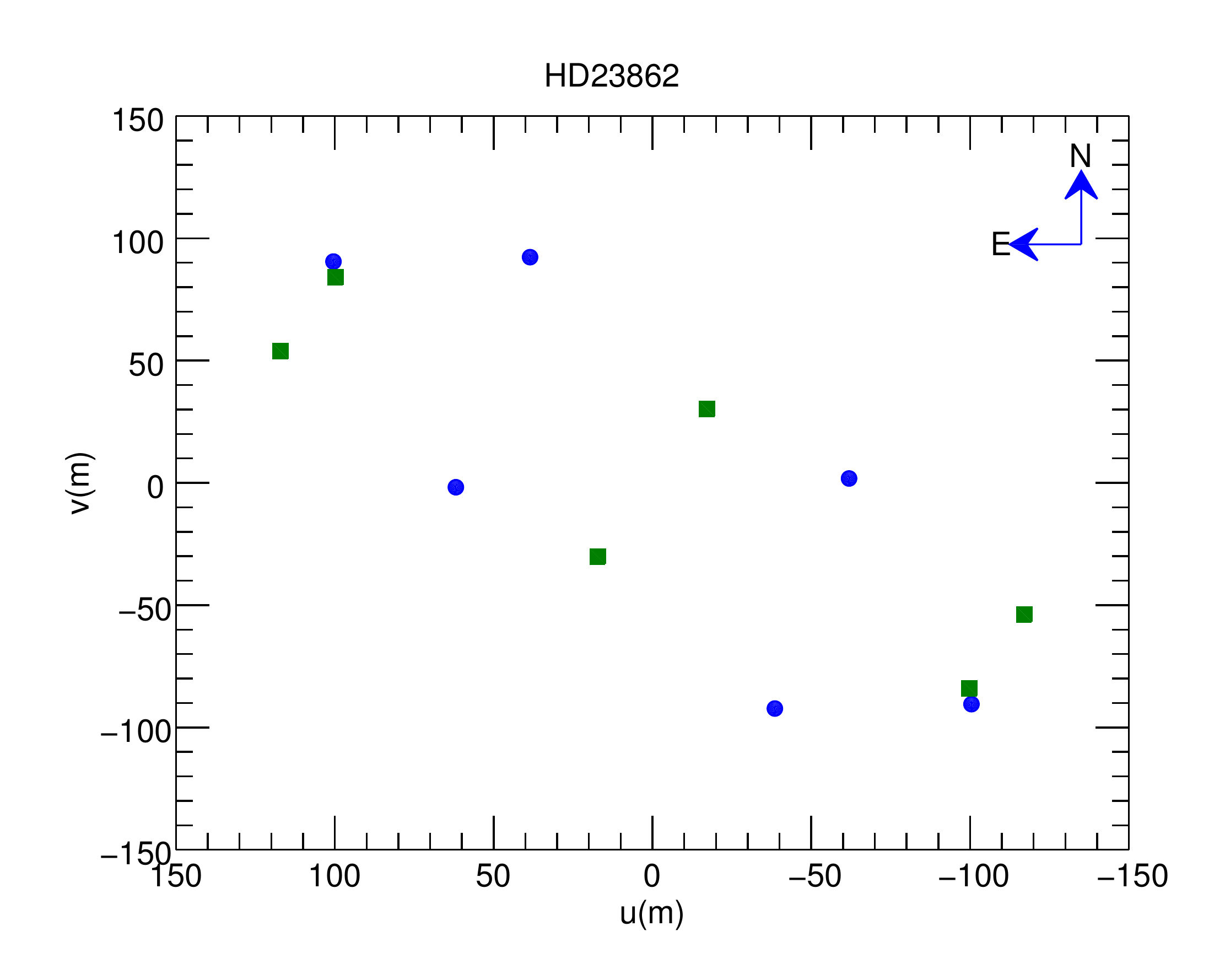}

   \includegraphics[width=0.49\hsize]{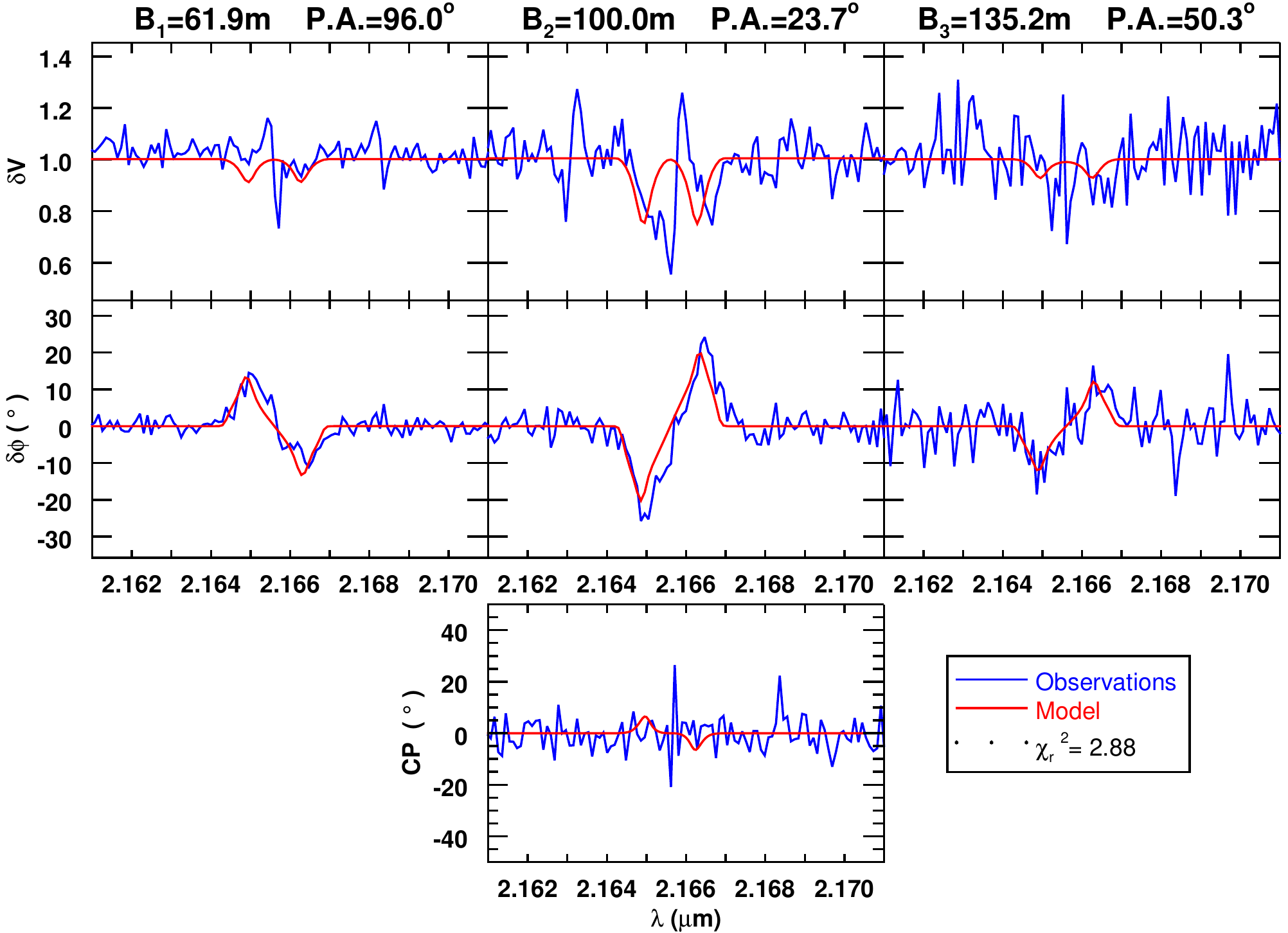}
   \includegraphics[width=0.49\hsize]{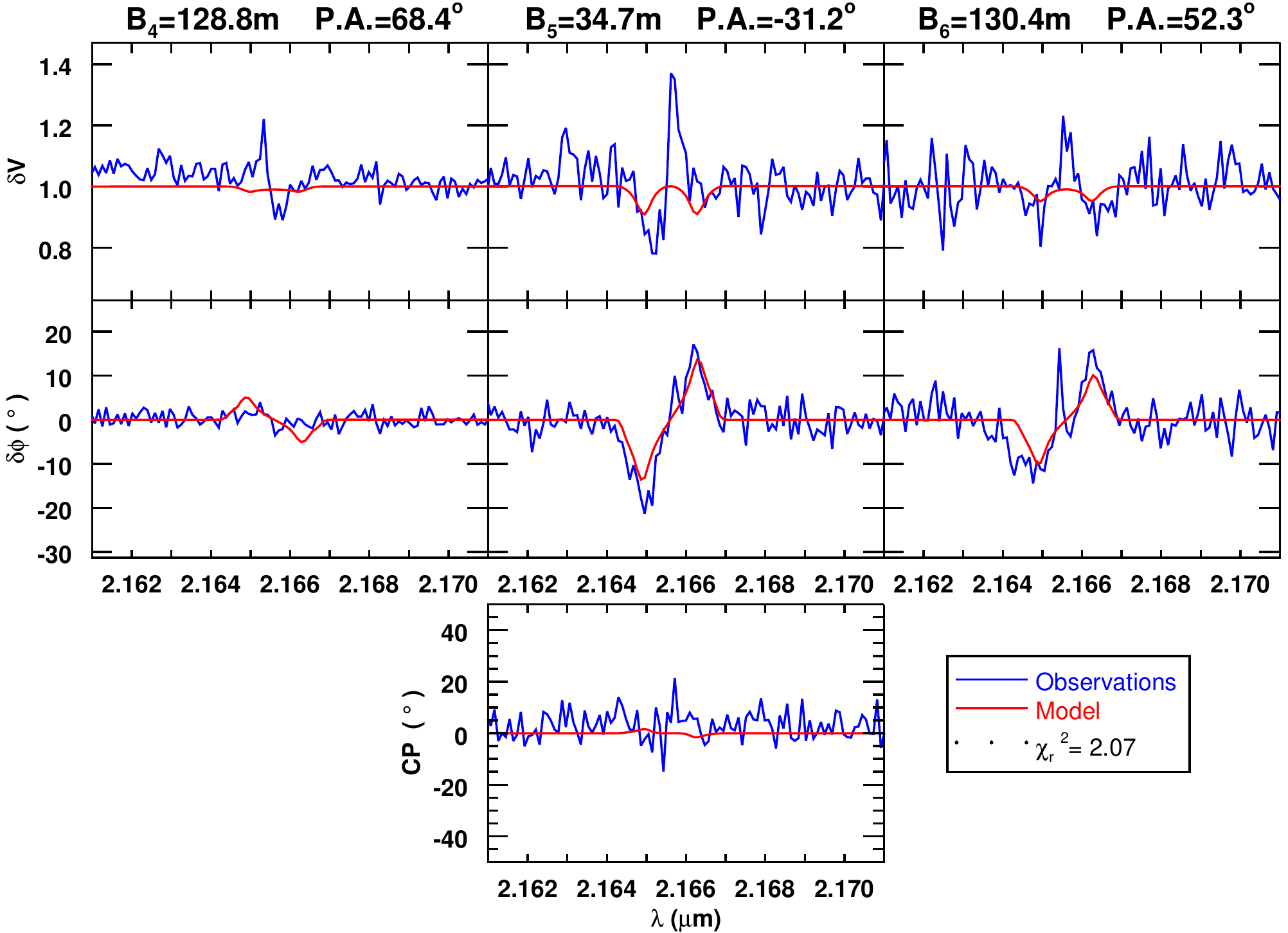}
   \caption{As in Fig.~\ref{fig:HD23630}, but for \object{HD\,23\,862}.}
   \label{fig:HD23862}
   \end{figure*}

   \begin{figure*}
   \centering
   \includegraphics[width=0.49\hsize]{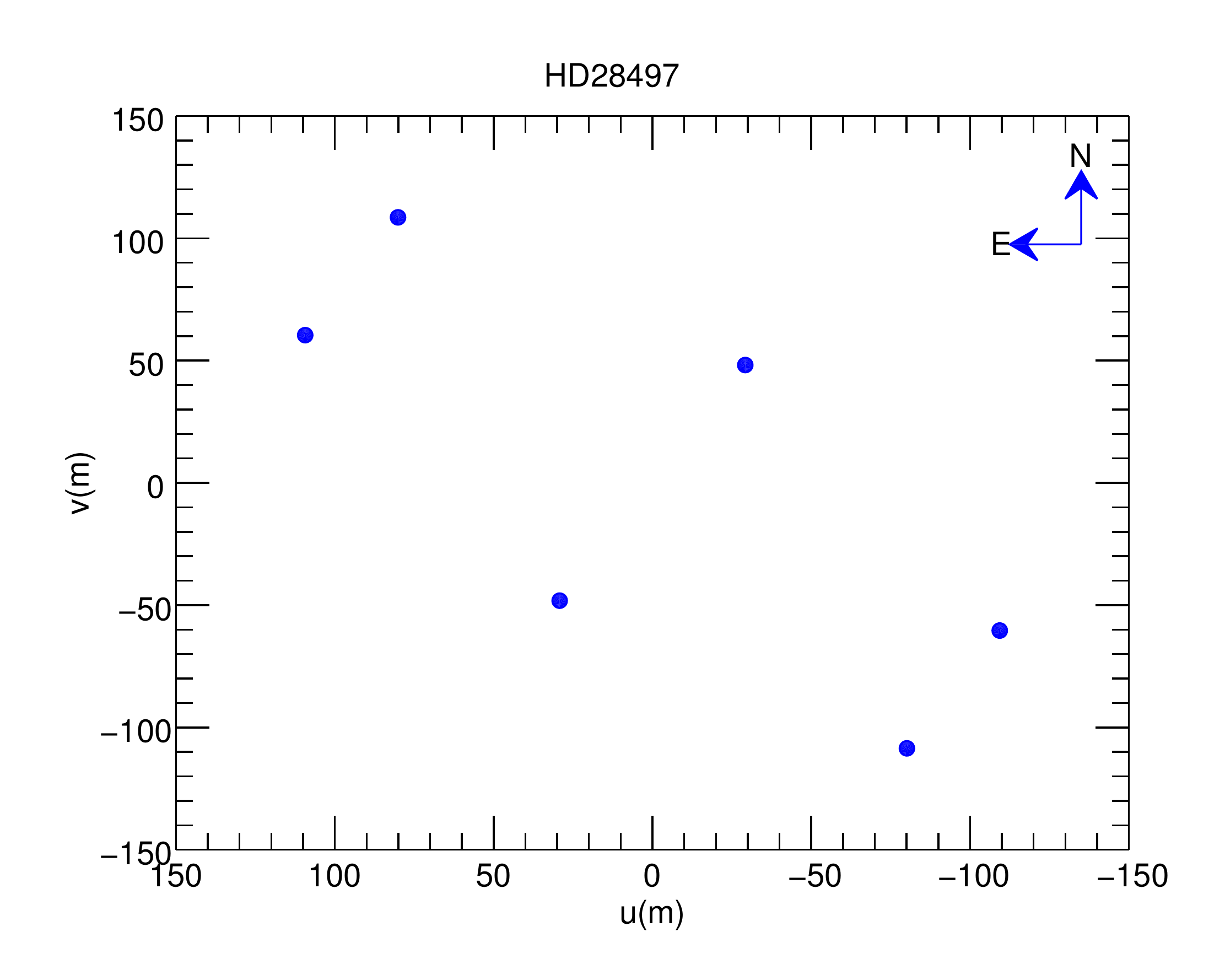}
   \includegraphics[width=0.49\hsize]{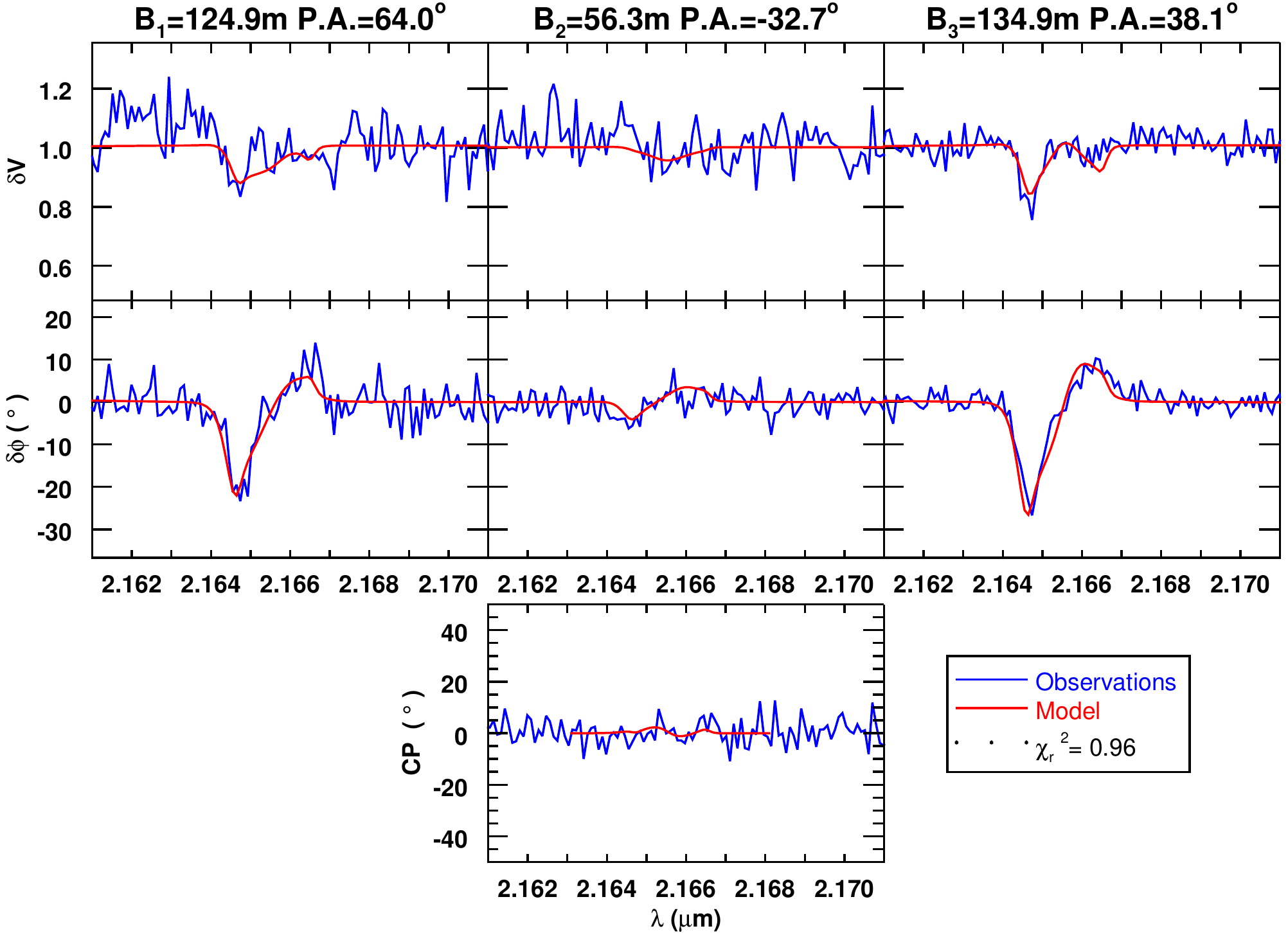}
   \caption{As in  Fig.~\ref{fig:HD23630}, but for \object{HD\,28\,497}.}
   \label{fig:HD28497}
   \end{figure*}

   \begin{figure*}
   \centering
   \includegraphics[width=0.49\hsize]{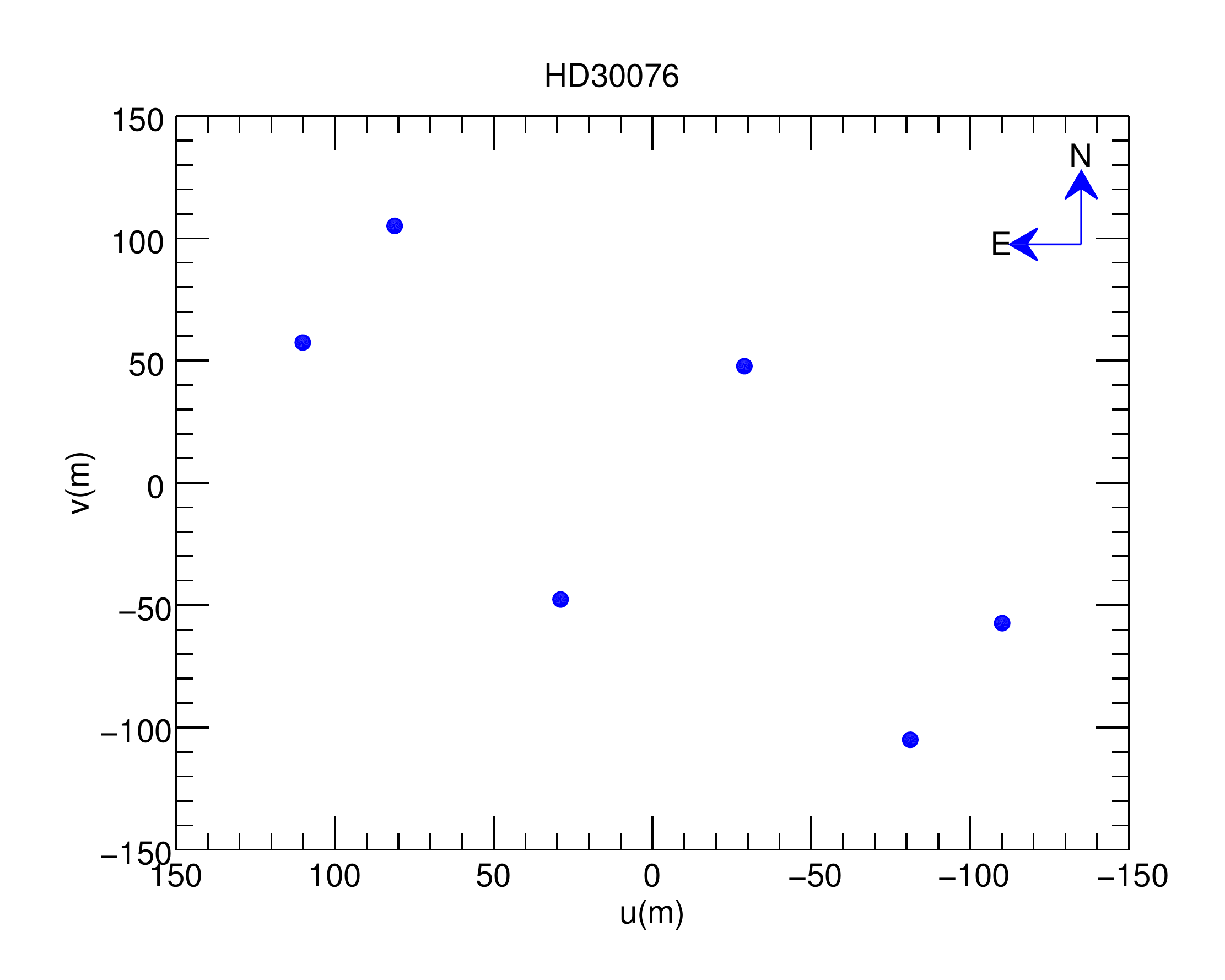}
   \includegraphics[width=0.49\hsize]{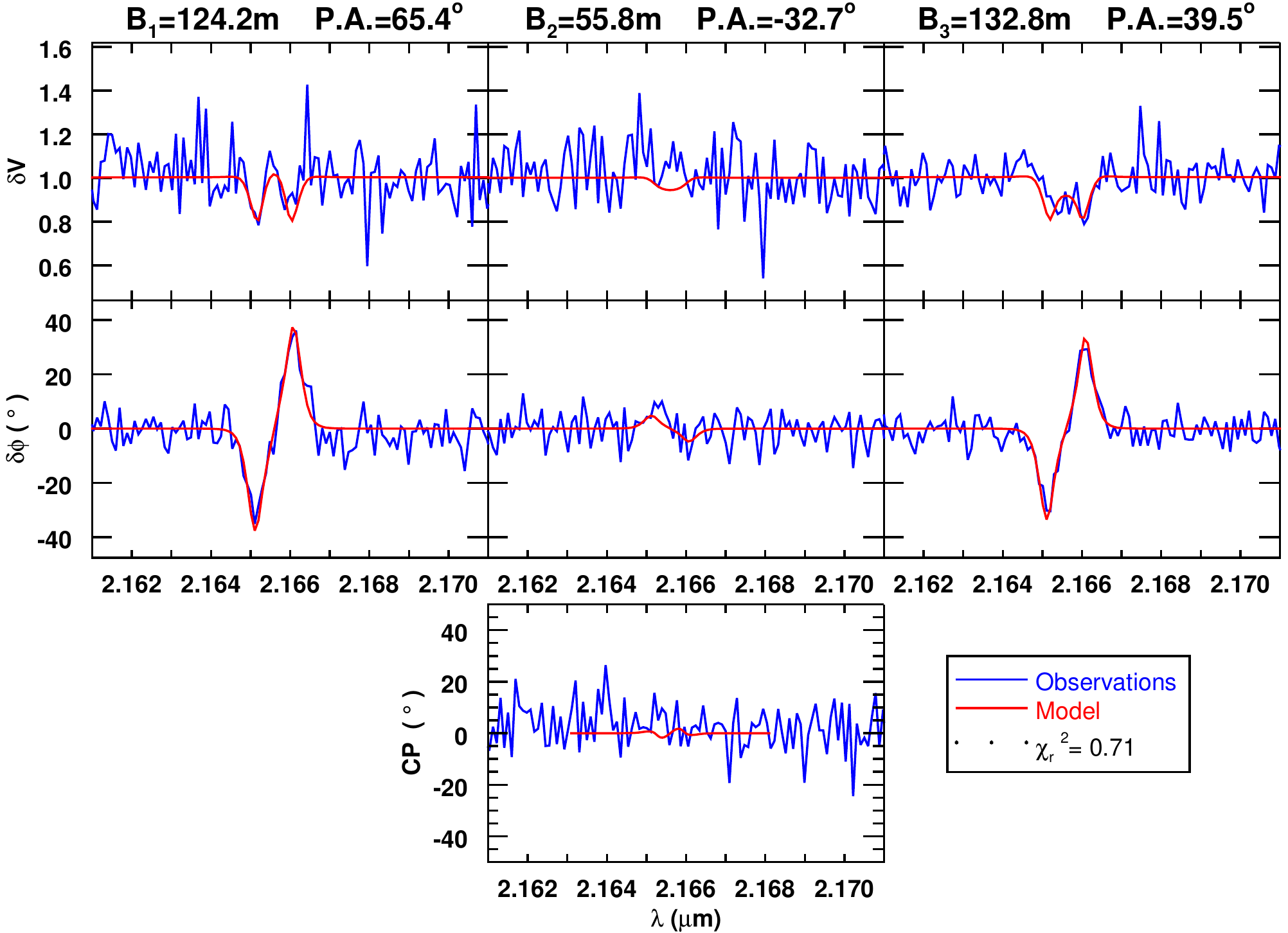}
   \caption{As in  Fig.~\ref{fig:HD23630}, but for \object{HD\,30\,076}.}
   \label{fig:HD30076}
   \end{figure*}

   \begin{figure*}
   \centering
   \includegraphics[width=0.49\hsize]{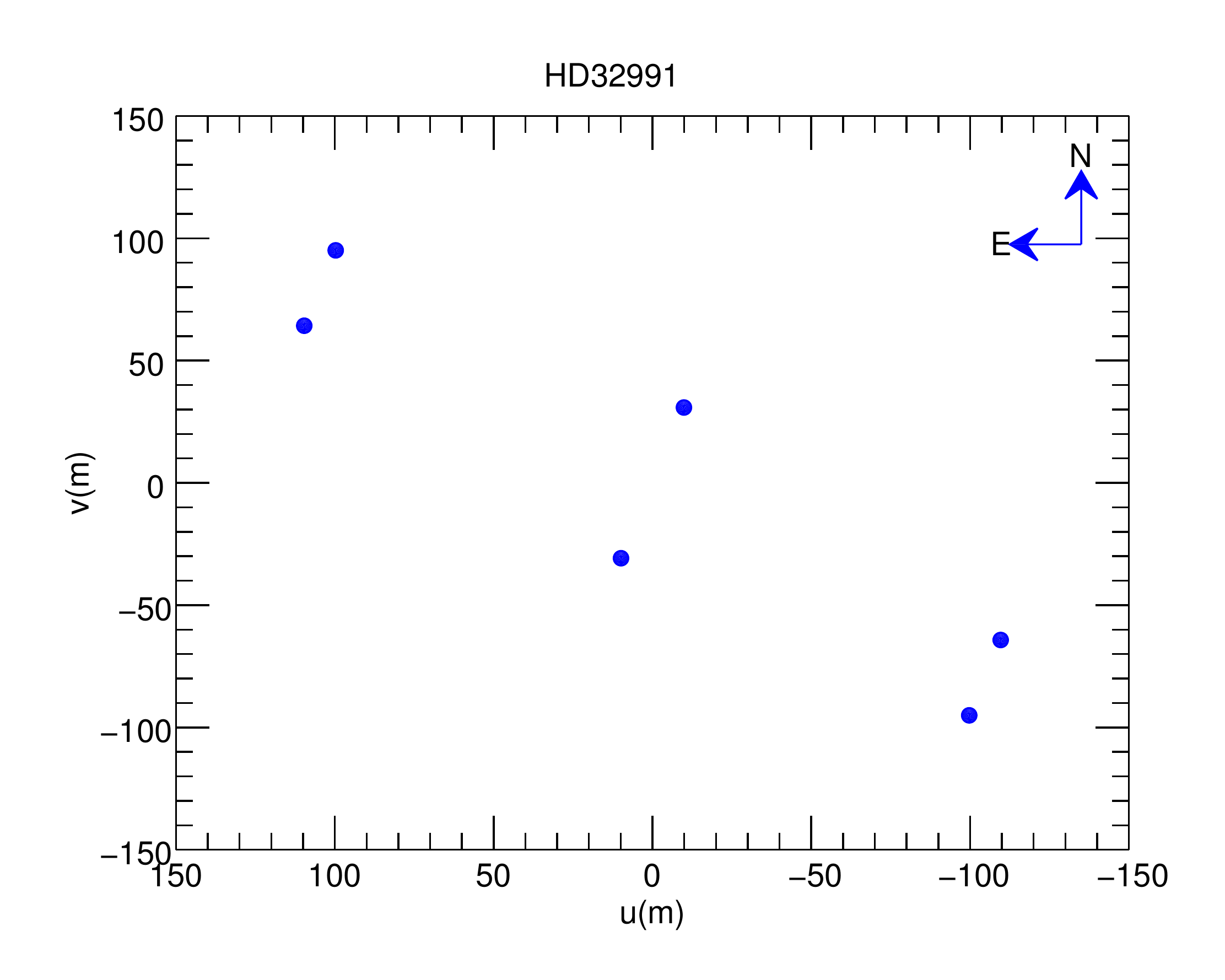}
   \includegraphics[width=0.49\hsize]{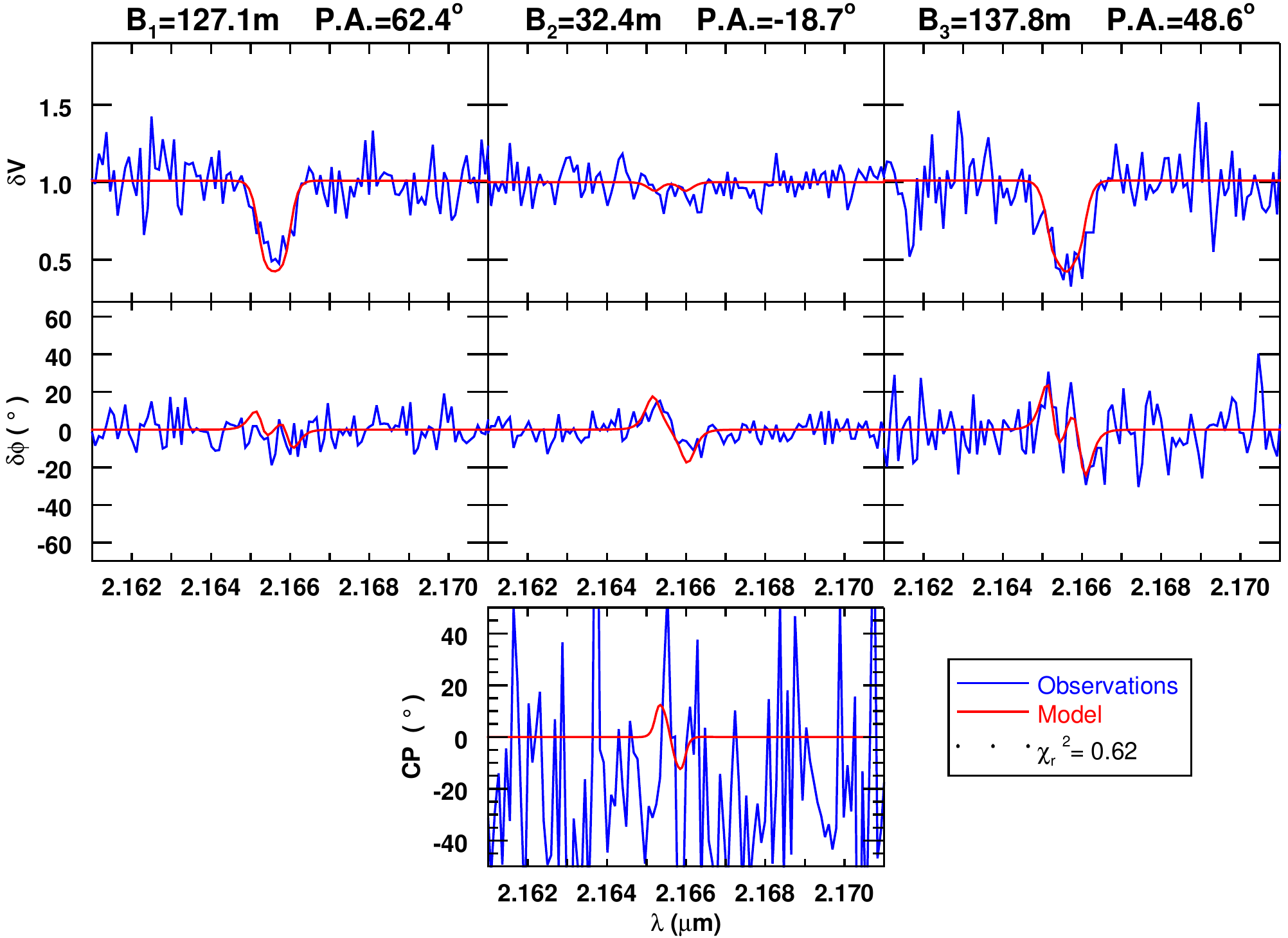}
   \caption{As in Fig.~\ref{fig:HD23630}, but for \object{HD\,32\,991}.}
   \label{fig:HD32991}
   \end{figure*}

   \begin{figure*}
   \centering
   \includegraphics[width=0.49\hsize]{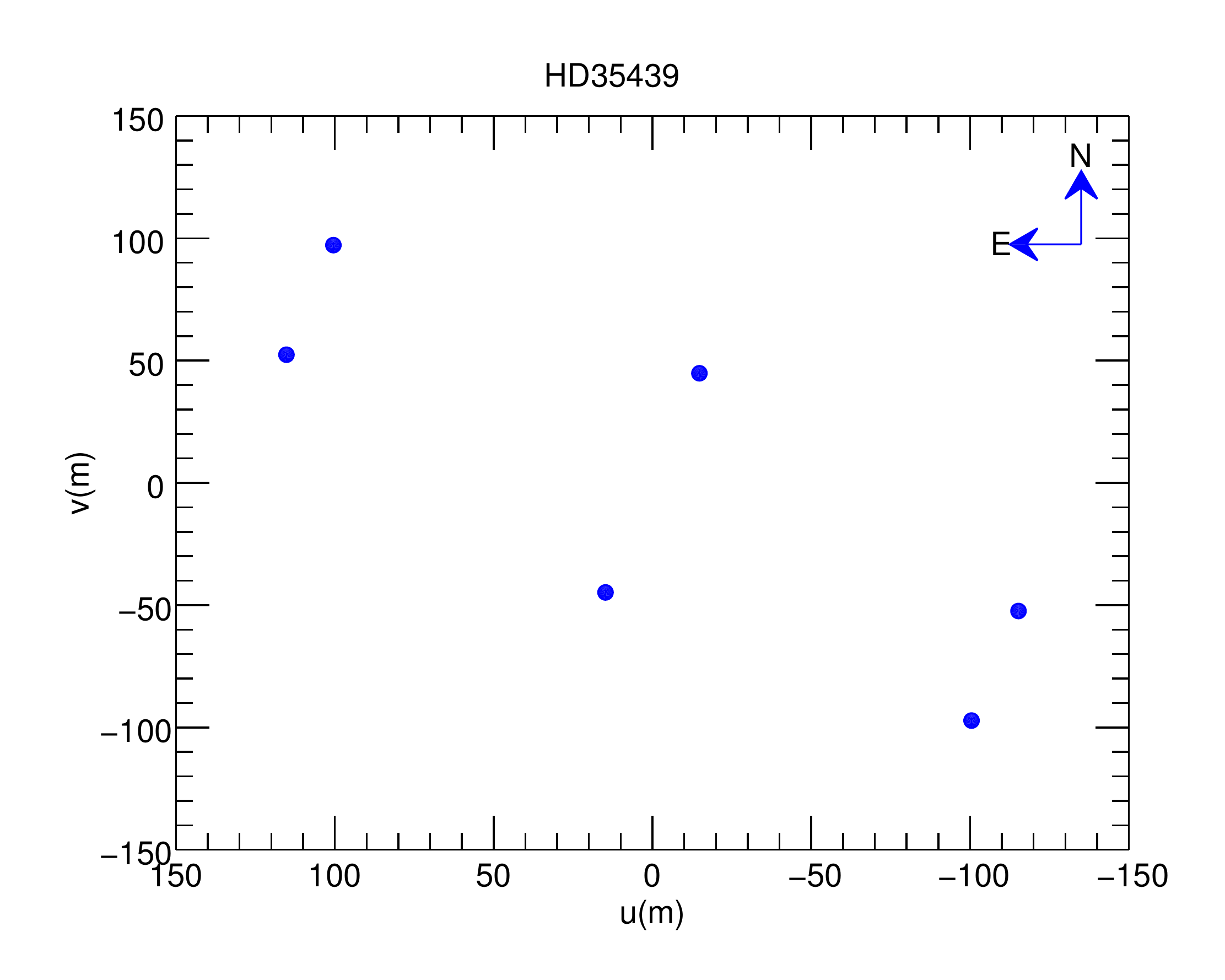}
   \includegraphics[width=0.49\hsize]{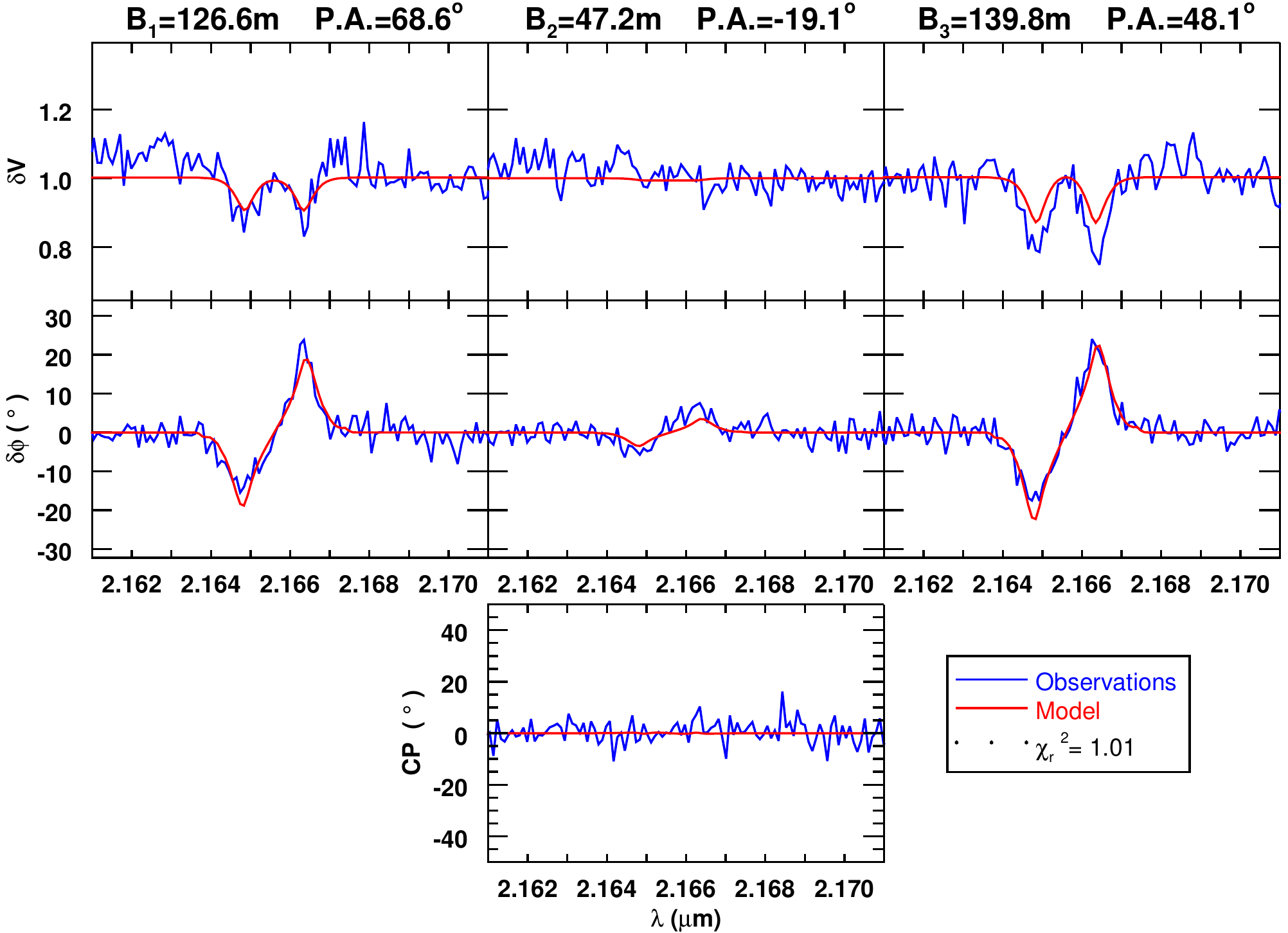}
   \caption{As in Fig.~\ref{fig:HD23630}, but for \object{HD\,35\,439}.}
   \label{fig:HD35439}
   \end{figure*}

   \begin{figure*}
   \centering
   \includegraphics[width=0.49\hsize]{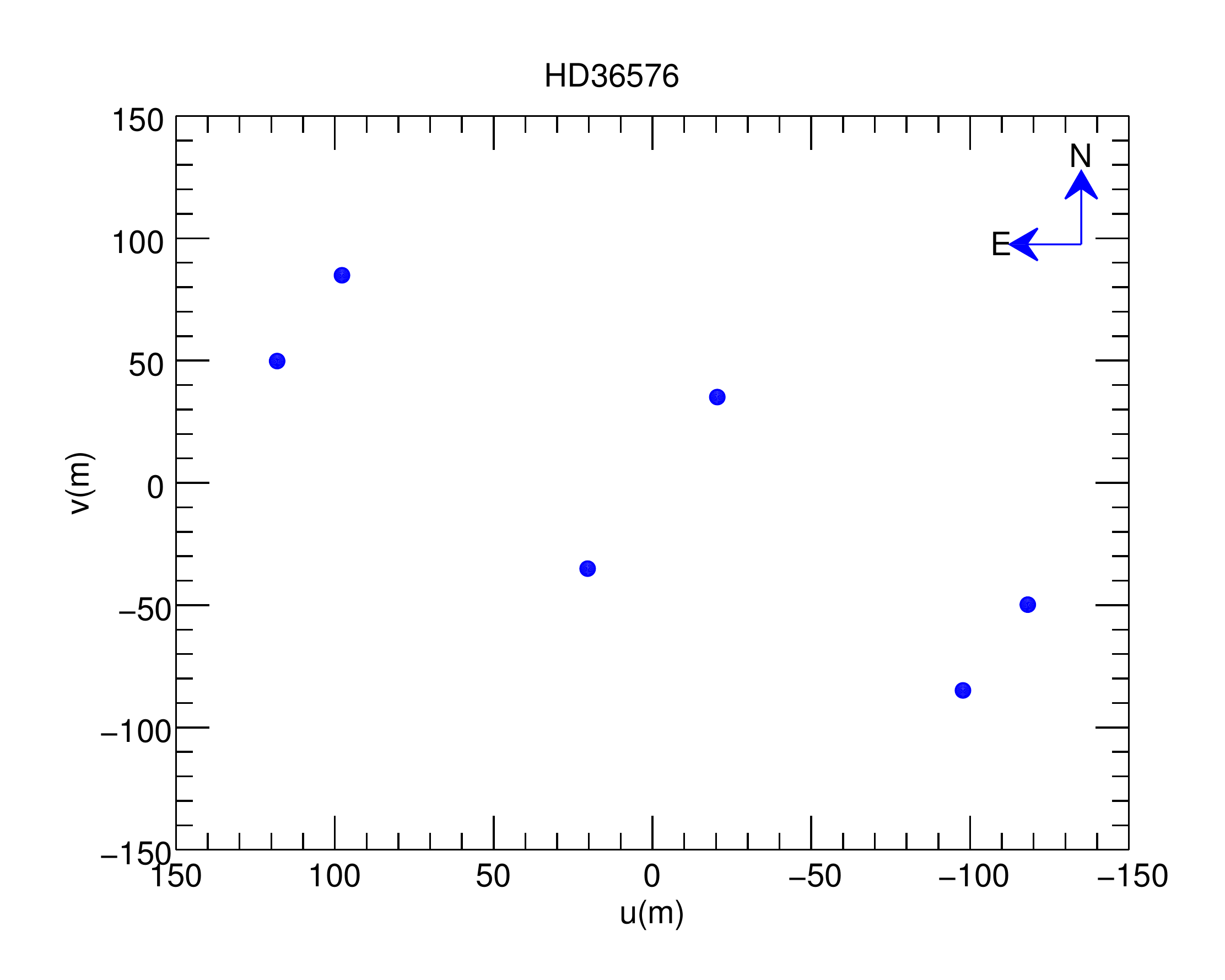}
   \includegraphics[width=0.49\hsize]{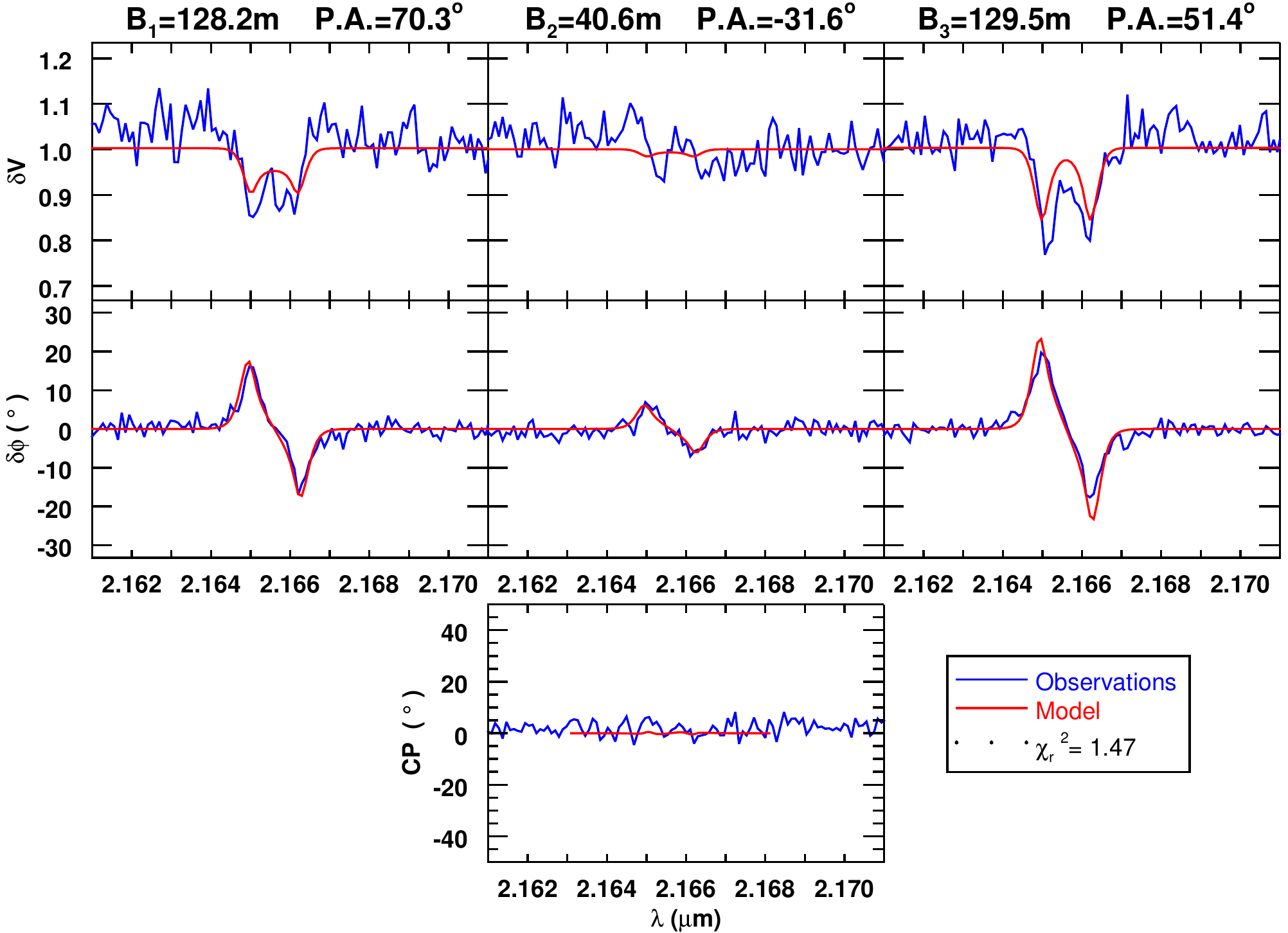}
   \caption{As in Fig.~\ref{fig:HD23630}, but for \object{HD\,36\,576}.}
   \label{fig:HD36576}
   \end{figure*}

   \begin{figure*}
   \centering
   \includegraphics[width=0.49\hsize]{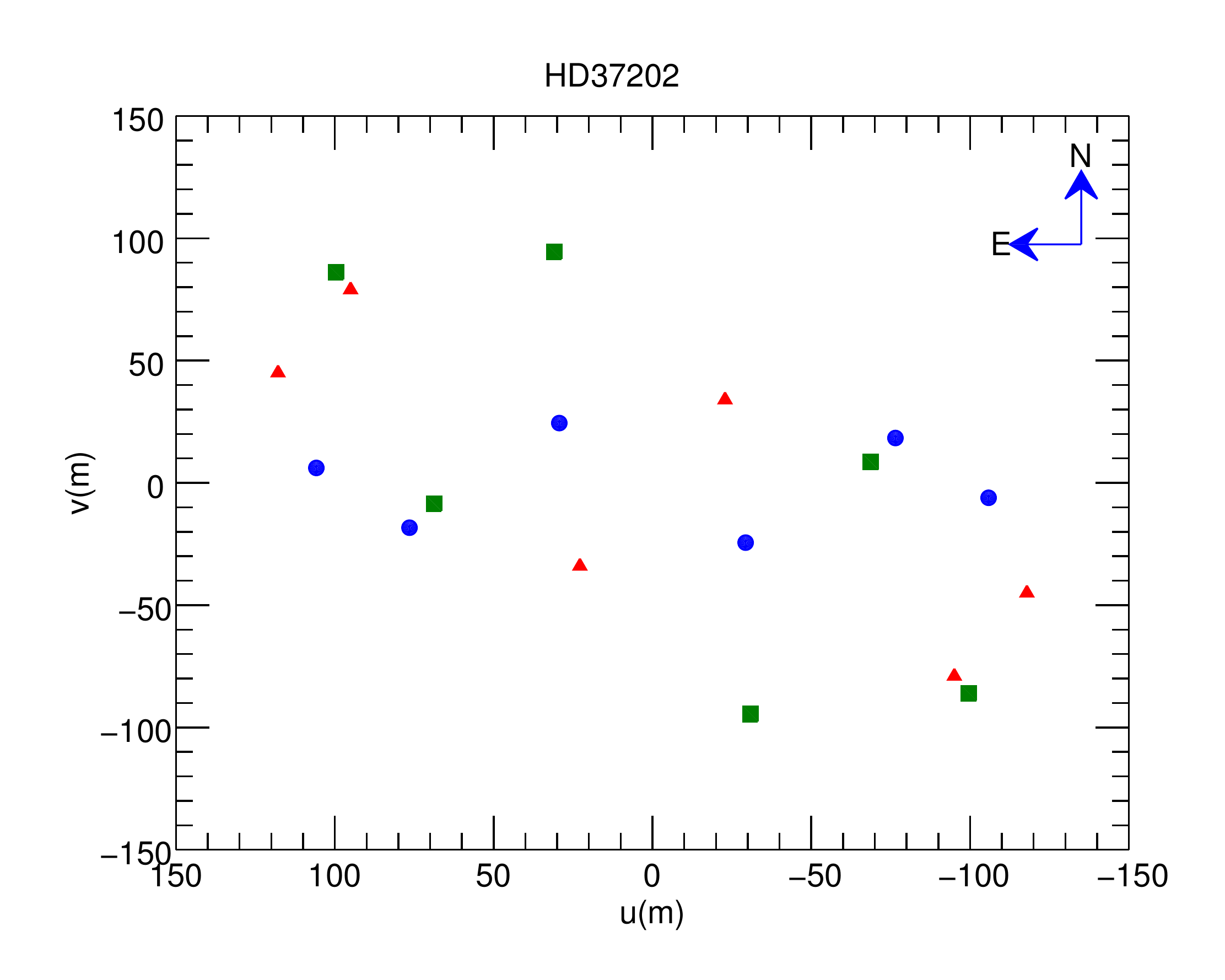}
   \includegraphics[width=0.49\hsize]{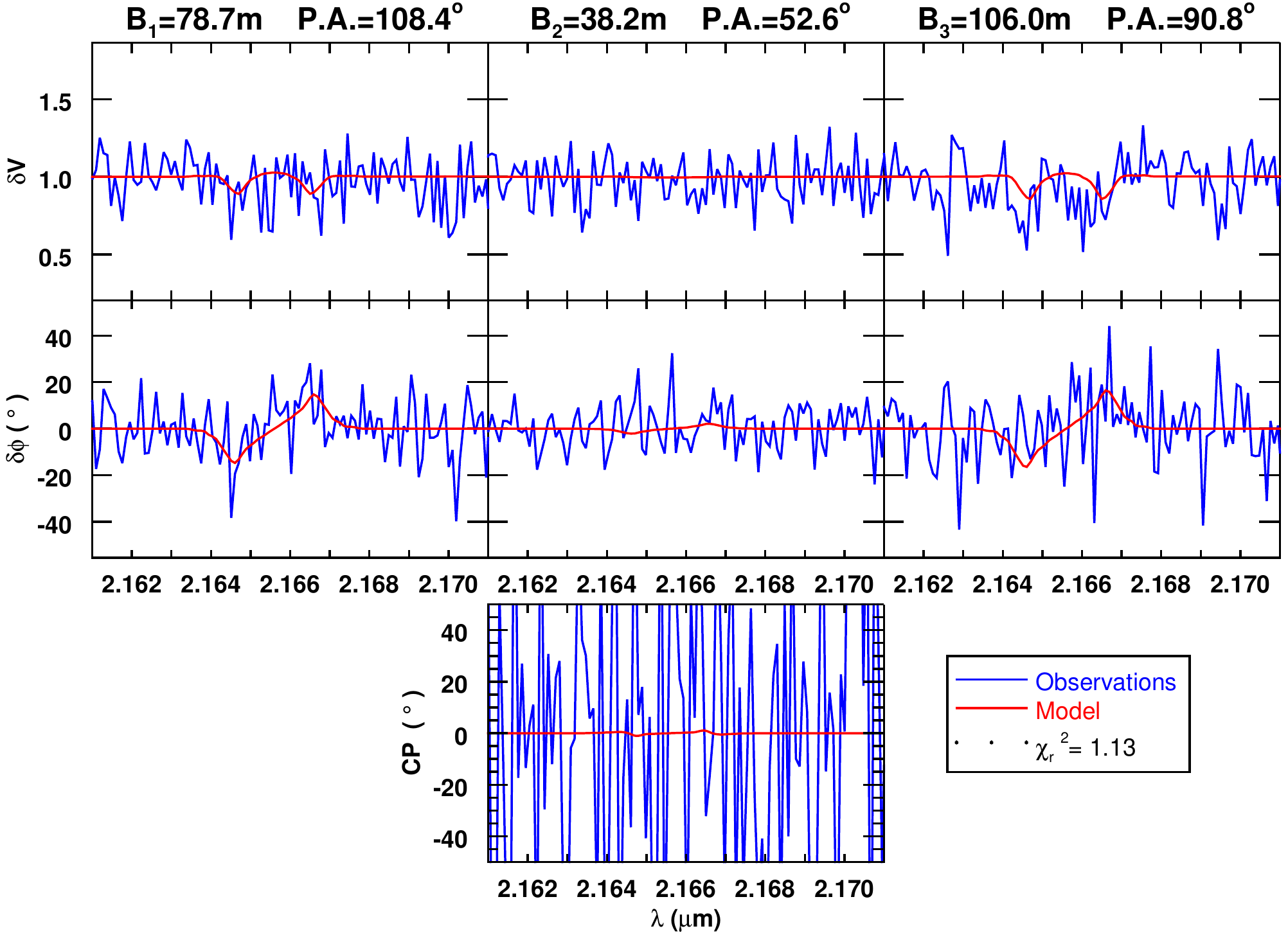}

   \includegraphics[width=0.49\hsize]{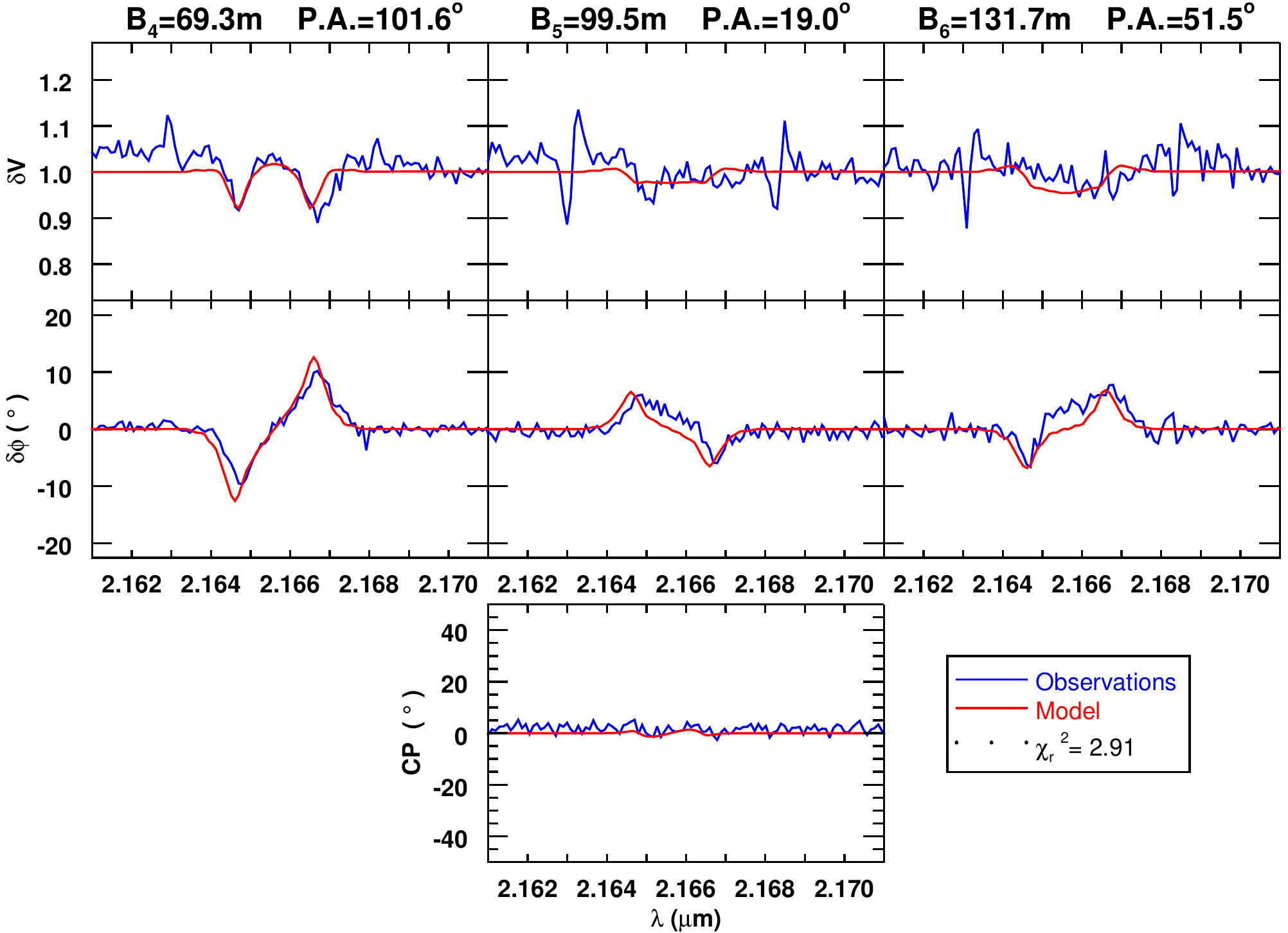}
   \includegraphics[width=0.49\hsize]{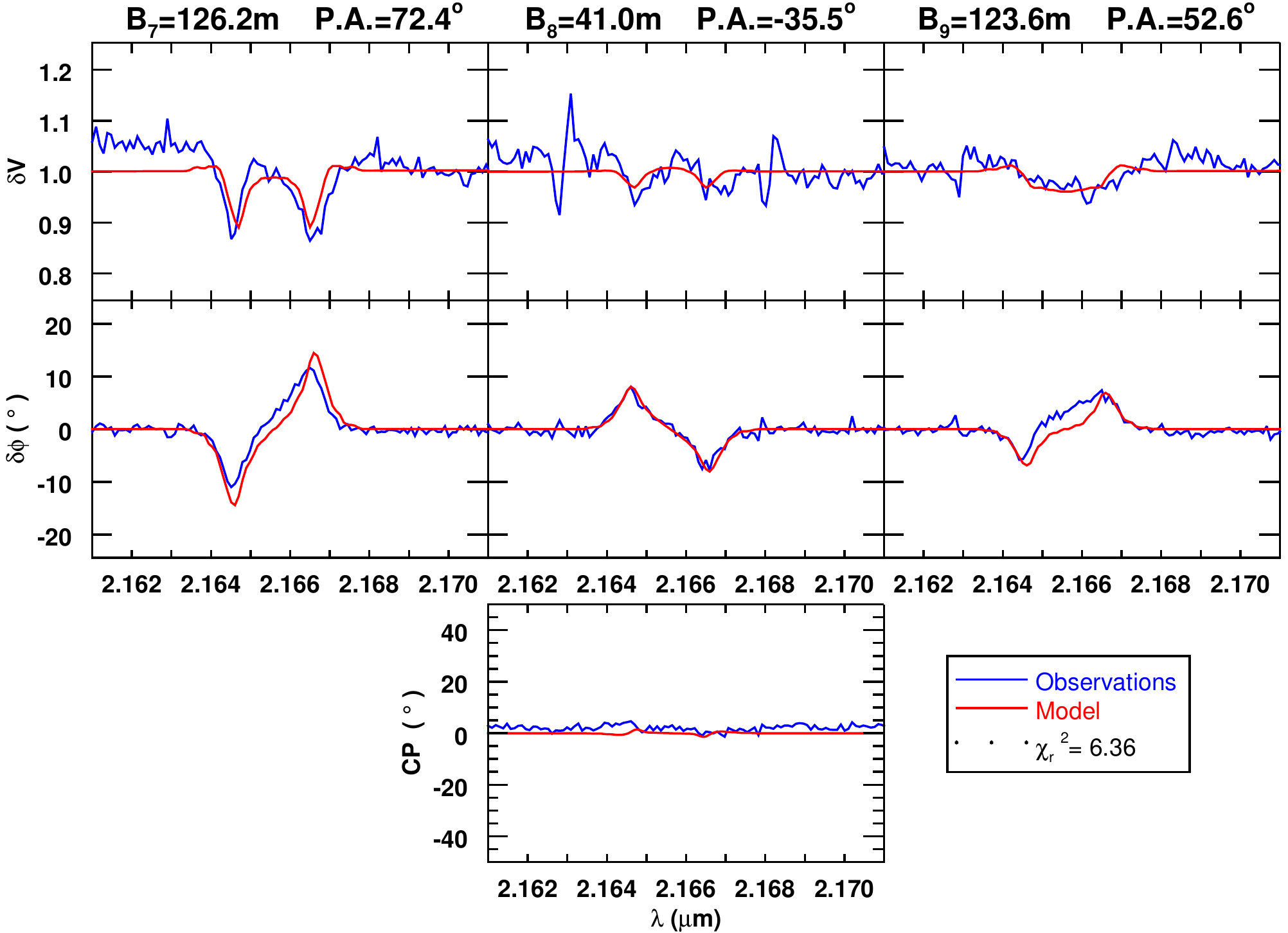}
   \caption{As in Fig.~\ref{fig:HD23630}, but for \object{HD\,37\,202}.}
   \label{fig:HD37202}
   \end{figure*}

   \begin{figure*}
   \centering
   \includegraphics[width=0.49\hsize]{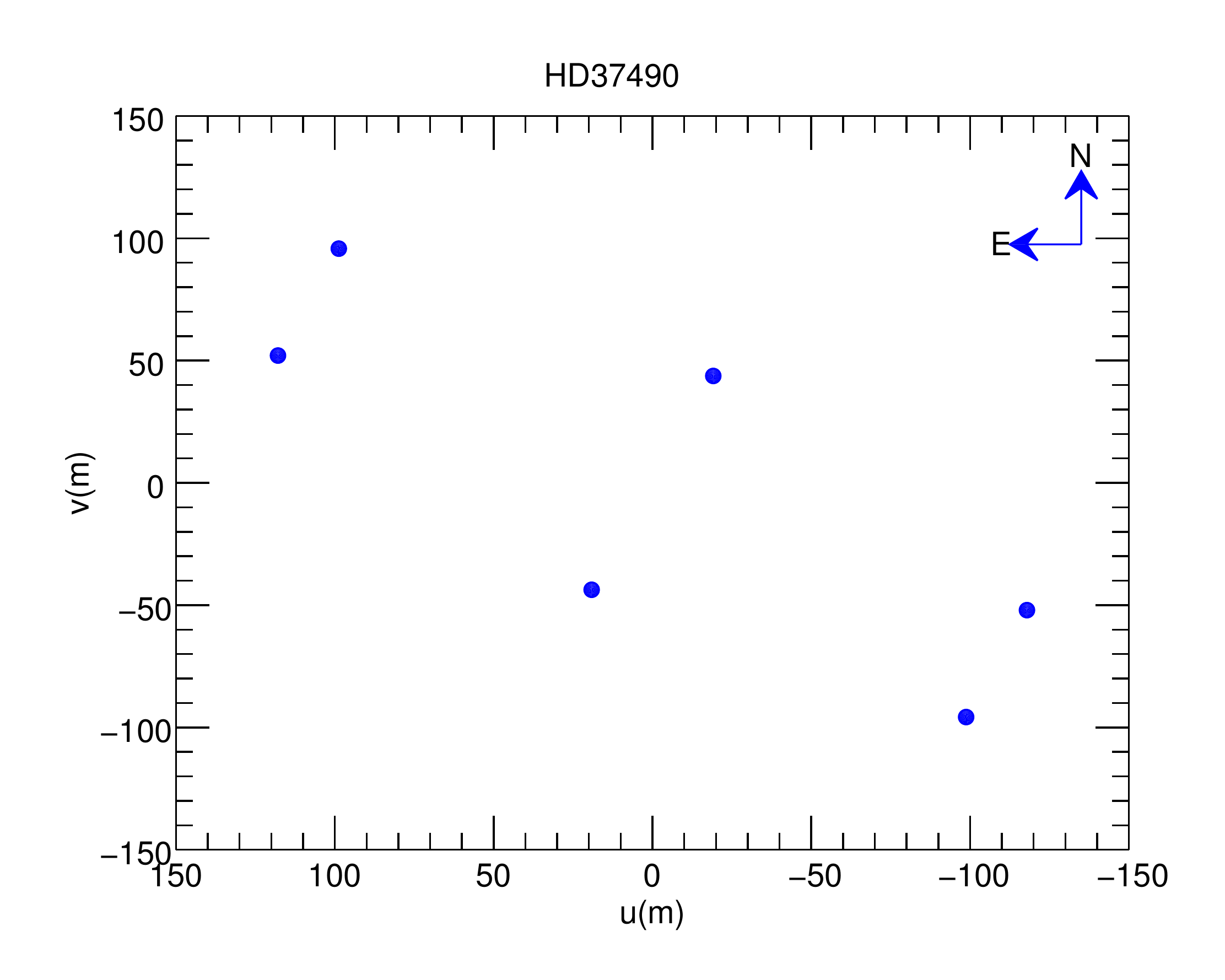}
   \includegraphics[width=0.49\hsize]{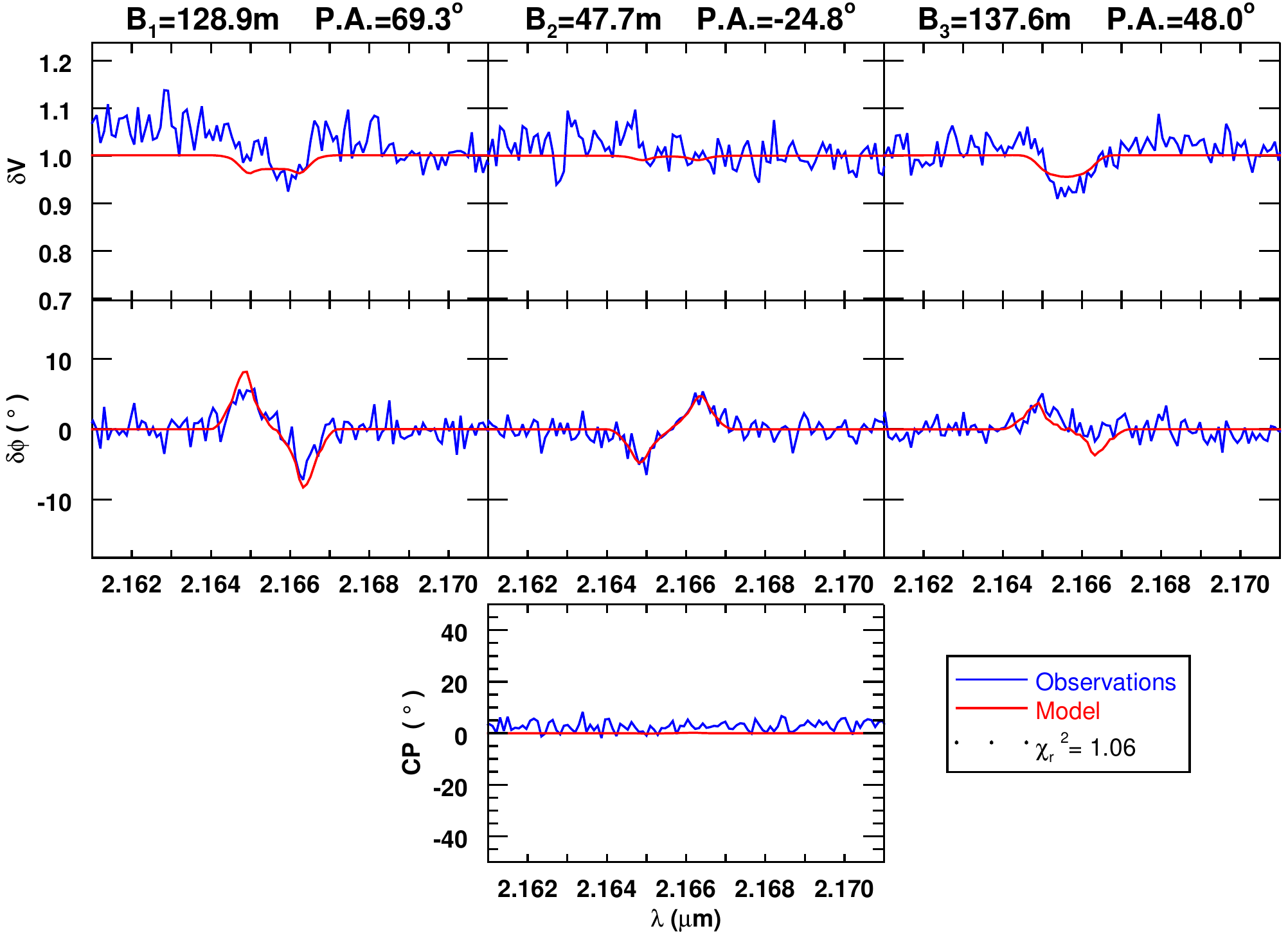}
   \caption{As in Fig.~\ref{fig:HD23630}, but for \object{HD\,37\,490}.}
   \label{fig:HD37490}
   \end{figure*}

   \begin{figure*}
   \centering
   \includegraphics[width=0.49\hsize]{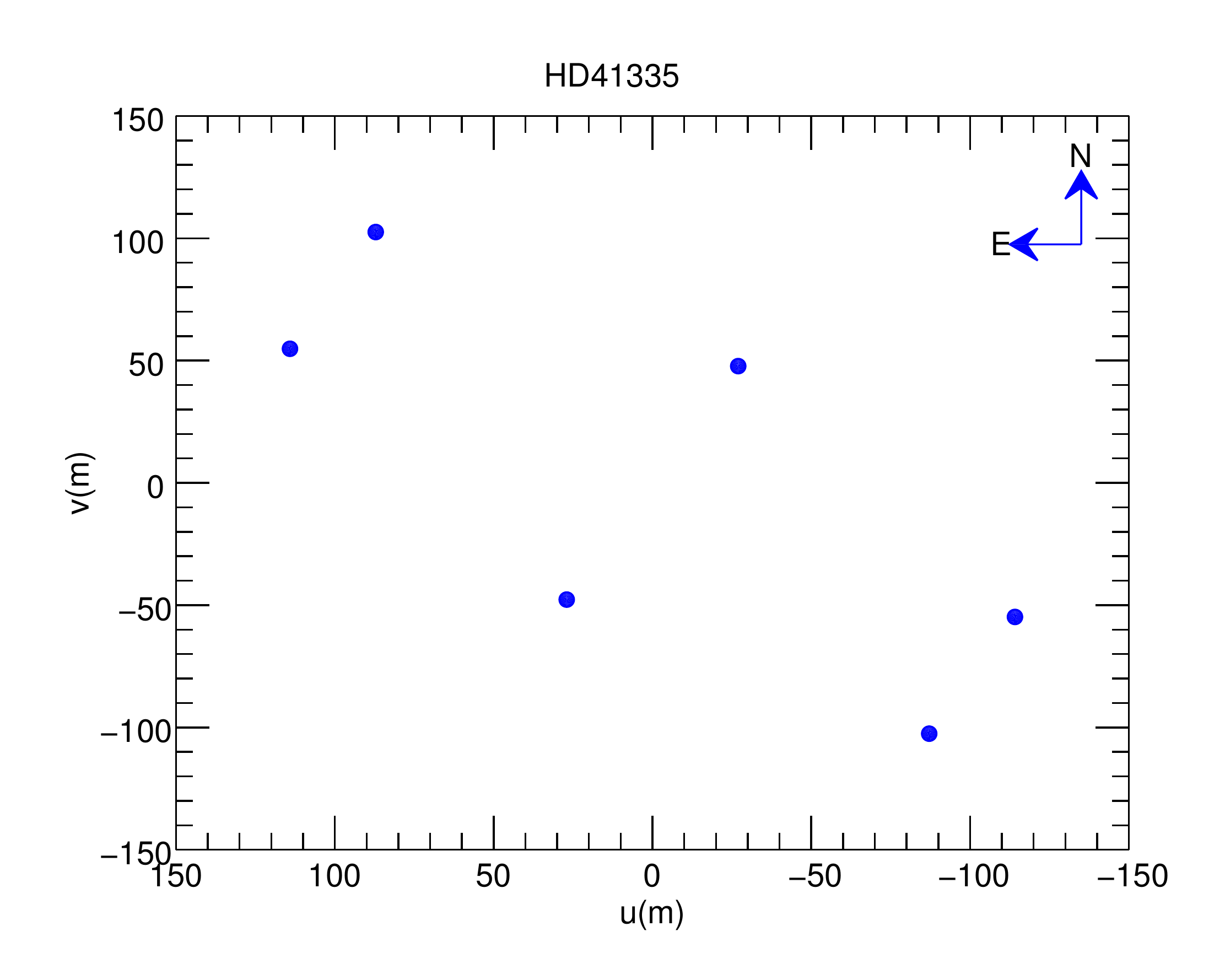}
   \includegraphics[width=0.49\hsize]{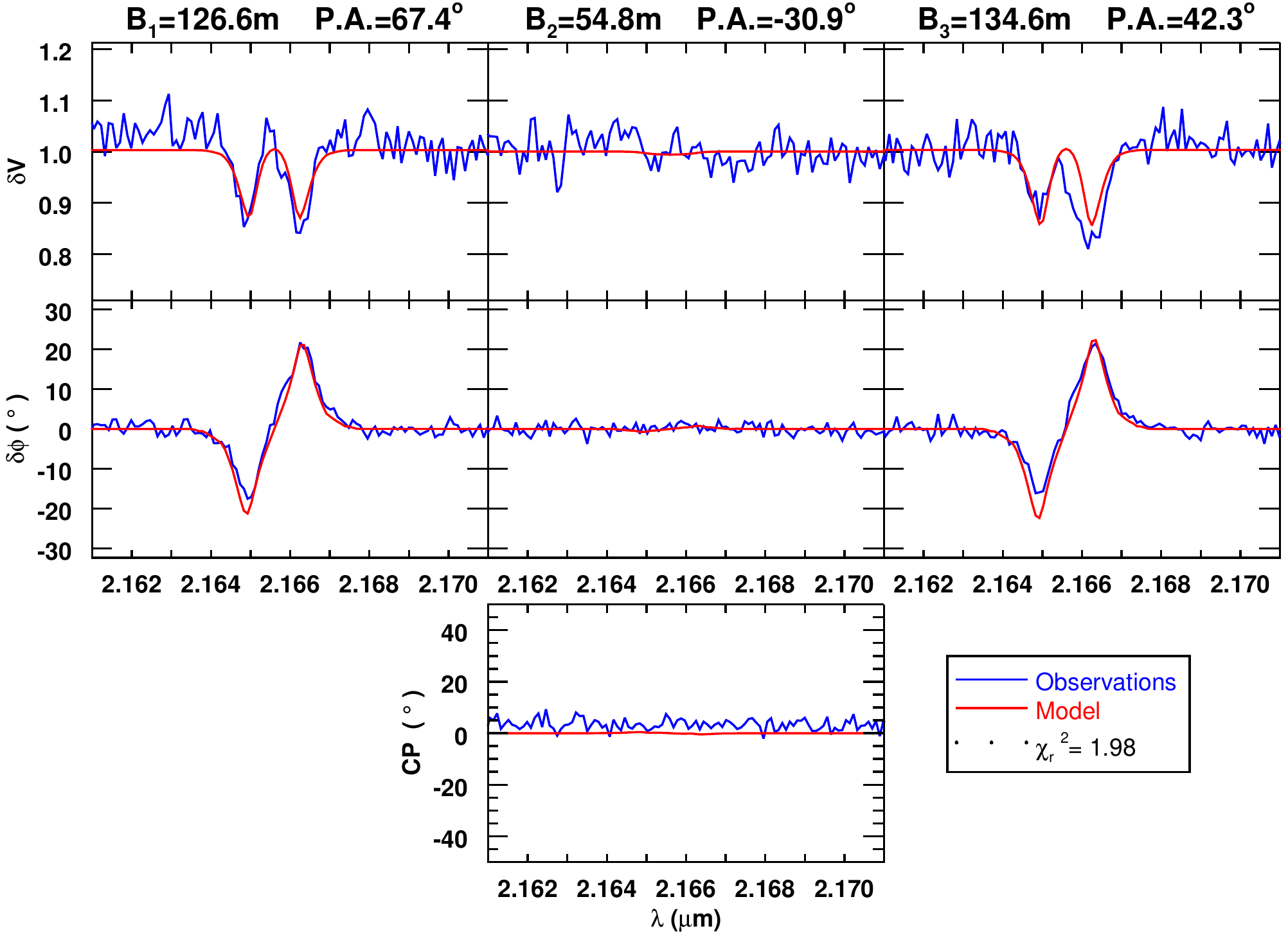}
   \caption{As in Fig.~\ref{fig:HD23630}, but for \object{HD\,41\,335}.}
   \label{fig:HD41335}
   \end{figure*}

   \begin{figure*}
   \centering
   \includegraphics[width=0.49\hsize]{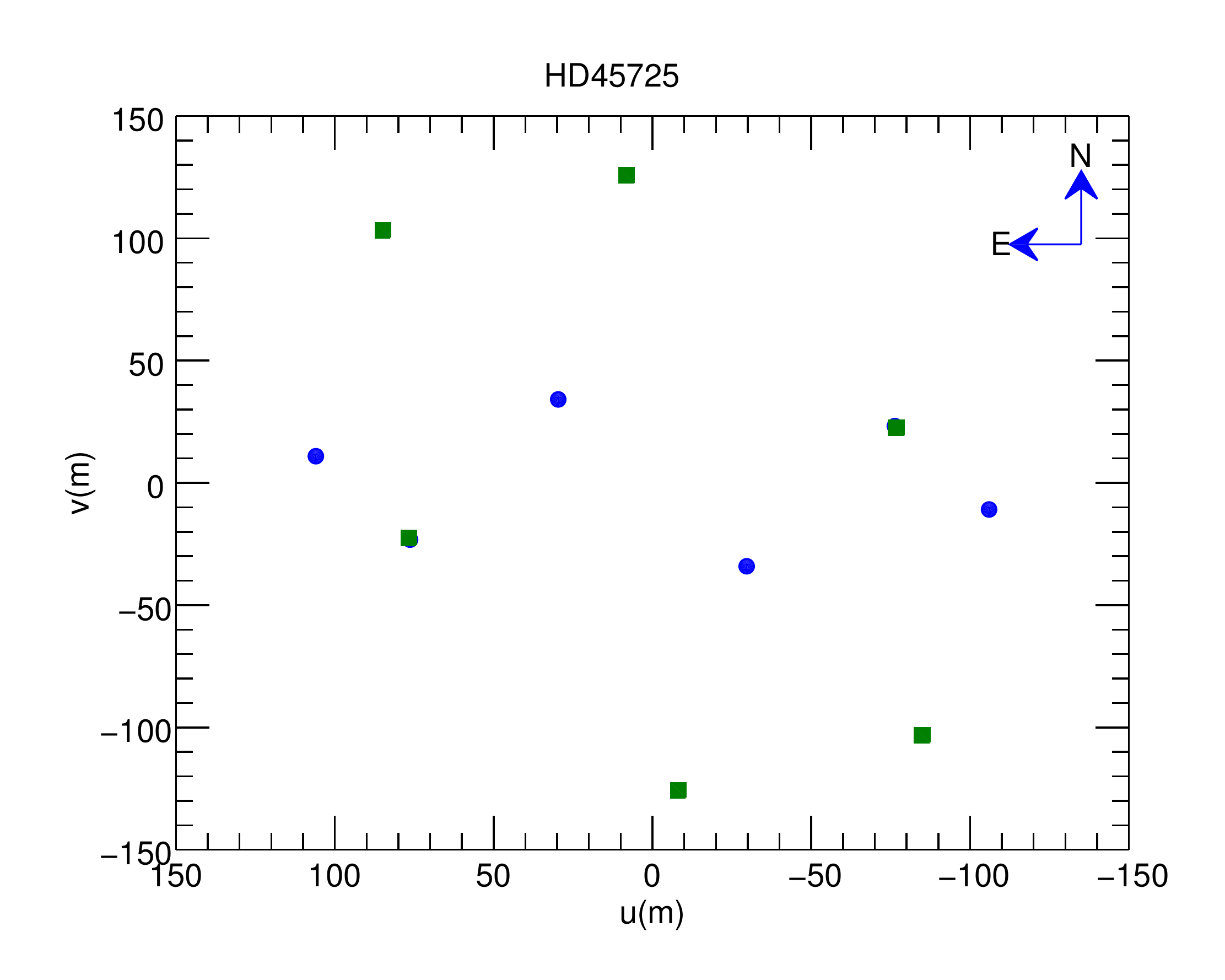}

   \includegraphics[width=0.49\hsize]{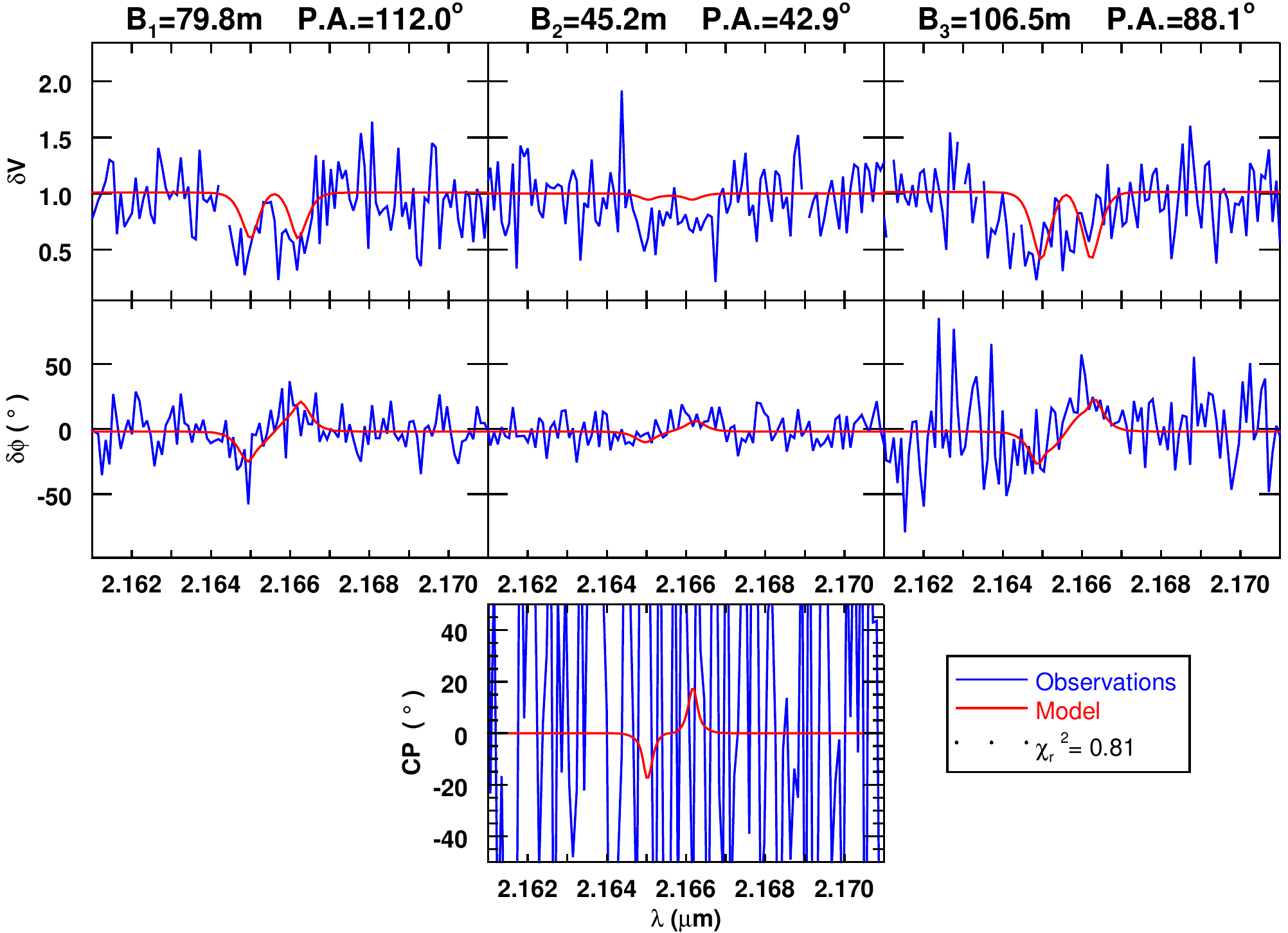}
   \includegraphics[width=0.49\hsize]{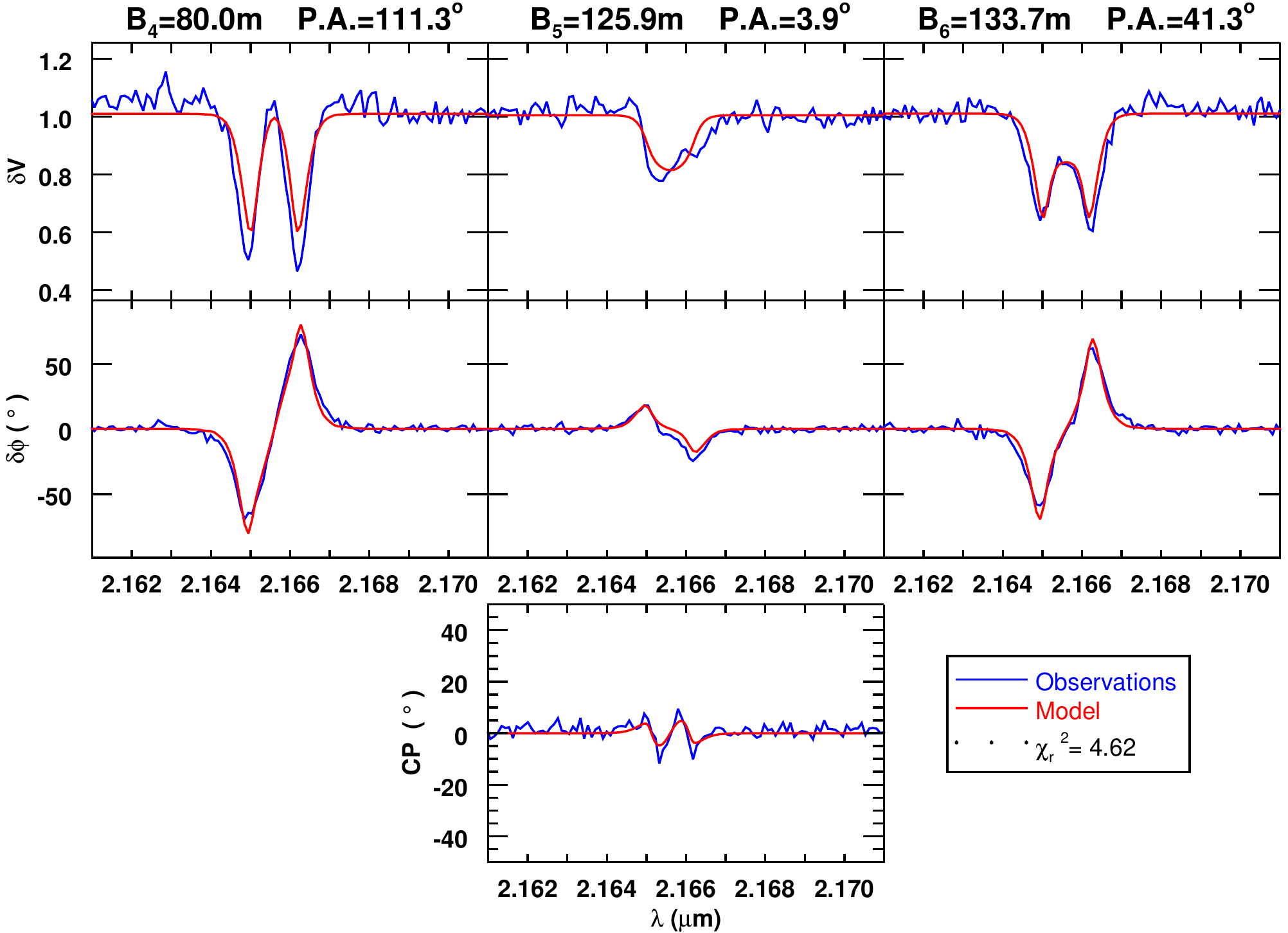}
   \caption{As in Fig.~\ref{fig:HD23630}, but for \object{HD\,45\,725}.}
   \label{fig:HD45725}
   \end{figure*}

   \begin{figure*}
   \centering
   \includegraphics[width=0.49\hsize]{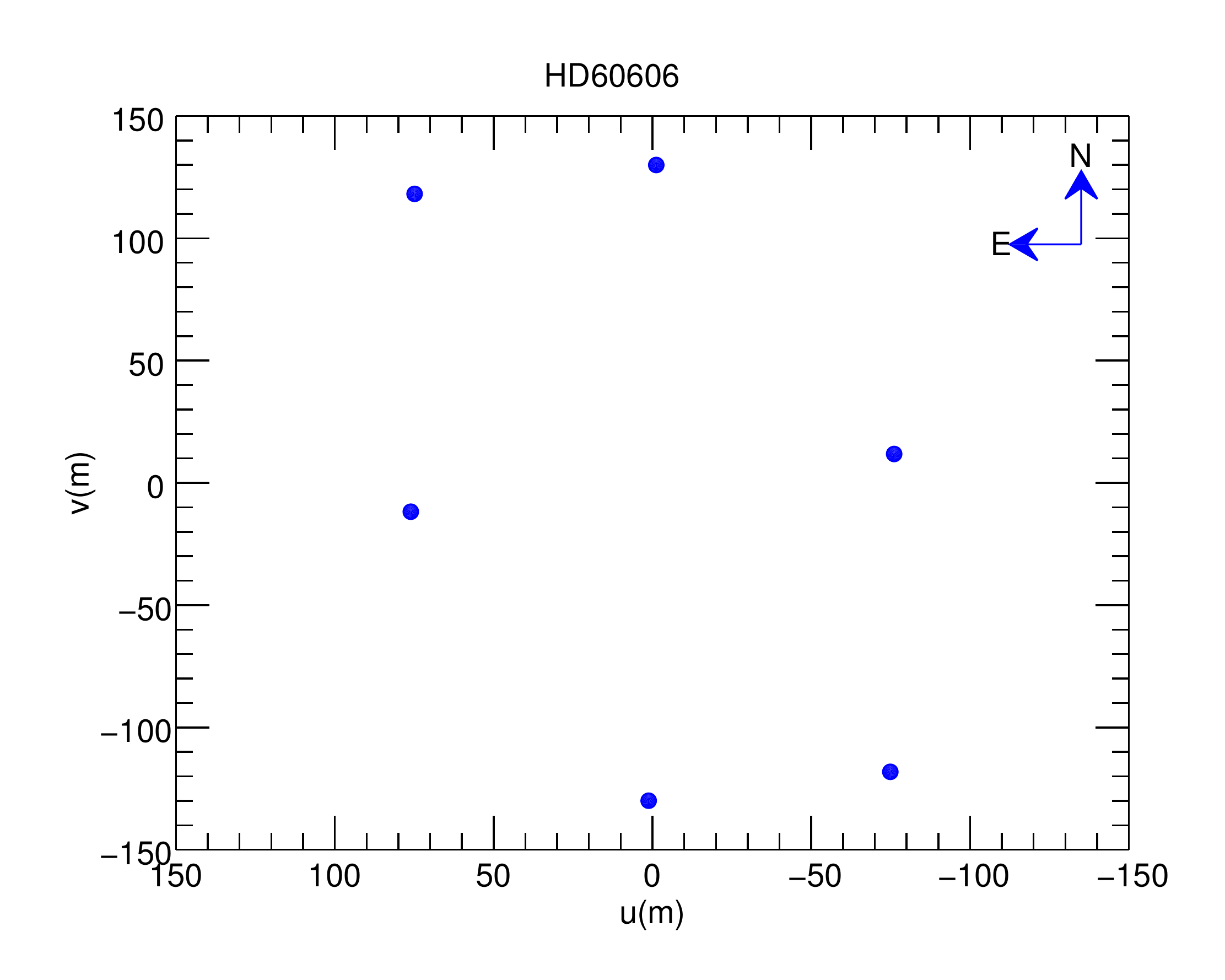}
   \includegraphics[width=0.49\hsize]{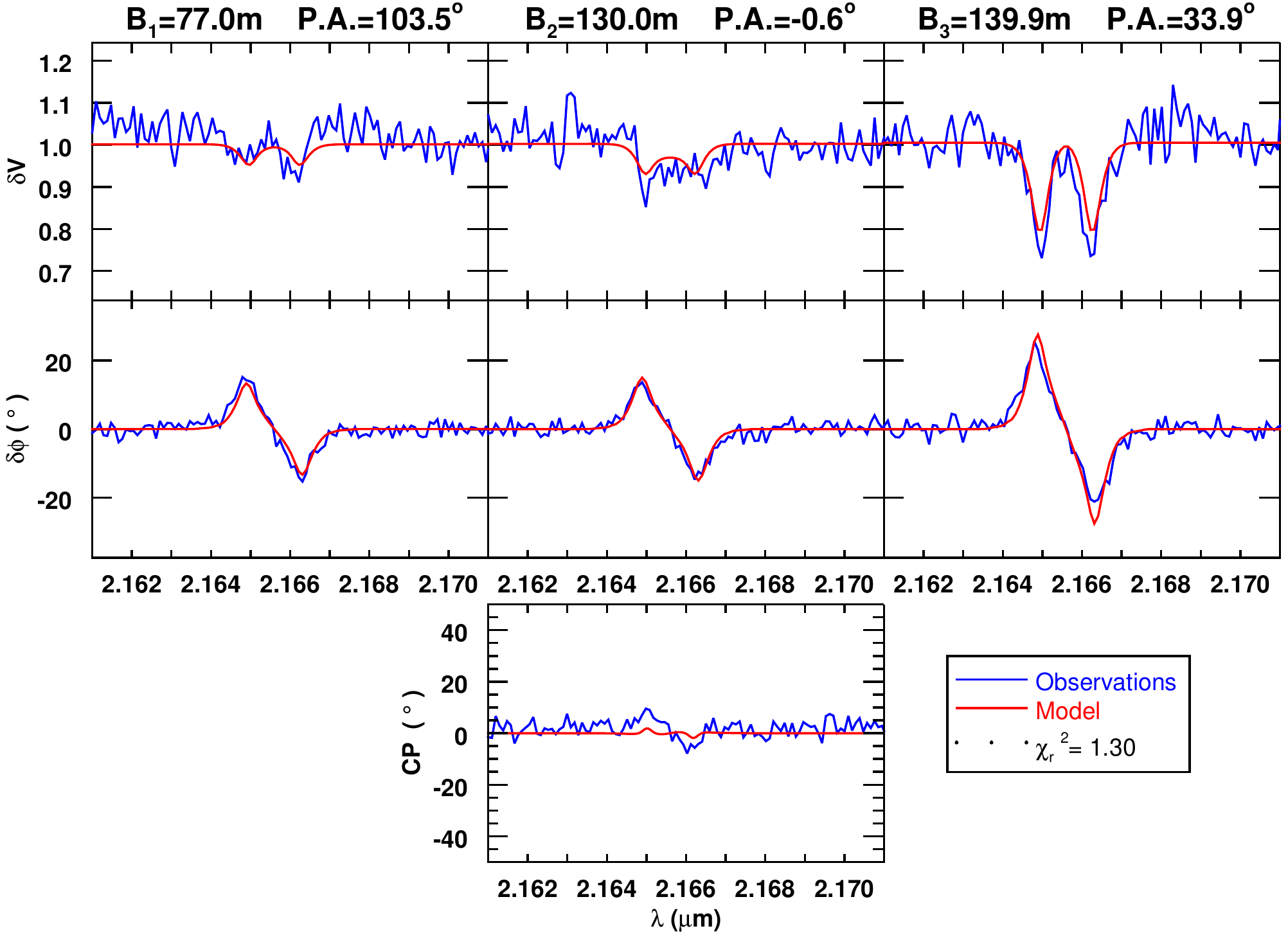}
   \caption{As in Fig.~\ref{fig:HD23630}, but for \object{HD\,60\,606}.}
   \label{fig:HD60606}
   \end{figure*}

   \begin{figure*}
   \centering
   \includegraphics[width=0.49\hsize]{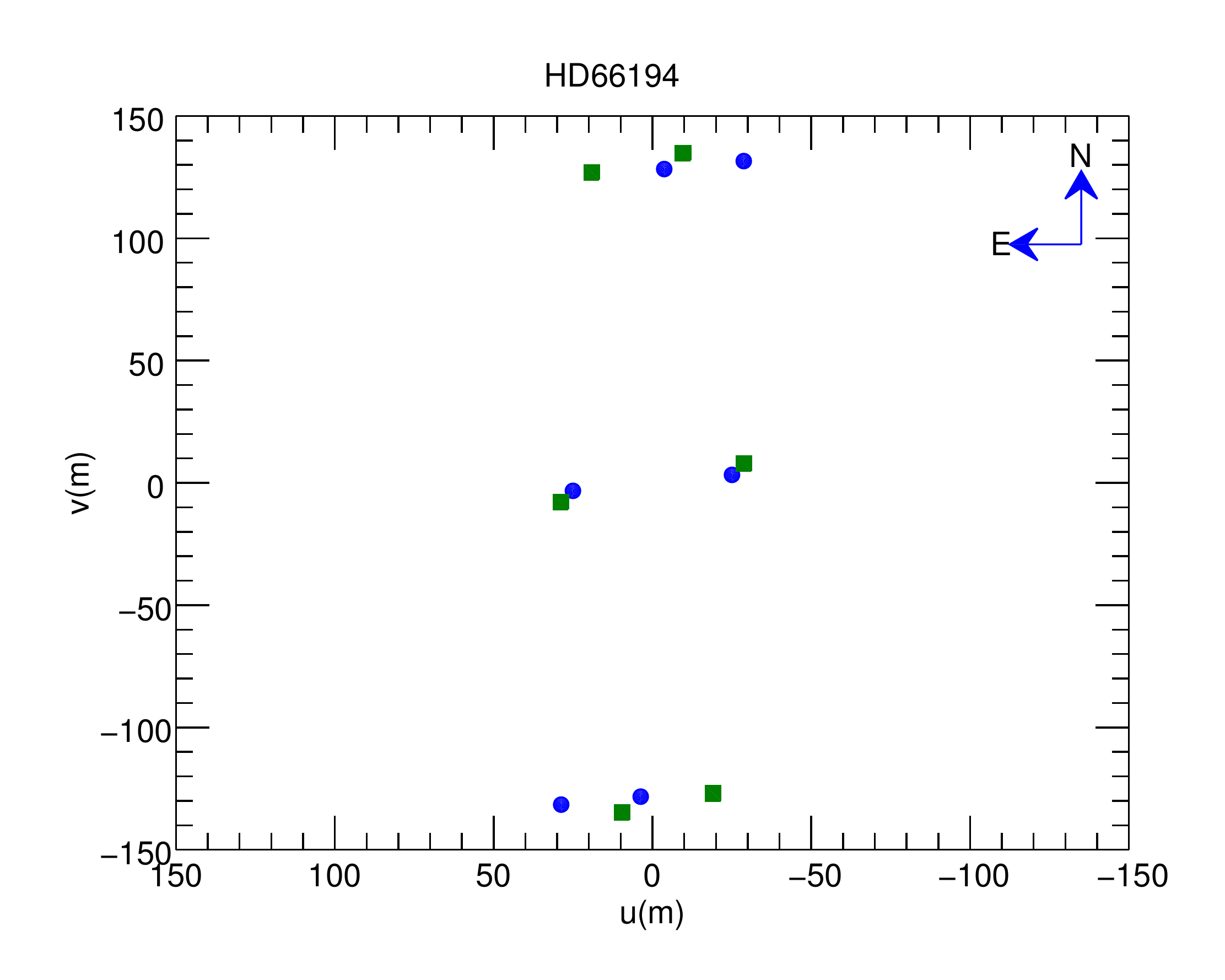}

   \includegraphics[width=0.49\hsize]{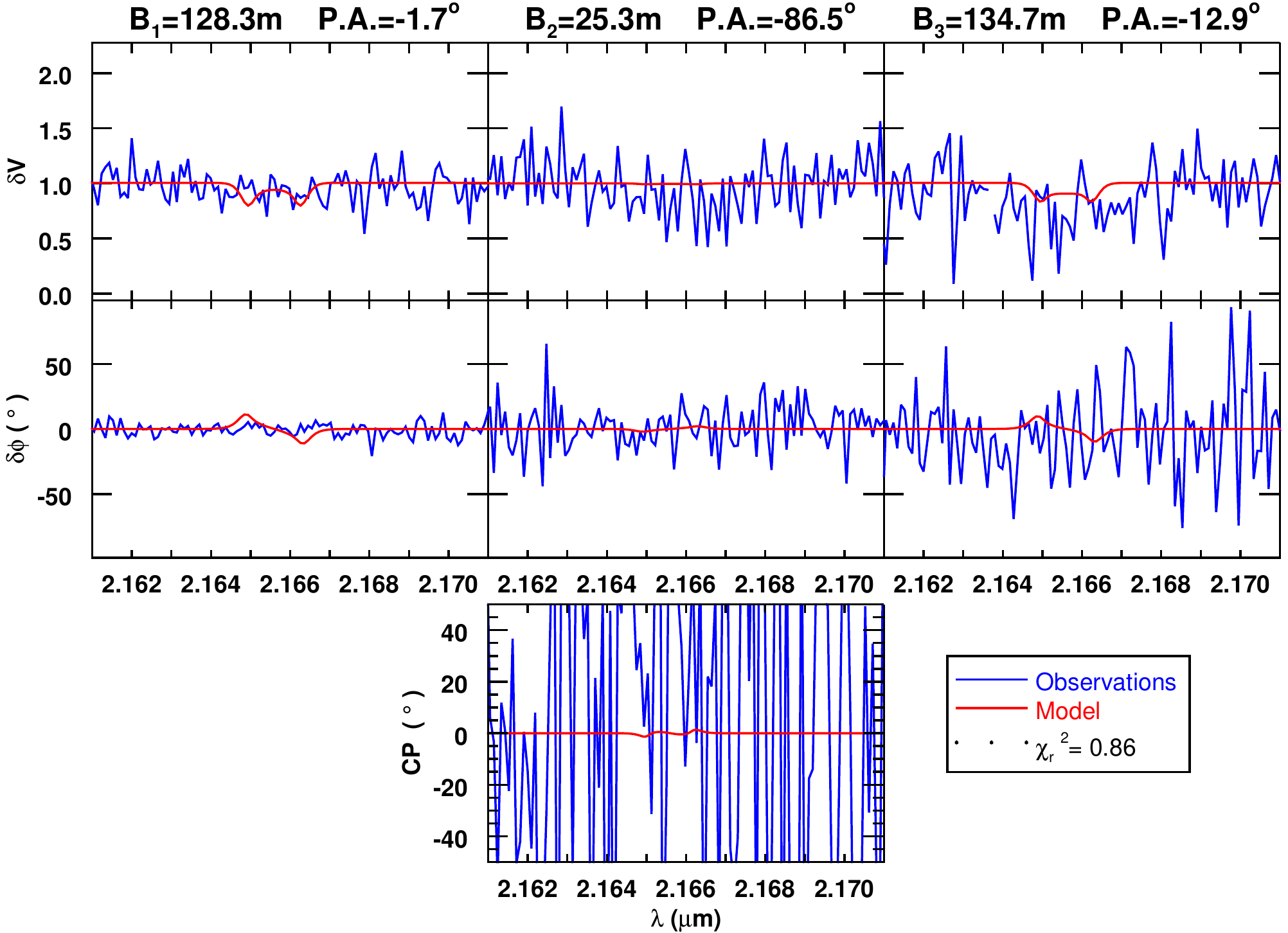}
   \includegraphics[width=0.49\hsize]{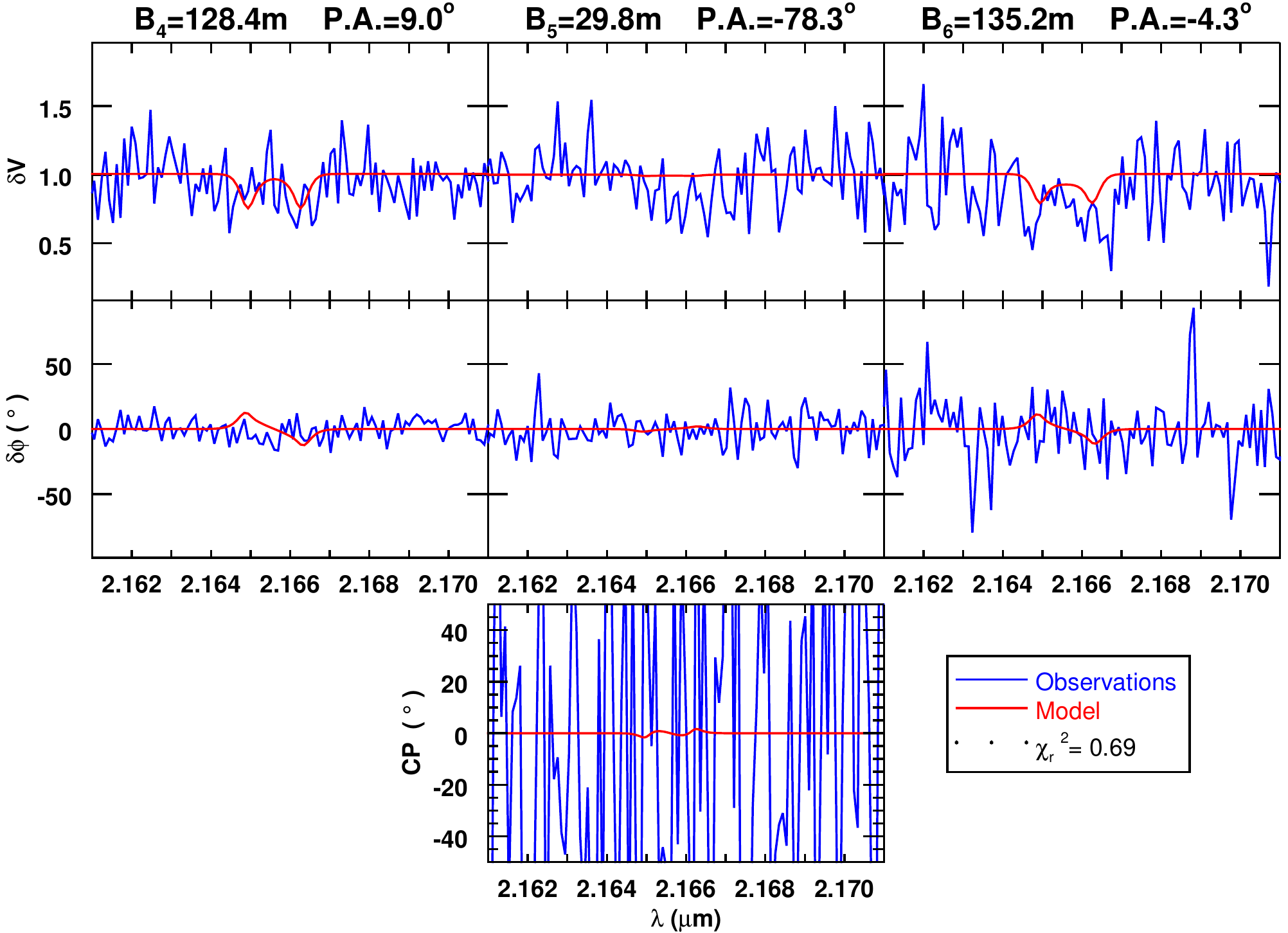}
   \caption{As in Fig.~\ref{fig:HD23630}, but for \object{HD\,66\,194}.}
   \label{fig:HD66194}
   \end{figure*}

   \begin{figure*}
   \centering
   \includegraphics[width=0.49\hsize]{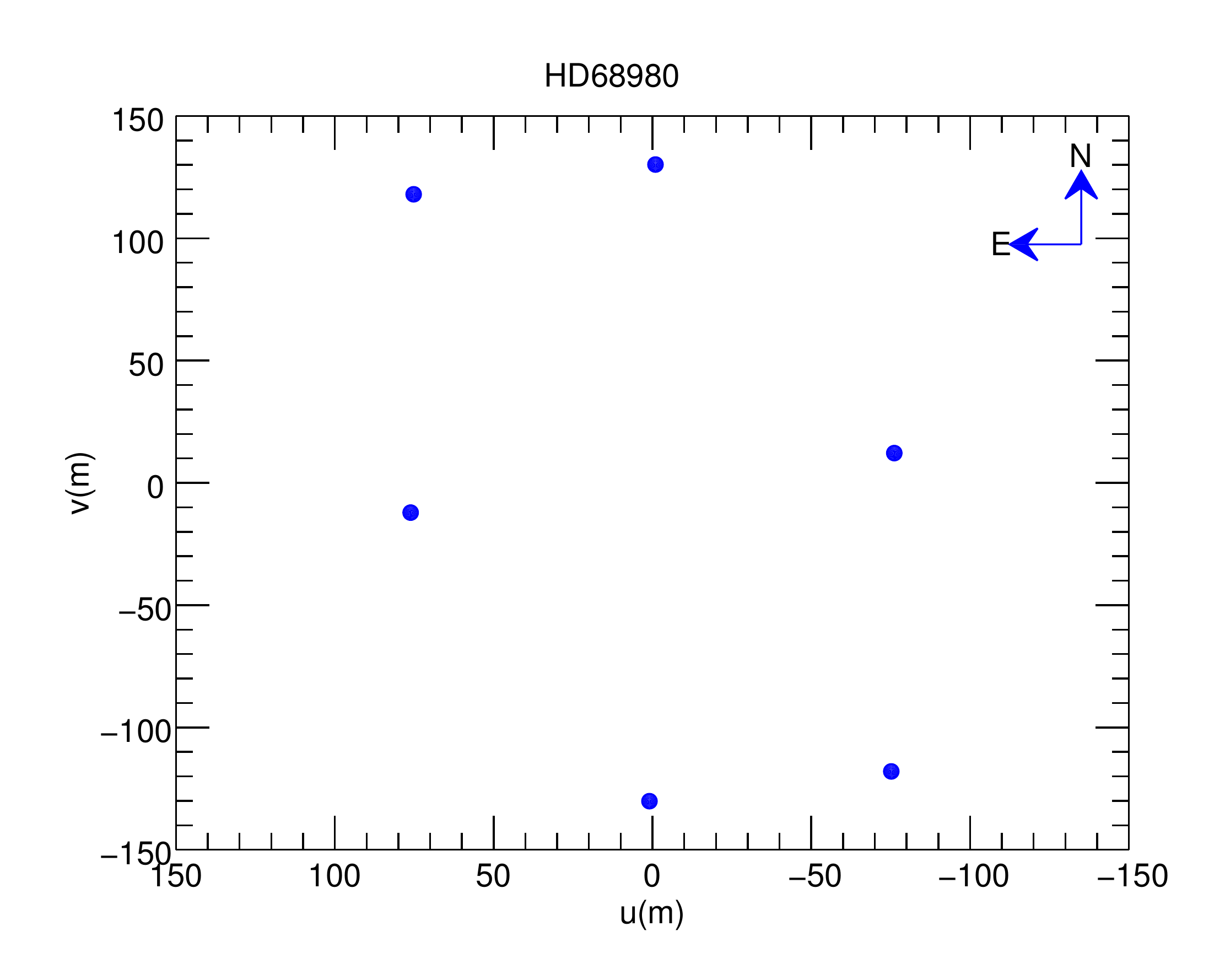}
   \includegraphics[width=0.49\hsize]{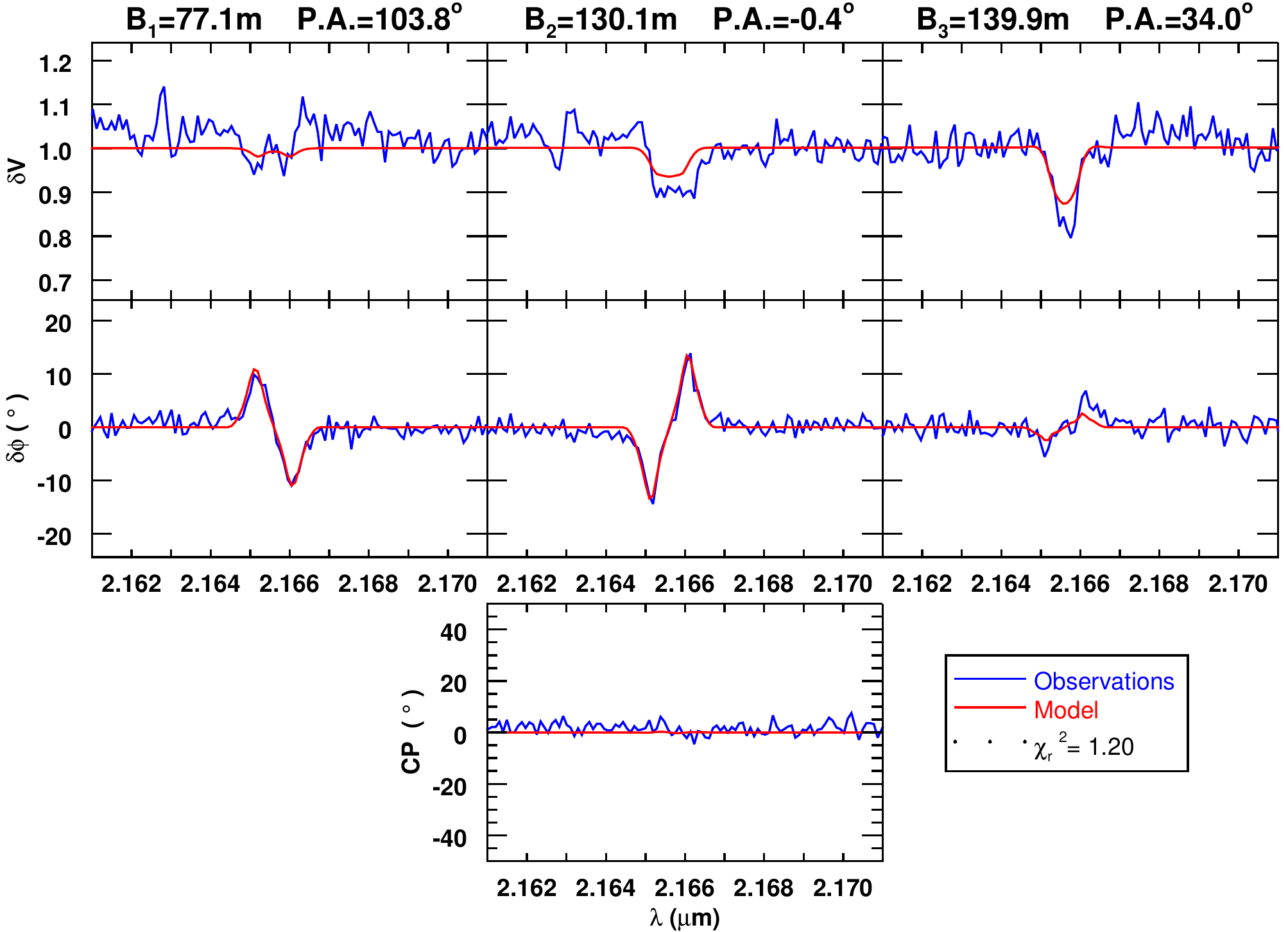}
   \caption{As in Fig.~\ref{fig:HD23630}, but for \object{HD\,68\,980}.}
   \label{fig:HD68980}
   \end{figure*}

   \begin{figure*}
   \centering
   \includegraphics[width=0.49\hsize]{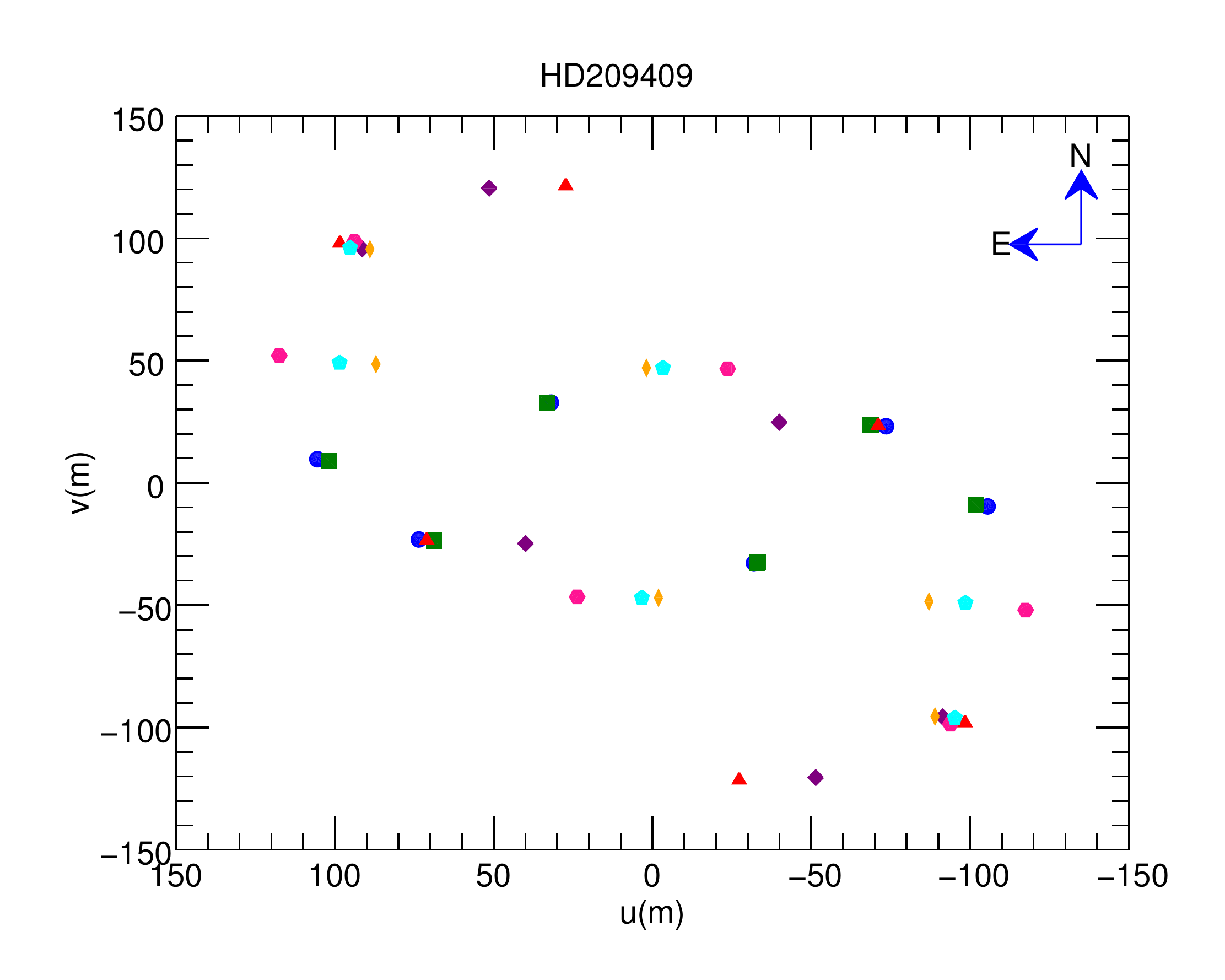}
   \includegraphics[width=0.49\hsize]{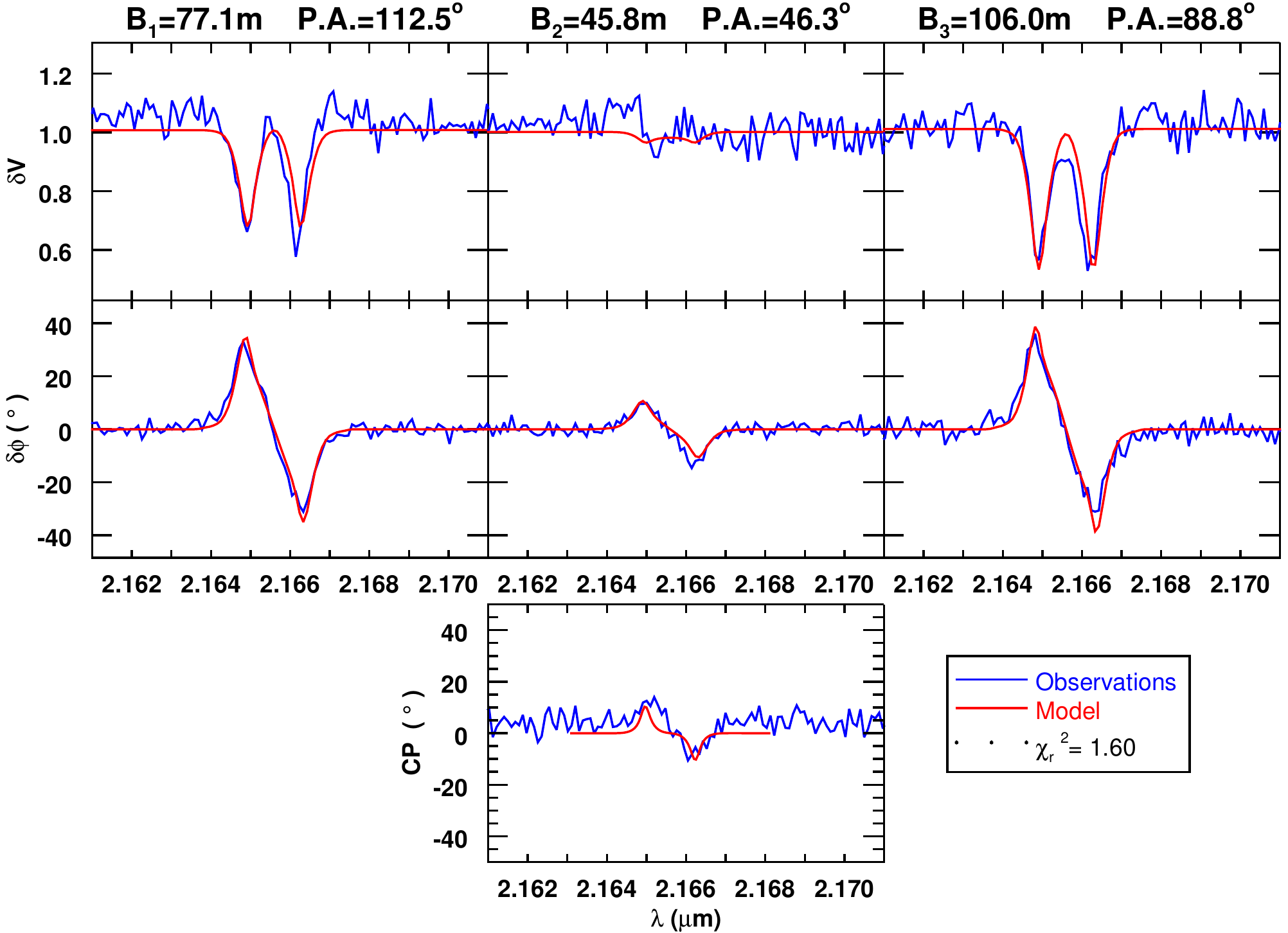}

   \includegraphics[width=0.49\hsize]{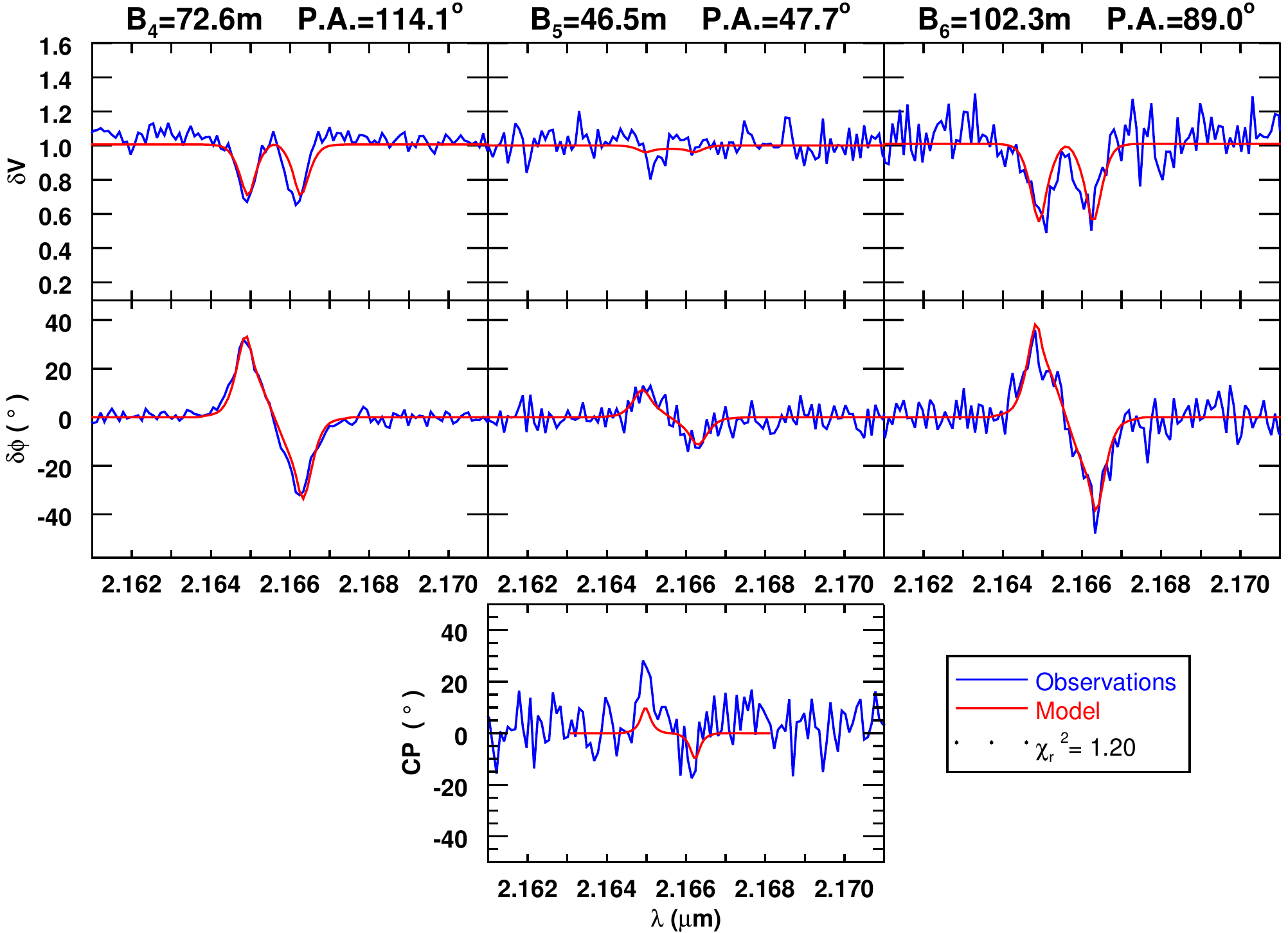}
   \includegraphics[width=0.49\hsize]{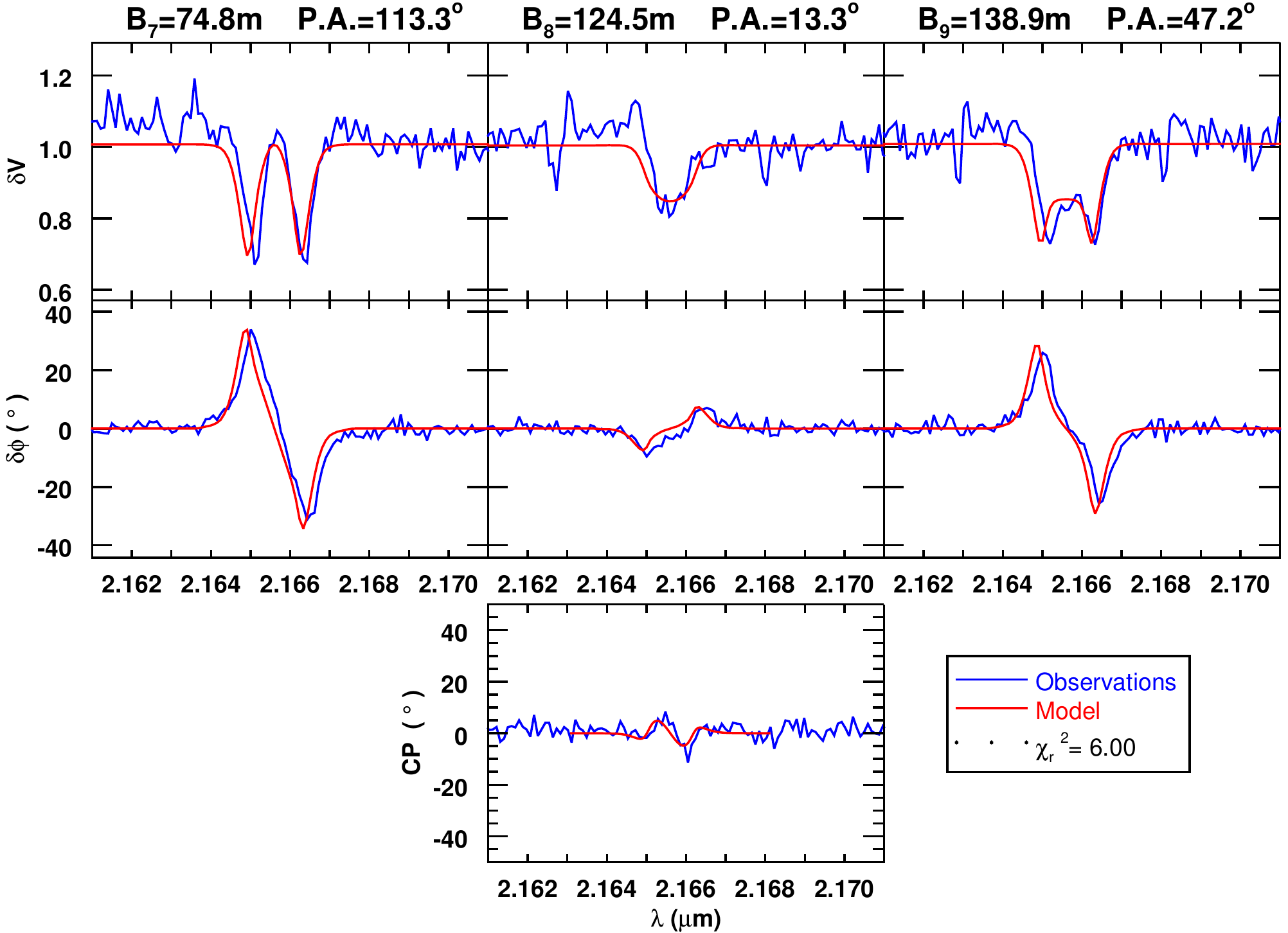}

   \includegraphics[width=0.49\hsize]{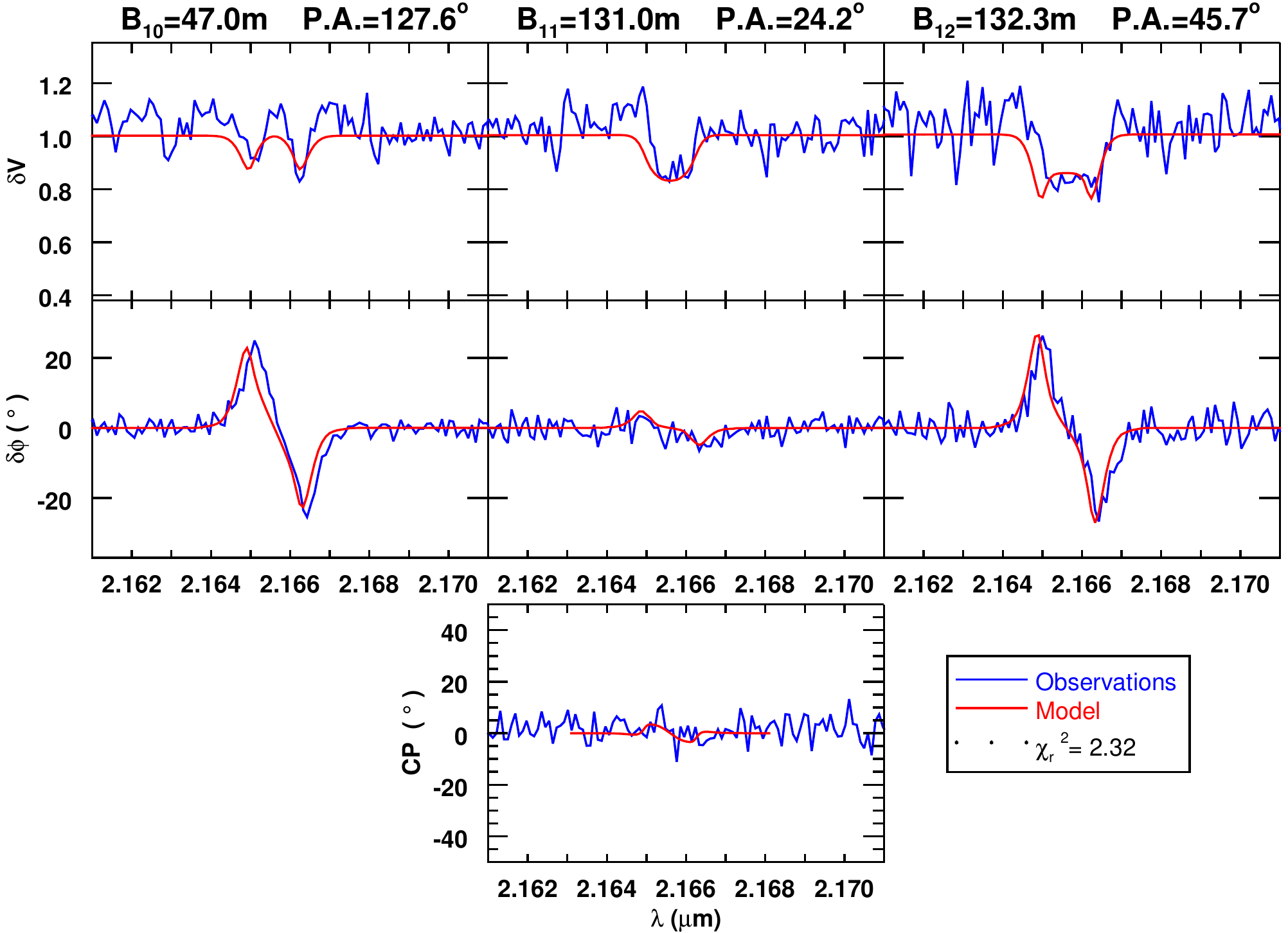}
   \includegraphics[width=0.49\hsize]{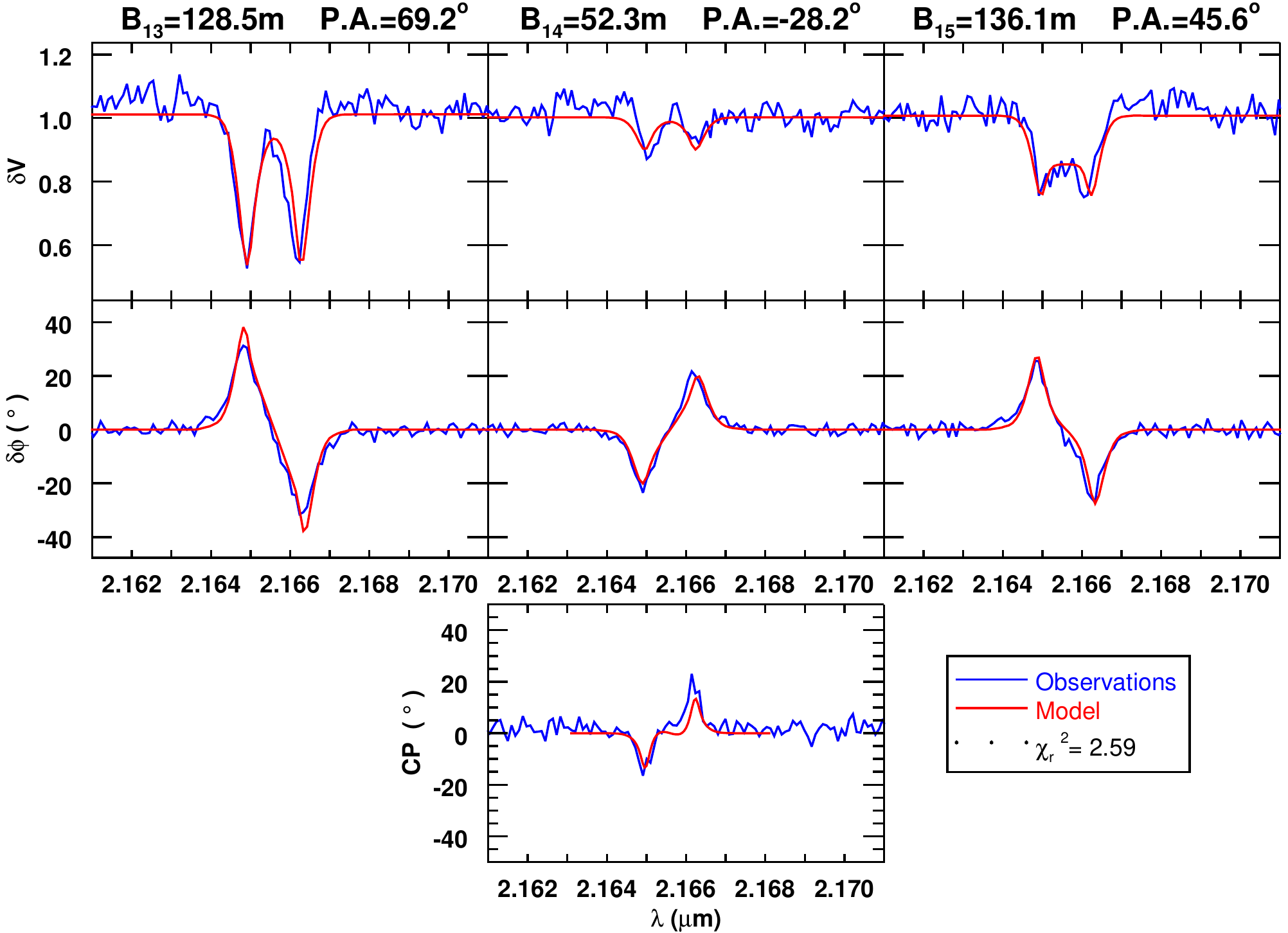}
   \caption{As in Fig.~\ref{fig:HD23630}, but for \object{HD\,209\,409}.}
   \label{fig:HD209409_1}
   \end{figure*}
   \begin{figure*}[!p]
   \centering
   \includegraphics[width=0.49\hsize]{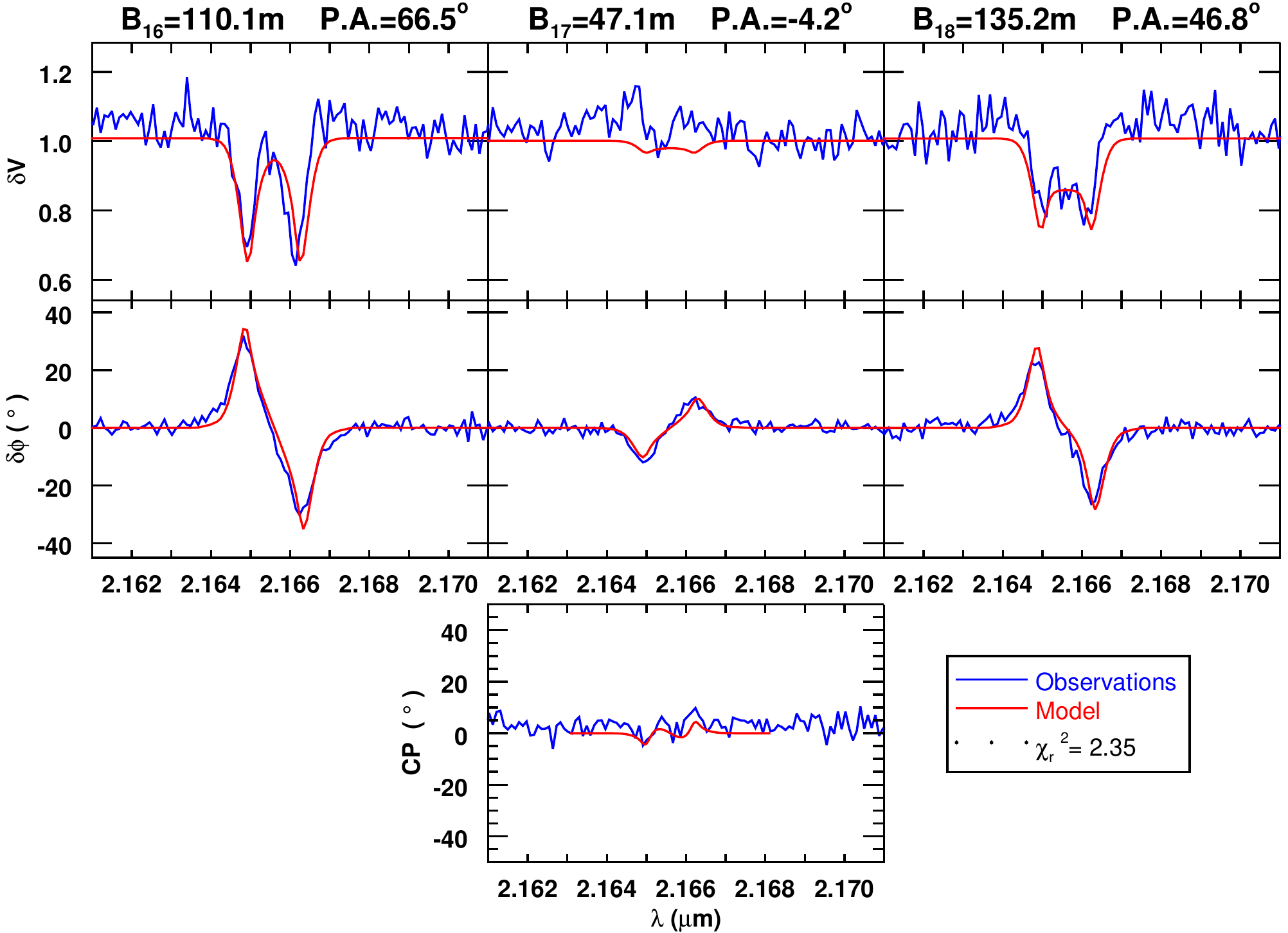}
   \includegraphics[width=0.49\hsize]{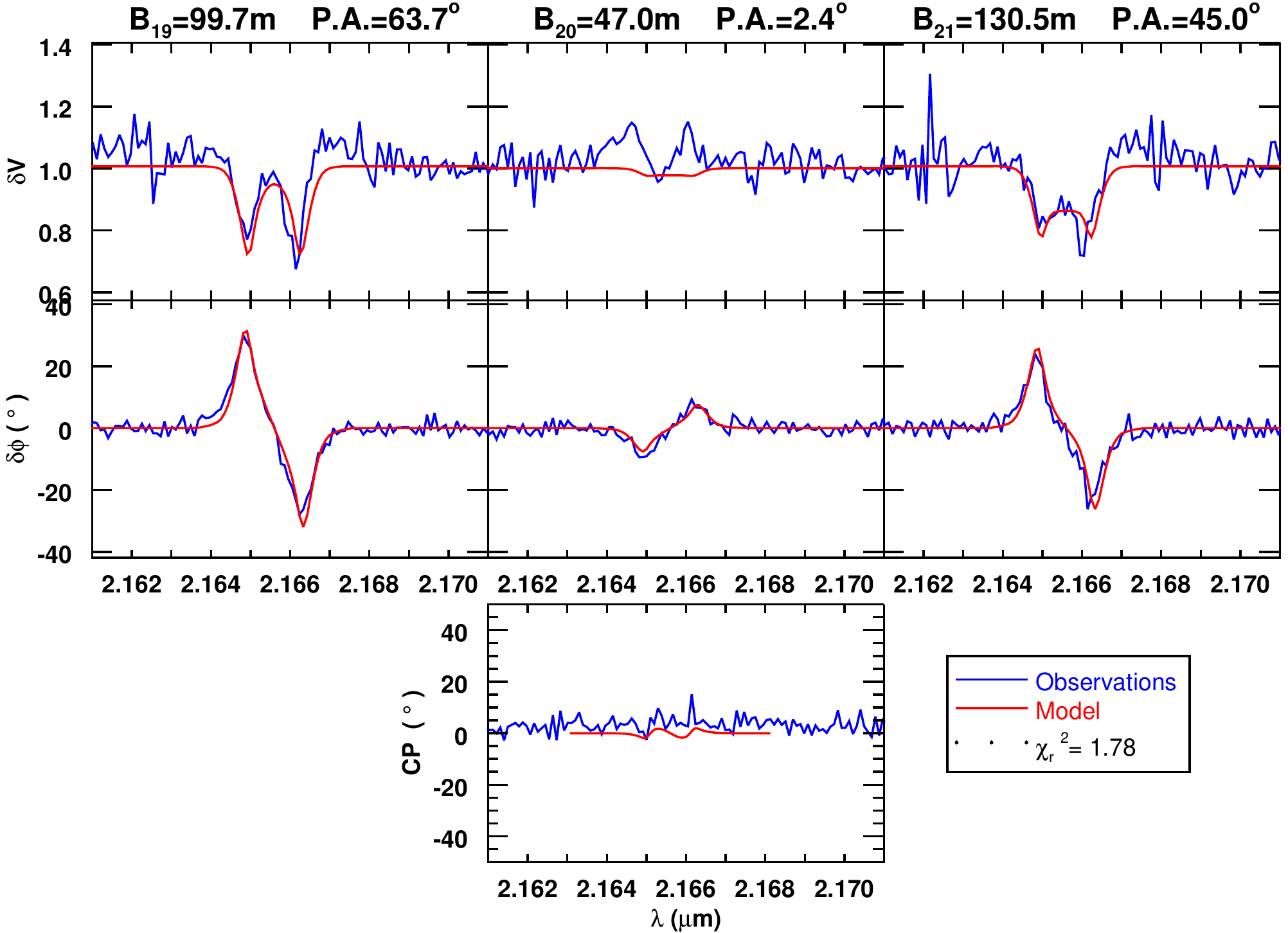}
   \caption{As in Fig.~\ref{fig:HD23630}, but for \object{HD\,209\,409} (cont.)}
   \label{fig:HD209409_2}
   \end{figure*}

   \begin{figure*}
   \centering
   \includegraphics[width=0.49\hsize]{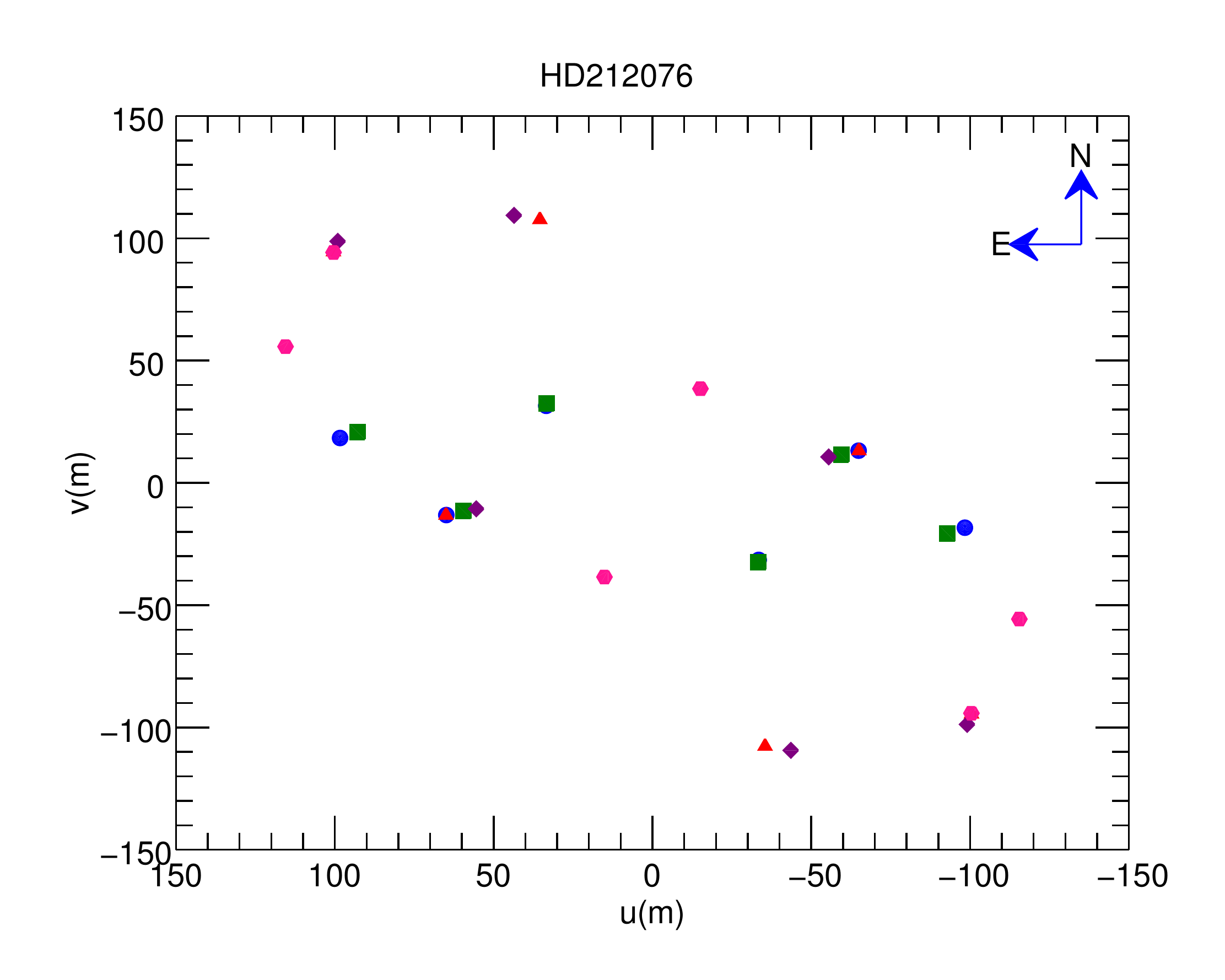}
   \includegraphics[width=0.49\hsize]{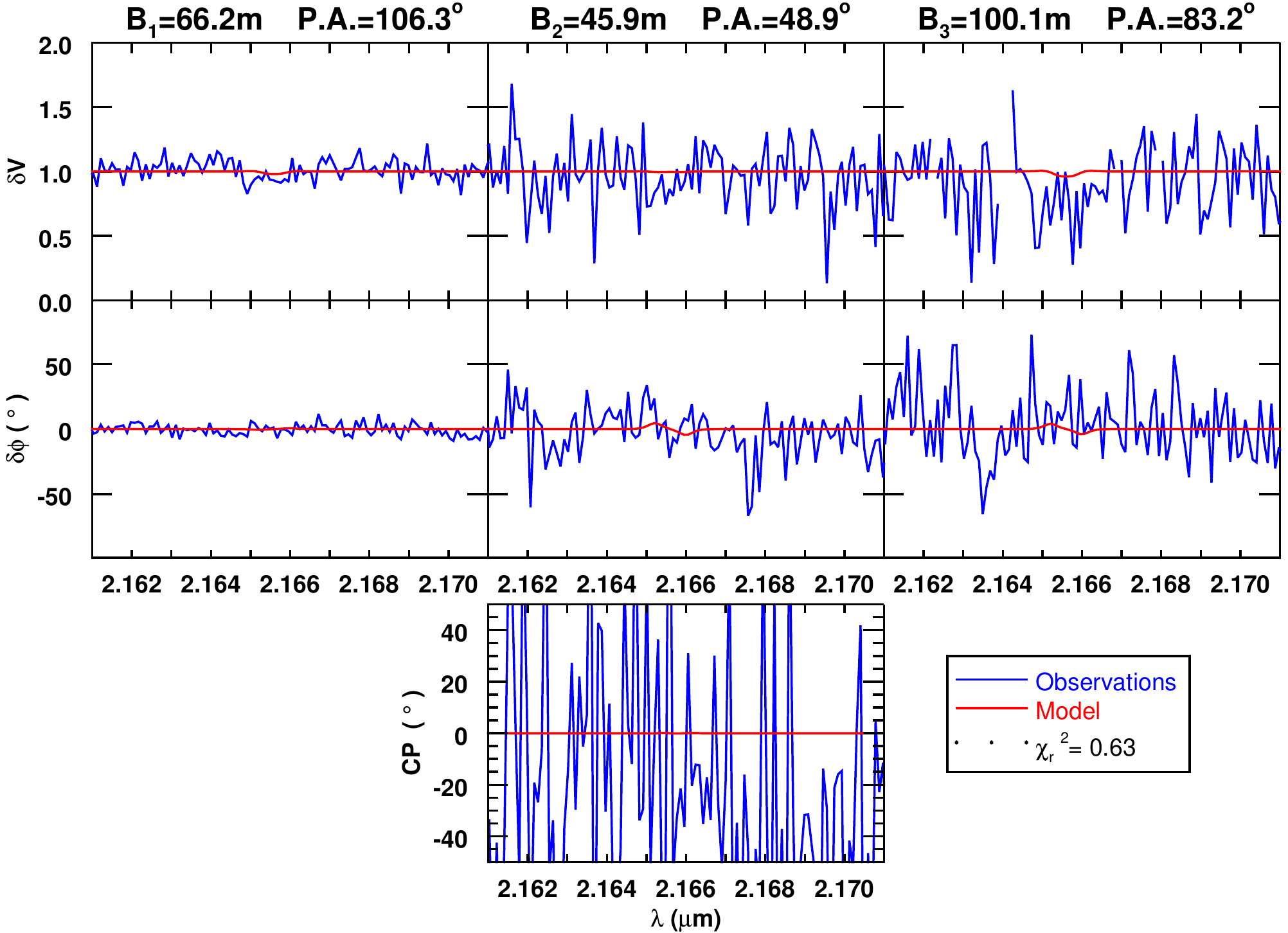}

   \includegraphics[width=0.49\hsize]{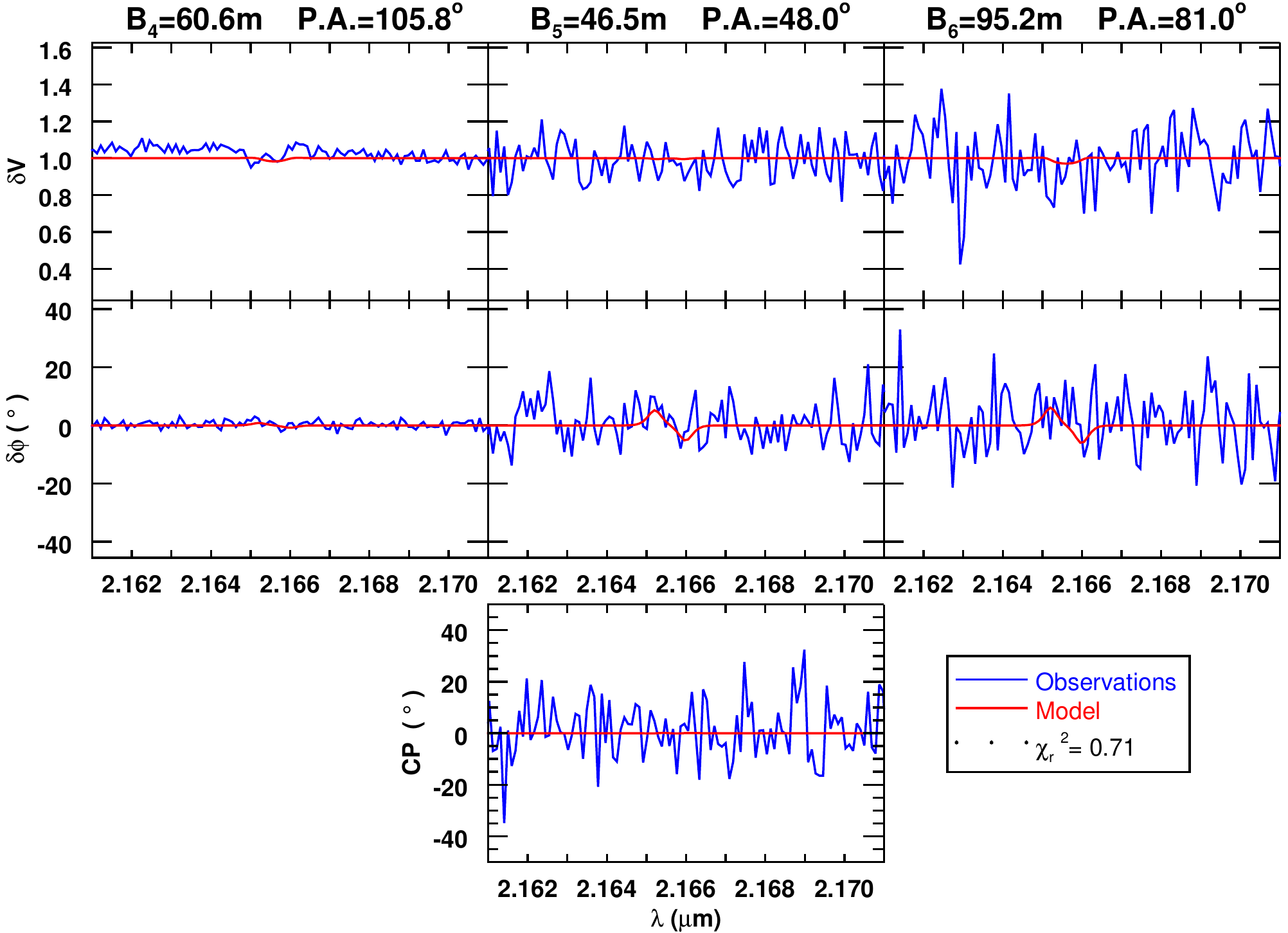}
   \includegraphics[width=0.49\hsize]{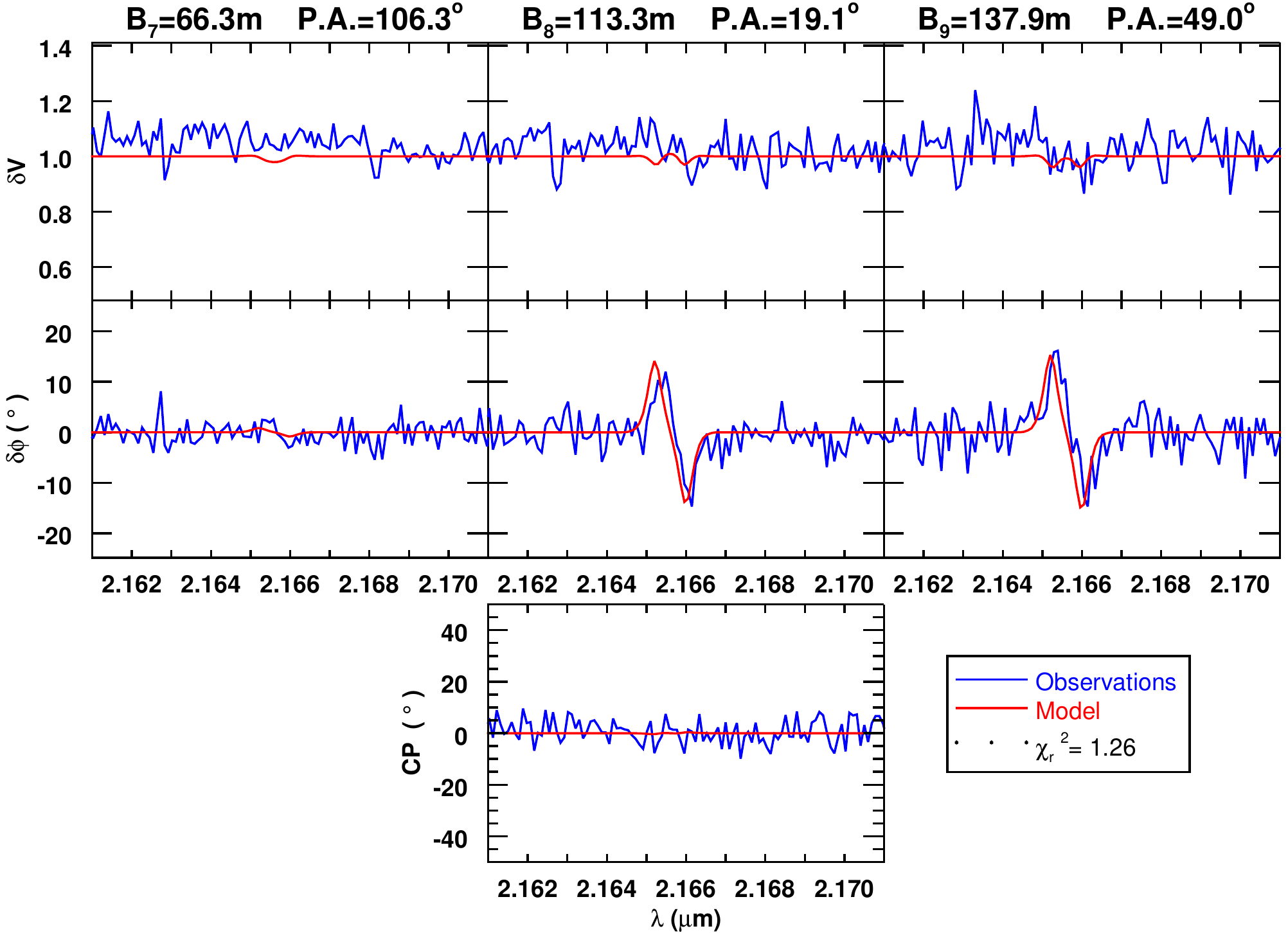}

   \includegraphics[width=0.49\hsize]{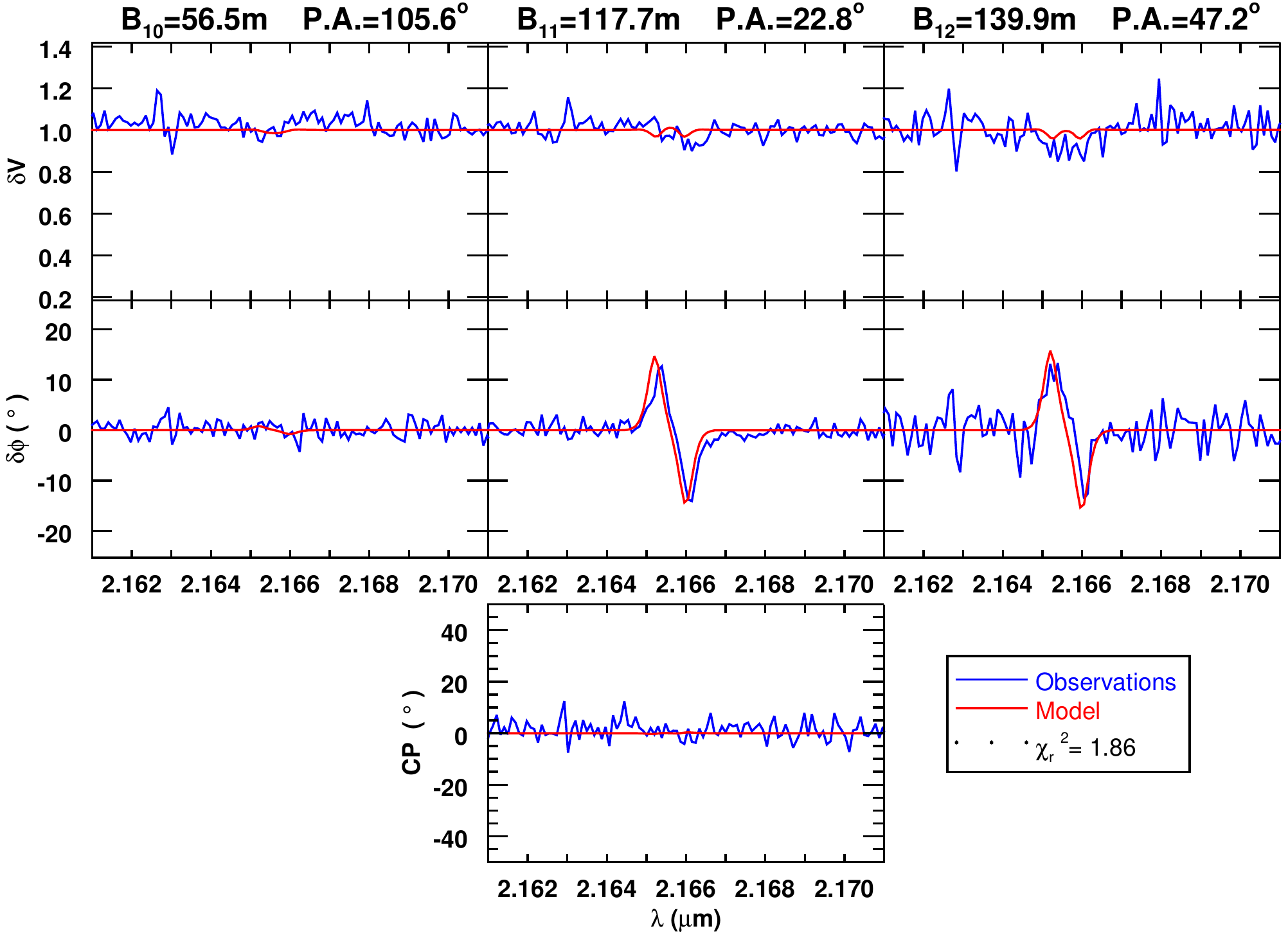}
   \includegraphics[width=0.49\hsize]{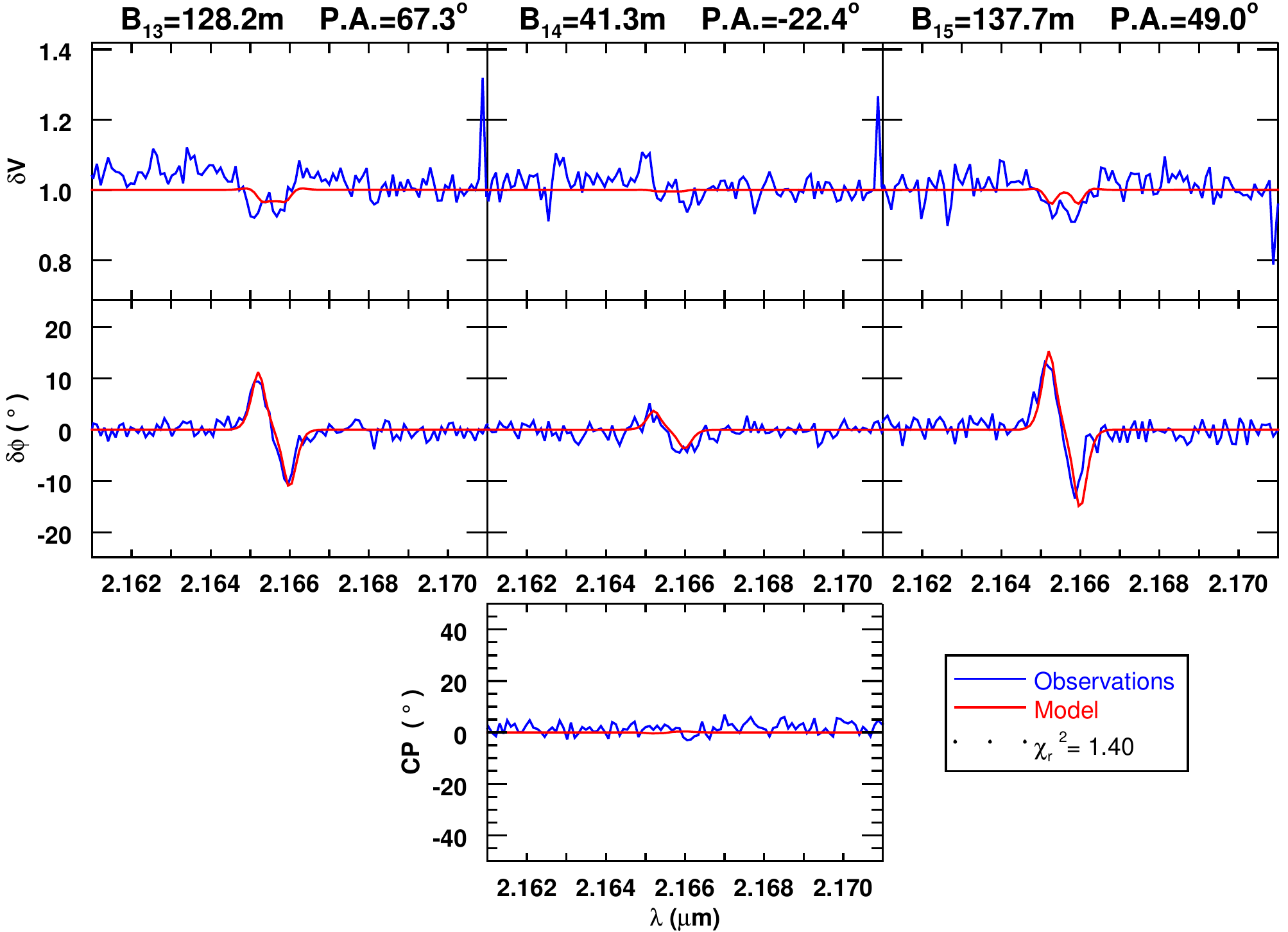}
   \caption{As in Fig.~\ref{fig:HD23630}, but for \object{HD\,212\,076}.}
   \label{fig:HD212076}
   \end{figure*}

   \begin{figure*}
   \centering
   \includegraphics[width=0.49\hsize]{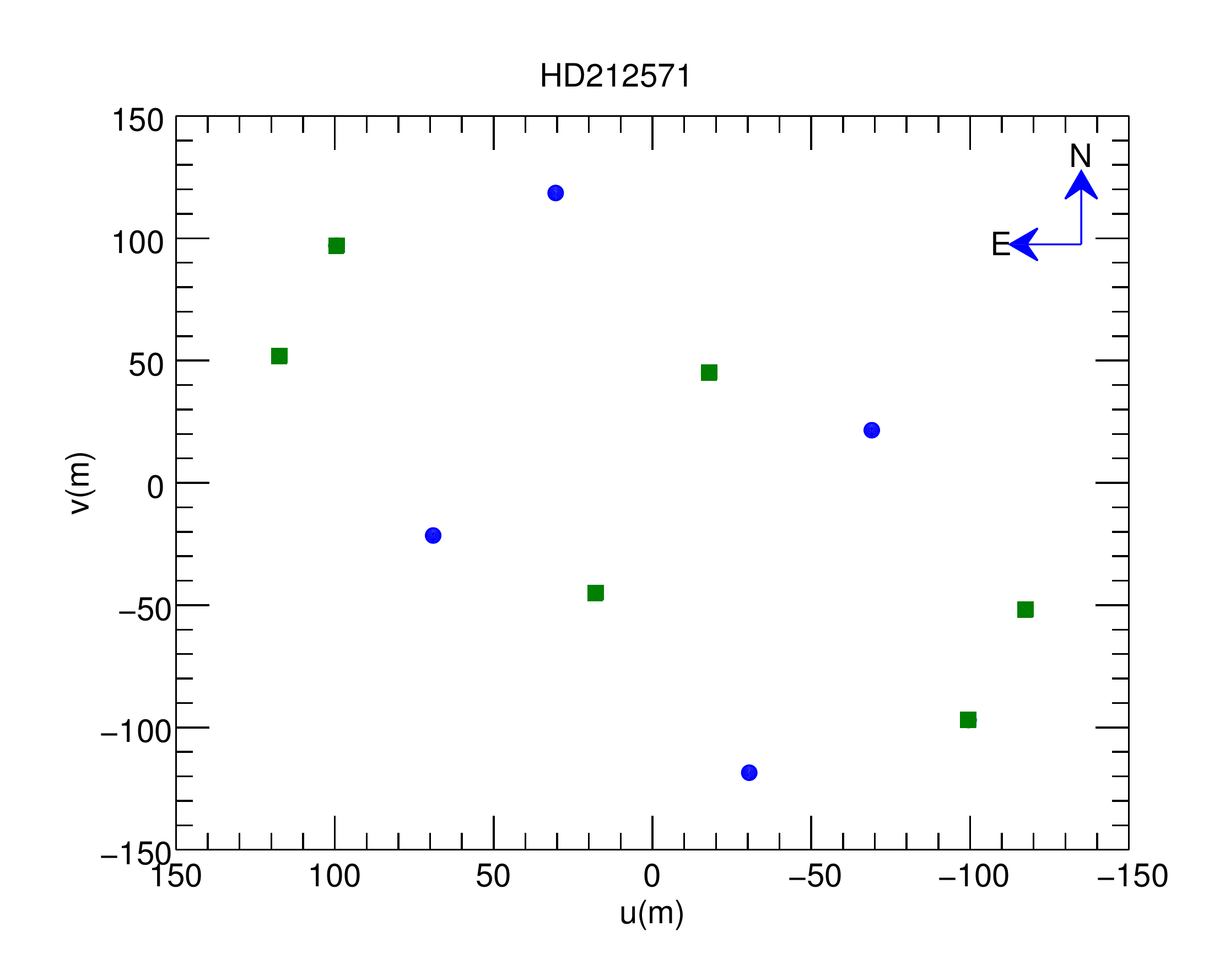}

   \includegraphics[width=0.49\hsize]{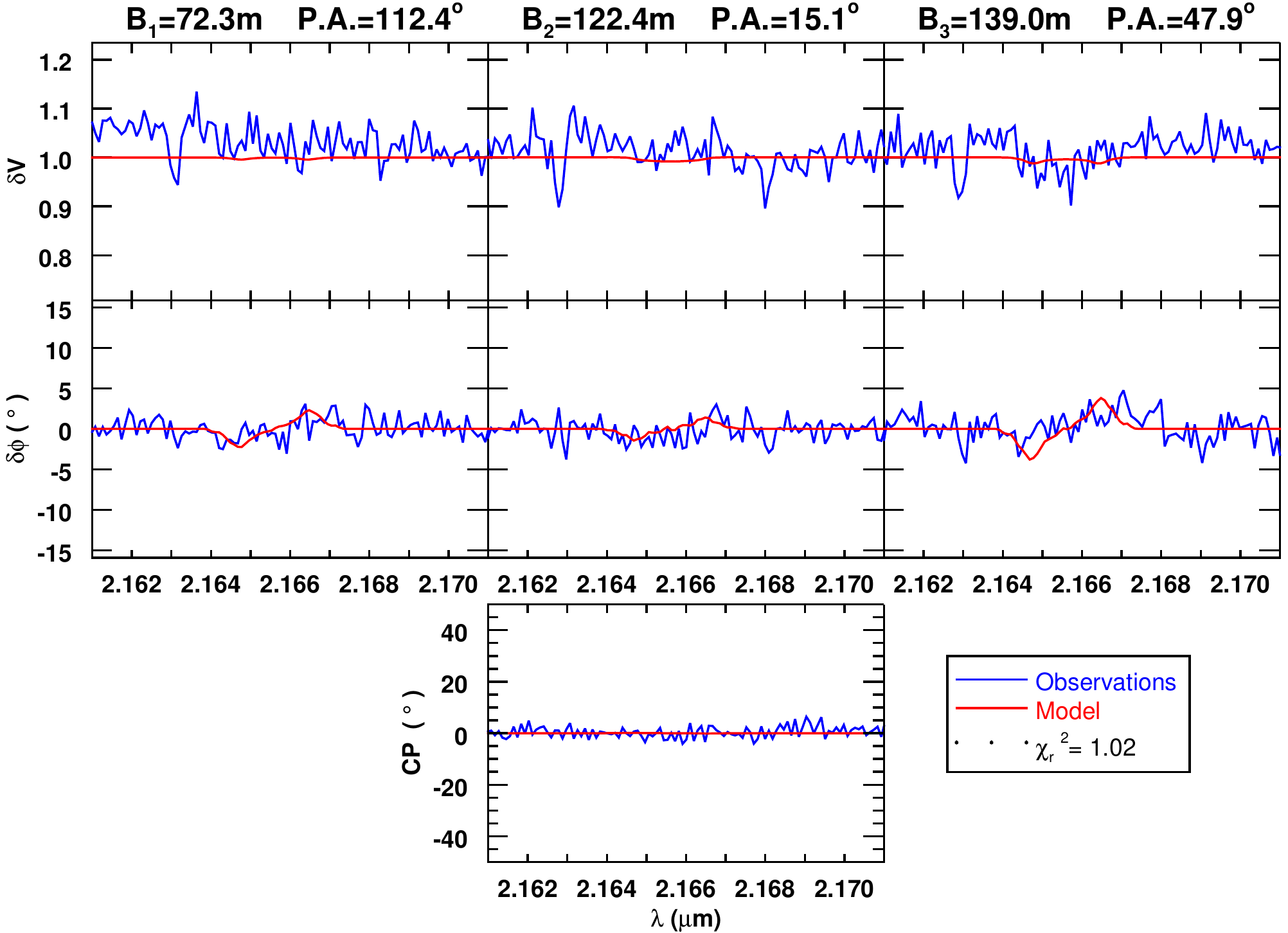}
   \includegraphics[width=0.49\hsize]{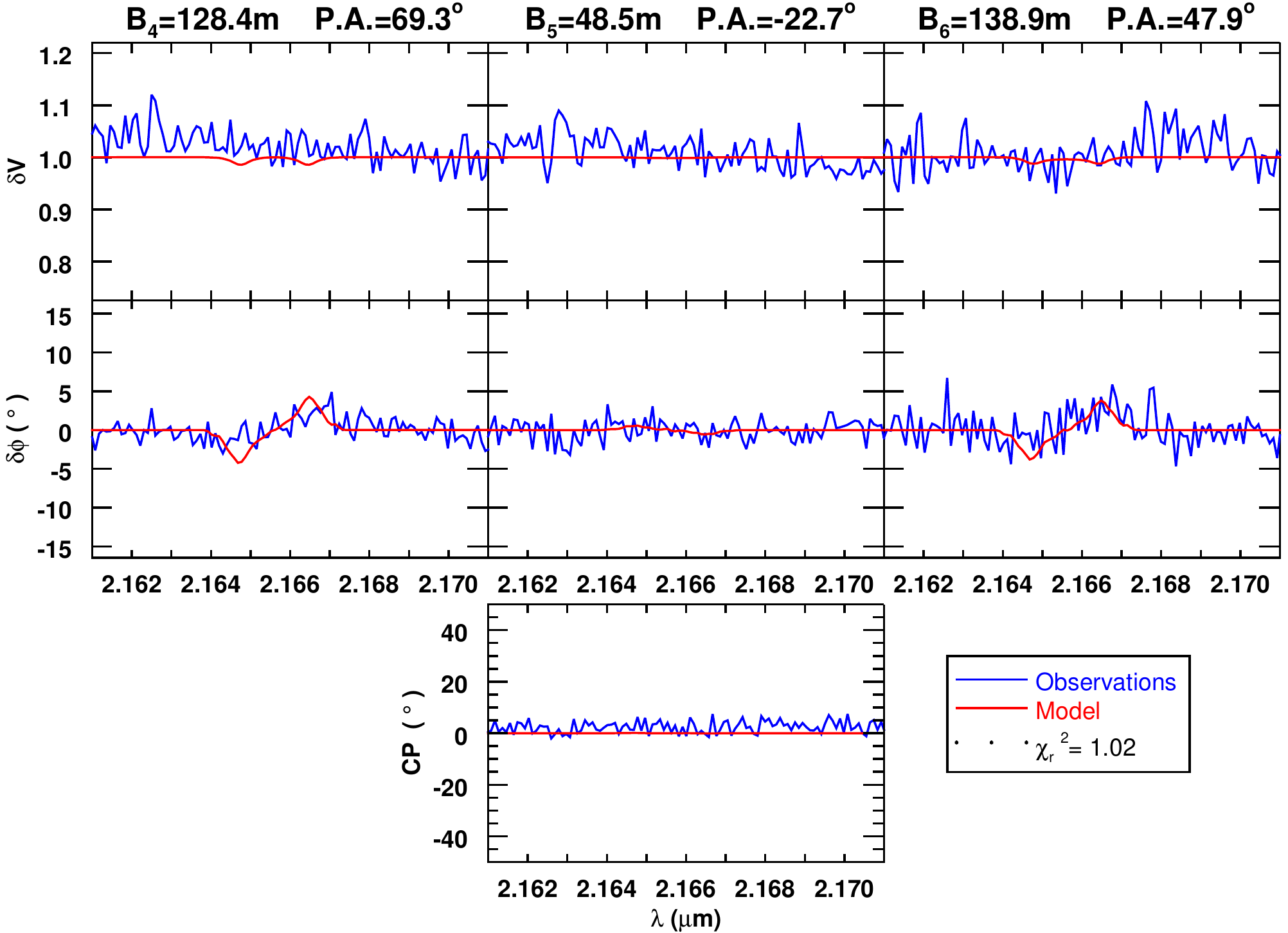}
   \caption{As in Fig.~\ref{fig:HD23630}, but for \object{HD\,212\,571}.}
   \label{fig:HD212571}
   \end{figure*}

   \begin{figure*}
   \centering
   \includegraphics[width=0.49\hsize]{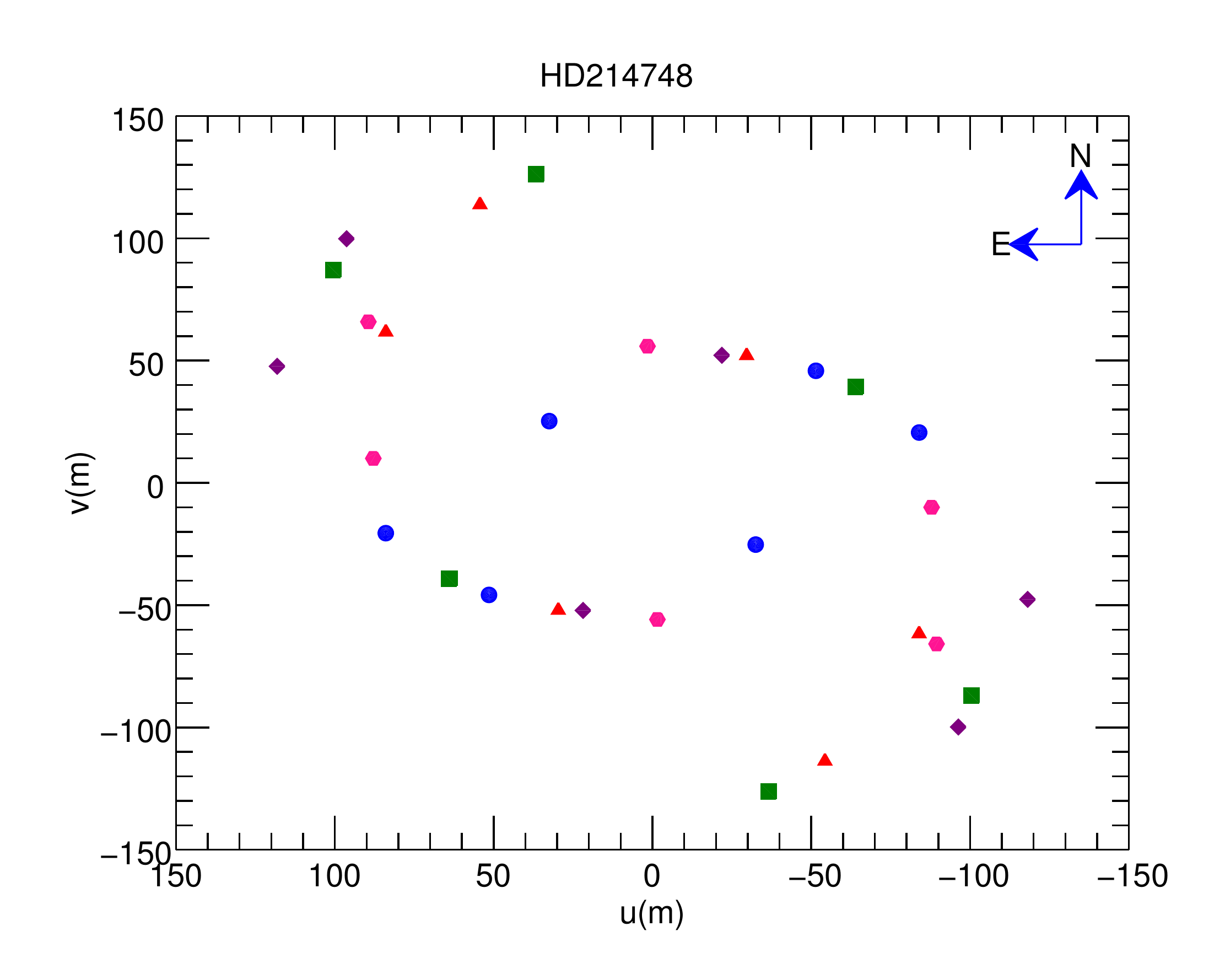}
   \includegraphics[width=0.49\hsize]{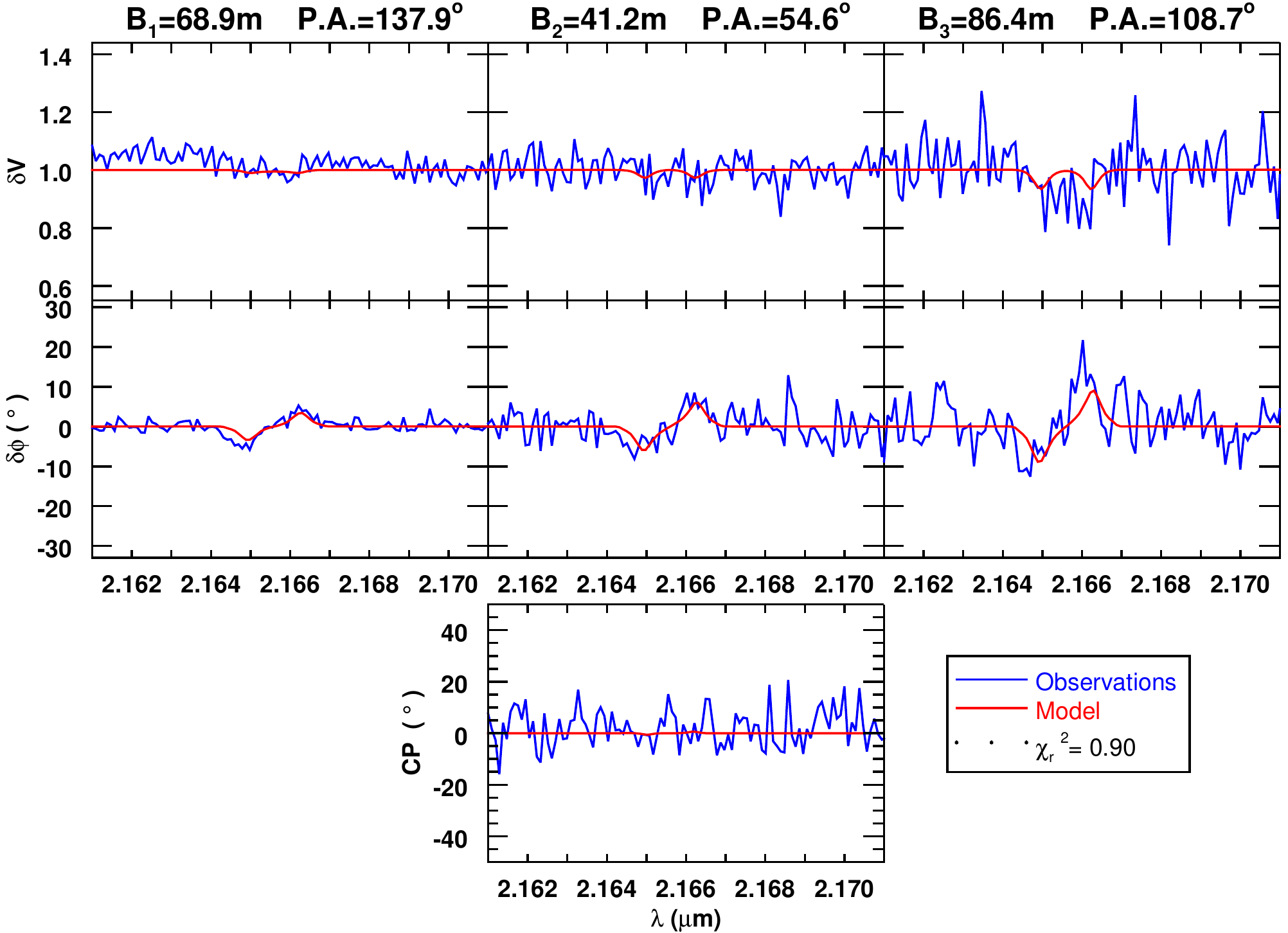}

   \includegraphics[width=0.49\hsize]{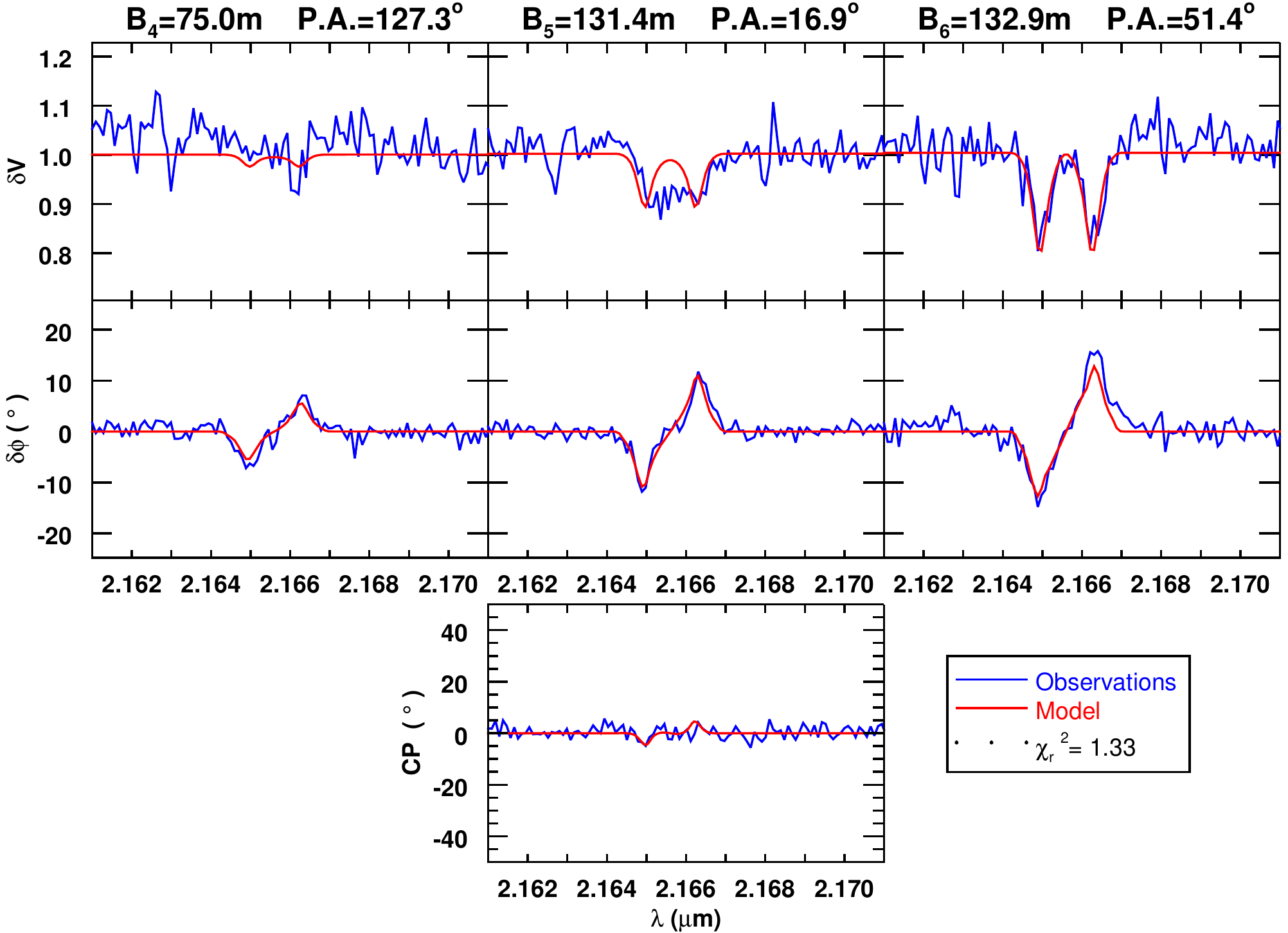}
   \includegraphics[width=0.49\hsize]{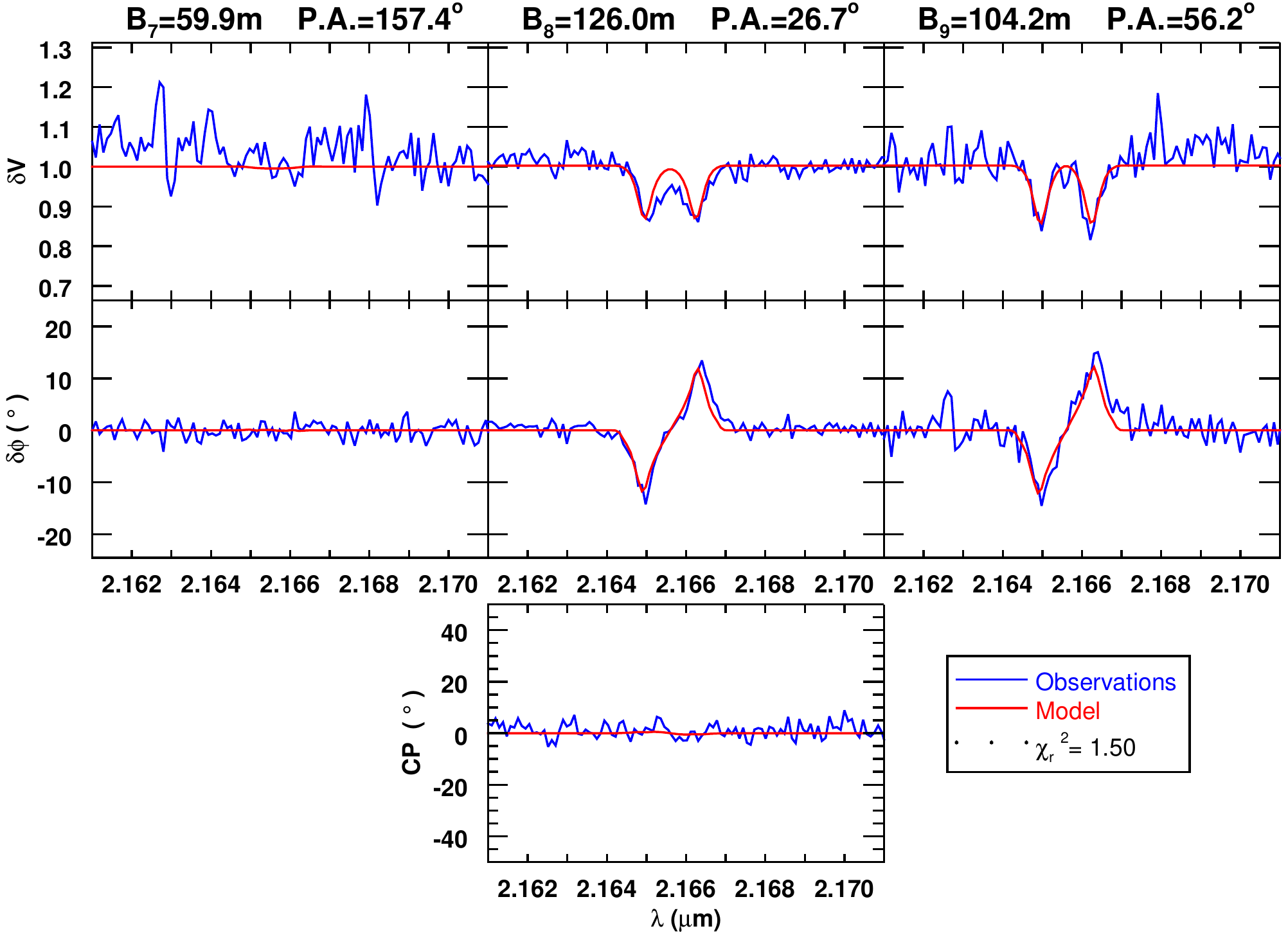}

   \includegraphics[width=0.49\hsize]{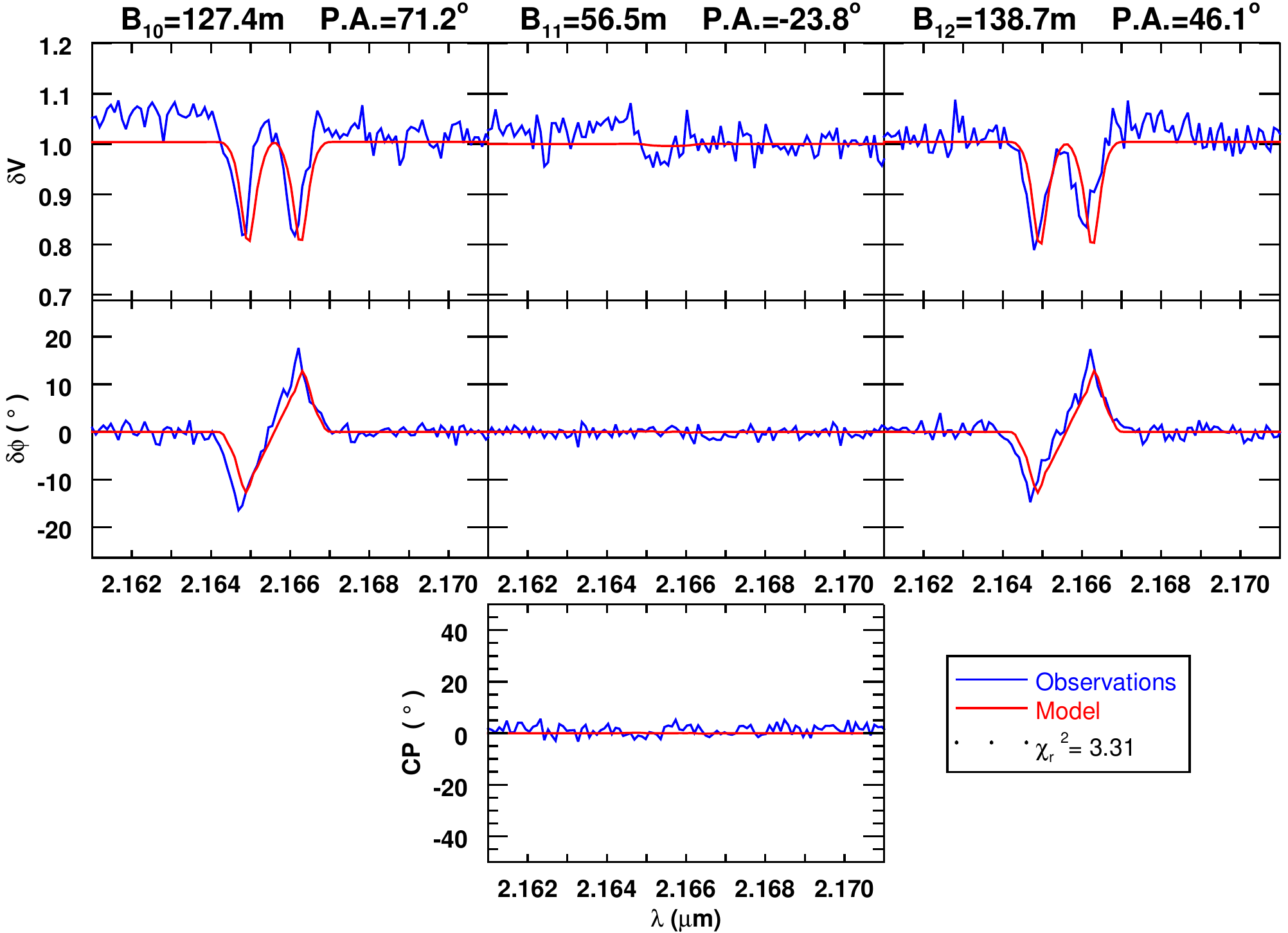}
   \includegraphics[width=0.49\hsize]{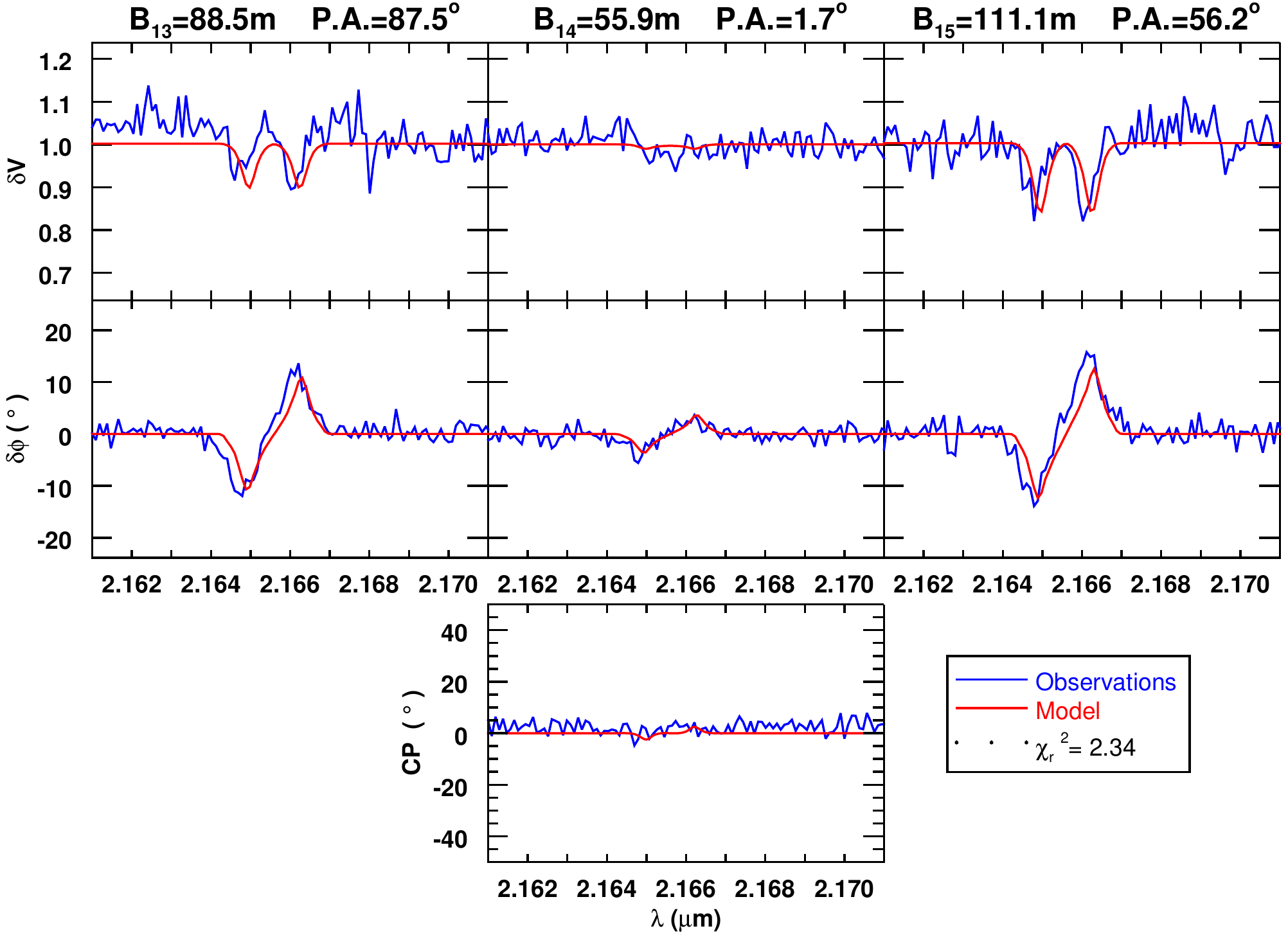}
   \caption{As in Fig.~\ref{fig:HD23630}, but for \object{HD\,214\,748}.}
   \label{fig:HD214748}
   \end{figure*}

\end{appendix}

\end{document}